\documentclass[a4paper,fleqn]{cas-dc}  % fleqn = left-aligned equations

\usepackage[numbers]{natbib}   % numbered citations
\usepackage{amsmath,amssymb,amsfonts}
\usepackage{graphicx}
\usepackage{tikz}
\usepackage{tcolorbox}
\usepackage{pdflscape}
\usepackage{rotating}
\usepackage{longtable}
\usepackage{multirow}
\usepackage{enumitem}
\usepackage{fontawesome}
\usepackage{tablefootnote}
\usepackage{colortbl}
\usepackage[export]{adjustbox}
\usepackage{framed}
\usepackage{svg}
\usepackage{subfig}
\usepackage{placeins}
\usepackage{array}
\usepackage{float}
\usepackage{hyperref}
\usepackage{balance}
\usepackage{tabularx}
\usepackage{booktabs}
\usepackage{ragged2e}
\usepackage{longtable}
\usepackage[table]{xcolor}  % enables \rowcolor in tables
\definecolor{lightgrayrow}{gray}{0.92}
\newcolumntype{L}[1]{>{\RaggedRight\arraybackslash}p{#1}}

\newcolumntype{Y}{>{\RaggedRight\arraybackslash}X}
\begin{document}
\begin{sloppypar}

\let\WriteBookmarks\relax
\def\floatpagepagefraction{1}
\def\textpagefraction{.001}

\shorttitle{Understanding the Issues in Microservices Systems}
\shortauthors{Waseem et al.}

\title[mode=title]{Understanding the Issues, Their Causes and Solutions in Microservices Systems: An Empirical Study\tnotemark[1]}

% ---- Authors ----
\author[TAU]{Muhammad Waseem}
\ead{muhammad.waseem@tuni.fi}

\author[WHU]{Peng Liang}\cormark[1]
\ead{liangp@whu.edu.cn}

\author[LUL]{Aakash Ahmad}
\ead{a.abbasi@derby.ac.uk}

\author[OULU]{Arif Ali Khan}
\ead{arif.khan@oulu.fi}

\author[RMIT]{Mojtaba Shahin}
\ead{mojtaba.shahin@rmit.edu.au}

\author[WHU]{Ali Rezaei Nasab}
\ead{rezaei.ali.nasab@gmail.com}

\author[JYU]{Tommi Mikkonen}
\ead{tommi.j.mikkonen@jyu.fi}

\author[TAU]{Pekka Abrahamsson}
\ead{pekka.abrahamsson@tuni.fi}

% ---- Affiliations ----
\affiliation[TAU]{organization={Faculty of Information Technology and Communication Sciences, Tampere University},
  city={Tampere}, country={Finland}}
\affiliation[WHU]{organization={School of Computer Science, Wuhan University},
  city={Wuhan}, country={China}}
\affiliation[LUL]{organization={University of Derby, College of Science and Engineering, School of Computing},
  city={Derby}, country={United Kingdom}}
\affiliation[OULU]{organization={M3S Empirical Software Engineering Research Unit, University of Oulu},
  city={Oulu}, country={Finland}}
\affiliation[RMIT]{organization={School of Computing Technologies, RMIT University},
  city={Melbourne}, country={Australia}}
\affiliation[JYU]{organization={Faculty of Information Technology, University of Jyväskylä},
  city={Jyväskylä}, country={Finland}}

% ---- Notes ----
\tnotetext[1]{Submitted to the \textit{Journal of Systems and Software} (JSS).}
\cortext[1]{Corresponding author: Peng Liang (\href{mailto:liangp@whu.edu.cn}{liangp@whu.edu.cn}).}

% ---- Abstract & Keywords ----
\begin{abstract}
Many small to large organizations have adopted the Microservices Architecture (MSA) style to develop and deliver their core businesses. Despite the popularity of MSA in the software industry, there is a limited evidence-based and thorough understanding of the types of issues (e.g., errors, faults, failures, and bugs) that microservices system developers experience, the causes of the issues, and the solutions as potential fixing strategies to address the issues. To ameliorate this gap, we conducted a mixed-methods empirical study that collected data from 2,641 issues from the issue tracking systems of 15 open-source microservices systems on GitHub, 15 interviews, and an online survey completed by 150 practitioners from 42 countries across 6 continents. Our analysis led to comprehensive taxonomies for the issues, causes, and solutions. The findings of this study inform that Technical Debt, Continuous Integration and Delivery, Exception Handling, Service Execution and Communication, and Security are the most dominant issues in microservices systems. Furthermore, General Programming Errors, Missing Features and Artifacts, and Invalid Configuration and Communication are the main causes behind the issues. Finally, we found 177 types of solutions that can be applied to fix the identified issues. \textcolor{black}{Based on our study results, we propose a future research framework that outlines key problem dimensions and actionable study strategies to support the engineering of emergent and next-generation microservices systems.}
\end{abstract}


\maketitle

% ---- Main matter ----
\section{Introduction}
\label{sec:introduction}
The software industry has recently witnessed the growing popularity of the Microservices Architecture (MSA) style as a promising design approach to develop applications that consist of multiple small, manageable, and independently deployable services \cite{fowler2014microservices}. Software development organizations may have adopted or planned to use the MSA style for various reasons. Specifically, some of them want to increase the scalability of applications using the MSA style, while others use it to quickly release new products and services to the customers, whereas it is argued that the MSA style can also help build autonomous development teams \cite{Davide2017Processes, jamshidi2018microservices}. From an architectural perspective, a microservices system (a system that adopts the MSA style) entails a significant degree of complexity both at the design phase as well as at runtime configuration \cite{newman2020building}. This implies that the MSA style brings unique challenges for software organizations, and many quality attributes may be (positively or negatively) influenced \cite{waseem2020systematic}. For example, service level security may be impacted because microservices are developed and deployed by various technologies (e.g., Docker containers \cite{combe2016docker}) and tools that are potentially vulnerable to security attacks \cite{newman2020building, yu2019survey}. Data management is also influenced because each microservice needs to own its domain data and logic \cite{wagner2018net}. This can, for example, challenge achieving and managing data consistency across multiple microservices.

Zimmermann argues that MSA is not entirely new from Service-Oriented Architecture (SOA) (e.g., ``microservices constitute one particular implementation approach to SOA -- service development and deployment'') \cite{zimmermann2017microservices}. Similarly, Márquez and Astudillo discovered that some existing design rationale and patterns from SOA fit the context for MSA~\cite{1-marquez2018actual}. However, an important body of literature (e.g., \cite{esposito2016challenges, jamshidi2018microservices}) has concluded that there are overwhelming differences between microservices systems, monolithic systems, and traditional service-oriented systems in terms of design, implementation, testing, and deployment. Gupta and Palvankar indicated that even having SOA experience can lead to suboptimal decisions (e.g., excessive service calls) in microservices systems \cite{Dinkar2020Pitfalls}. Hence, microservices systems may have an \textit{additional} and \textit{specific} set of \textbf{issues}. Borrowing the idea from \cite{ren2020understanding}, we define \textbf{issues} in this study as errors, faults, failures, and bugs that occur in a microservices system and consequently impact its quality and functionality. Hence, there is a need to leverage existing methods or derive new practices, techniques, and tools to address the specific and additional issues in microservices systems.

Recently, a number of studies have investigated particular issues (e.g., code smell~\cite{taibi2018definition}, debugging~\cite{ZhouIEEE}, performance~\cite{WuNoms}) in microservices systems. Despite these efforts, there is no in-depth and comprehensive study on the nature of different types of issues that microservices developers face, the potential causes of these issues, and possible fixing strategies for these issues. Jamshidi \textit{et al}. believed that this can be partially attributed to the fact that researchers have limited access to industry-scale microservices systems~\cite{jamshidi2018microservices}. The empirical knowledge on the nature of issues occurring in microservices systems can be useful from the following perspectives: (i) understanding common issues in the design and implementation of microservices systems and how to avoid them, (ii) identifying trends in the types of issues that arise in microservices systems and how to address them effectively, (iii) experienced microservices developers can be allocated to address the most frequent and challenging issues, (iv) novice microservices developers can quickly be informed of empirically-justified issues and avoid common mistakes, and (v) the industry and academic communities can synergies theory and practice to develop tools and techniques for the frequently reported issues in microservices systems.

\begin{figure*}[!htbp]
 \centering
 \includegraphics[scale=0.7]{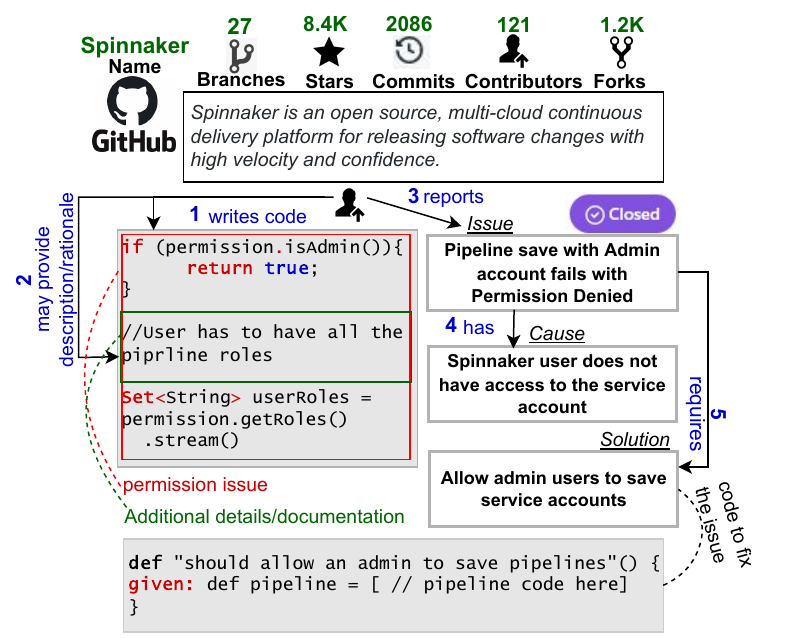}
 \caption{An example of the issue, cause, and their solution}
 \label{fig:IssuesBackground}
\end{figure*}

\textbf{Motivating Example}: We now contextualize the issues, causes, and solutions based on an example illustrated in Figure \ref{fig:IssuesBackground}. The example is taken from the Spinnaker project, an open-source microservices project hosted on GitHub (see Table \ref{tab:selectedProjects}), and annotated with numbering to represent a sequence among the reported issue, its cause(s), and the solution(s) to resolve the issue. Figure~\ref{fig:IssuesBackground} shows auxiliary information about the Spinnaker project, such as project description, stars, and contributors. As shown in the example, a contributor, typically a microservices developer, writes code and may provide additional details of the code through comments. Once the code is compiled, the contributor reports permission denied \textit{issue} highlighting ``\textit{pipeline save with Admin account fails with permission denied}''. As the next step, the same or other contributor highlights the \textit{cause} for such issue as ``\textit{Spinnaker user does not have access to the service account}''. As the last step, an individual or a community of developers provides a \textit{solution} such as ``\textit{Allow the admin users to save the accounts}'' to resolve the issue. We are only interested in analyzing issues that have been marked as closed to ensure that a solution to resolve the issue exists. As shown in Table \ref{tab:selectedProjects}, the Spinnaker project has 4,595 closed issues and 121 contributors.

This work aims to \textit{systematically and comprehensively study and categorize the issues that developers face in developing microservices systems, the causes of the issues, and the solutions (if any)}. To this end, we conducted a mixed-methods empirical study following the guideline proposed by Easterbrook and his colleagues \cite{easterbrook2008selecting}. We collected the data by (i) mining 2,641 issues from the issue tracking systems of 15 open-source microservices systems on GitHub, (ii) conducting 15 interviews, and (iii) deploying an online survey completed by 150 practitioners to develop the taxonomies of the issues, their causes and solutions in microservices systems. The key findings of this study are:
\begin{enumerate}
 \item The \textit{issue} taxonomy consists of 19 categories, 54 subcategories, and 402 types, indicating the diversity of issues in microservices systems. The top three categories of issues are Technical Debt, Continuous Integration and Delivery, and Exception Handling.
 \item The \textit{cause} taxonomy consists of 8 categories, 26 subcategories, and 228 types, in which General Programming Errors, Missing Features and Artifacts, and Invalid Configuration and Communication are the most frequently reported causes.
 \item The \textit{solution} taxonomy consists of 8 categories, 32 subcategories, and 177 types of solutions, in which the top three categories of solutions for microservices issues are Fix Artifacts, Add Artifacts, and Modify Artifacts.
 \item The overall \textit{survey findings} confirm the taxonomies of the issues, their causes and solutions in microservices systems and also indicate no major statistically significant differences in practitioners’ perspectives on the developed taxonomies.
\end{enumerate} 

The empirically validated taxonomies of issues, causes, and solutions in this study may greatly benefit software developers. For instance, developers can gain valuable insights into the diverse range of issues that they may encounter. These taxonomies can also help developers understand the causes of issues and implement preventive measures, enabling them to make informed decisions about addressing the issues during the development process. Furthermore, the proposed taxonomies in this study can facilitate knowledge sharing, training, and ensure a shared understanding among developers about the issues, causes, and solutions in microservices systems.

This paper has extended our previous work \cite{waseem2021nature} by adding two new research questions (\textbf{RQ3} and \textbf{RQ4}) and expanding and enhancing the results of \textbf{RQ1} and \textbf{RQ2} with increased volume and variety of data and a mixed-methods approach. Specifically, we explored 10 more open-source microservices systems on GitHub (now 15 projects), interviewed 15 practitioners, and conducted an online survey with 150 microservices practitioners for getting their perspectives on the proposed taxonomies of the issues, their causes and solutions in microservices systems, as well as the mapping between the issues, causes, and solutions.

Our study makes the following key \textbf{contributions}: 
\begin{enumerate}
 
 \item We developed the taxonomies of the issues, their causes and solutions in microservices systems based on a qualitative and quantitative analysis of 2,641 issue discussions among developers on GitHub, 15 interviews, and an online survey completed by 150 practitioners.
 
 \item We provided the mapping between the issues, causes, and solutions in microservices systems with promising research directions on microservices systems that require more attention.

 \item \textcolor{black}{We proposed a future research framework that synthesizes the major problem dimensions and outlines actionable study strategies (Table \ref{tab:future_framework}), thereby translating our empirical findings into a concrete research agenda for the microservices community.}

 \item We made the dataset of this study available online \cite{replpack}, which includes the data collection and analysis from GitHub and microservices practitioners, as well as detailed hierarchies of the taxonomies of issues, causes, and solutions, to enable replication of this study and conduct future research.
\end{enumerate}

The remainder of the paper is structured as follows. Section \ref{sec:methodology} details the research methodology employed. Section \ref{sec:results} presents the results of our study. Section \ref{Discussion} discusses the relationship between the issues, causes, and solutions, along with the implications and \textcolor{black}{evidence-driven future research agenda}. Section \ref{sec:threats} clarifies the threats to the validity of our results. Section \ref{RelatedWork} reviews related work, and Section \ref{sec:conclusion} draws conclusions and outlines avenues for future work.

\section{Methodology}
\label{sec:methodology}
\begin{figure*}[!htbp]
 \centering
 \includegraphics[width=0.51\textwidth]{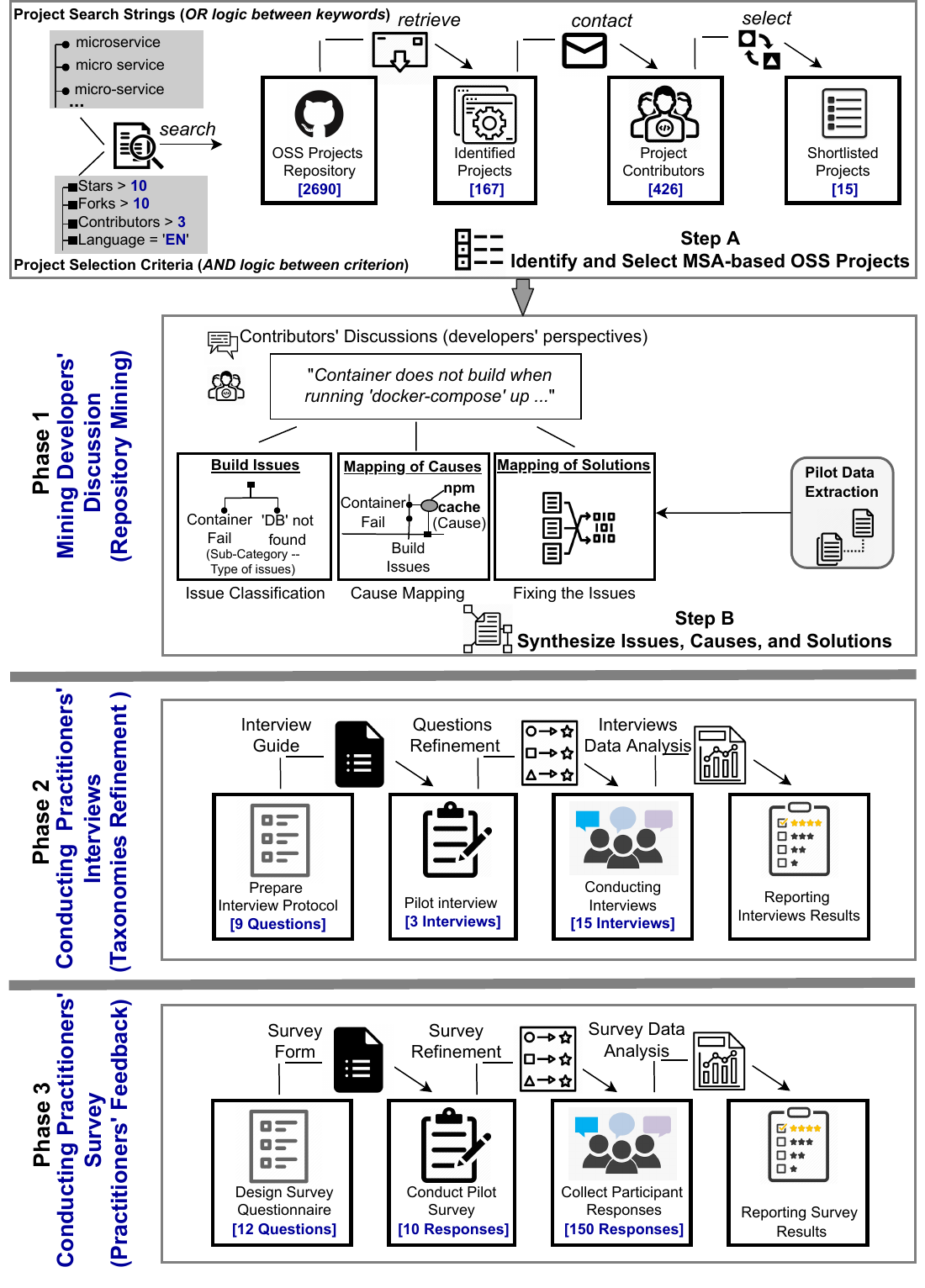}
 \caption{An overview of the research method}
 \label{fig:researchmethod}
\end{figure*}

The research methodology of this study consists of three phases, as illustrated in Figure \ref{fig:researchmethod}. Given the nature of this research and the formulated research questions (see Section \ref{RQs}) – issues, causes, and solutions in microservices projects, we decided to use a mixed-methods study. Our study collected data from microservices projects hosted on GitHub, interviews, and a web-based survey. During \textbf{Phase 1}, we derived the taxonomies of 386 types of issues, 217 types of causes, and 177 types of solutions by mining and analyzing microservices practitioners’ discussions in the issue tracking systems of 15 open-source microservices projects hosted on GitHub. During \textbf{Phase 2}, we interviewed 15 microservices practitioners to extend and verify the taxonomies and identified additional 14 types of issues, 20 types of causes, and 22 types of solutions. During \textbf{Phase 3}, we surveyed 150 microservices practitioners using a Web-based survey to validate the outcomes of Phase 1 and Phase 2, i.e., using practitioners’ perspectives and feedback to validate the extracted types of issues, their causes and solutions.

\subsection{Research Questions}
\label{RQs}
We formulated the following research questions (RQs).

\begin{tcolorbox} [sharp corners, boxrule=0.1mm,]
\small
\textbf{RQ1}: What issues do occur in the development of microservice systems?
\end{tcolorbox}

\textbf{\underline{Rationale}}: \textbf{RQ1} aims to systematically identify and taxonomically classify the types of issues that occur in microservices systems. The answer to \textbf{RQ1} provides a comprehensive understanding of the issues (e.g., the most frequent issues) of microservices systems.

\begin{tcolorbox}[sharp corners, boxrule=0.1mm,]
\textbf{RQ2}: What are the causes of issues that occur in microservices systems?
\end{tcolorbox}
\textbf{\underline{Rationale}}: The aim of \textbf{RQ2} is to investigate and classify the root causes behind the issues identified in RQ1 and map causes to issues. The answer to \textbf{RQ2} helps practitioners avoid common issues in microservices systems.

\begin{tcolorbox}[sharp corners, boxrule=0.1mm,]
\textbf{RQ3}: What solutions are proposed to fix issues that occur in microservices systems?
\end{tcolorbox}
\textbf{\underline{Rationale}}: The aim of \textbf{RQ3} is to identify the solutions for the issues according to their causes and to develop the taxonomy of solutions. The answer to \textbf{RQ3} helps to understand the fixing strategies for addressing microservices issues.

\begin{tcolorbox}[sharp corners, boxrule=0.1mm,]
\textbf{RQ4}: What are the practitioners' perspectives on the taxonomies of the identified issues, causes, and solutions in microservices systems?
\end{tcolorbox}
\textbf{\underline{Rationale}}: The taxonomies of issues, causes, and solutions developed in RQ1, RQ2, and RQ3 are based on 15 open-source microservices systems and interviewing 15 practitioners. \textbf{RQ4} aims to evaluate the taxonomies of issues, causes, and solutions built in RQ1, RQ2, and RQ3 by conducting a relatively large-scale online survey.

\subsection {Phase 1 - Mining Developer Discussions}
\label{sec:RMPhase1}
This phase aims to systematically identify and synthesize the issues, their causes and solutions in open-source microservices systems on GitHub. For an objective and fine granular presentation of methodological details, this phase is divided into two steps, each elaborated below, based on the illustrative view in Figure \ref{fig:researchmethod}.

\subsubsection{Step A – Identify \& Select MSA-based OSS Projects}
The specified RQs require us to identify and select MSA-based OSS projects, representing a repository of developer discussions and knowledge, to extract the issues, their causes, and solutions. This means that the RQs guided the development of search strings based on the recommendations and steps from \cite{SystematicSearchMap2018} for string composition to retrieve developer discussions on microservices projects deployed on GitHub \cite{surana2020tool}. We formulated the search string using the format [\textit{keyword-1} [OR logic] \textit{keyword-2} … [OR logic] … \textit{keyword-N}], where keywords represented the synonyms as [‘\textit{micro service}’ OR ‘\textit{micro-service}’ OR ‘\textit{microservice}’ OR ‘\textit{Micro service}’ OR ‘\textit{Micro-service}’ OR ‘\textit{Microservice}’]. To extract MSA issues, we selected GitHub, which is one of the most popular and rapidly growing platforms for social coding and community-driven collaborative development of OSS. GitHub represents a modern genre of software forges that unifies traditional methods of development (e.g., version control, code hosting) with features of socio-collaborative development (e.g., issue tracking, pull requests) \cite{kalliamvakou2016depth}. The variety and magnitude of OSS available on GitHub also inspired our choice to investigate the largest OSS platform in the world, with approximately 40 million users and 28 million publicly available project repositories.

Based on the search string, we searched for the title and description of the OSS projects deployed on the GHTorrent dump hosted on Google Cloud. The search helped us retrieve a total of 2,690 potentially relevant MSA-based OSS projects for investigation. To shortlist and eventually select the projects pertinent to the outlined RQs, we applied multi-criteria filtering \cite{GitHubFilter2020}, considering a multitude of aspects such as the popularity or perceived significance of a project in the developers’ community (represented as total stars), adoption by or interests of developers (total forks), and the total number of developers involved (total contributors) for the project. As shown in Figure \ref{fig:researchmethod} (Step B), we only selected the projects that have (i) more than 10 stars and forks, (ii) an English description, and (iii) three or more contributors. This led to a shortlist of 167 microservices projects. To eliminate the instances of potential false positives, i.e., avoiding bias in construct validity, such as misleading project names and mockup code, we contacted the top three contributors of each project via their emails to clarify about:

\begin{enumerate}
 \item \textbf{Correct Interpretation of the Project}: Please confirm if our interpretation of your project (Project URL and Name as an identifier) is appropriate for its design and implementation based on MSA. Also, please help us clarify if this project (e.g., tool, framework, solution) supports the development of microservices systems or if this project is developed using MSA.
 \item \textbf{MSA-based Characteristics and/or Features of the Project}: What features and/or characteristics of the project reflect MSA being used in the project?
 \item \textbf{Optional Question}: Could you please help us identify (the names, URLs, etc. of) any other OSS projects that are designed or developed using MSA? 
\end{enumerate}

We contacted a total of 426 contributors, with 39 of them responded to our query. Based on their confirmation, we selected 15 MSA-based OSS projects (see Table \ref{tab:selectedProjects}). These 15 selected projects, based on the available data at the time of project selection, represent the highest number of issues, making them highly relevant to our study.

{\renewcommand{\arraystretch}{1}
\begin{table*}
 \centering
\scriptsize
 \caption{Identified open-source microservices systems}
\label{tab:selectedProjects}
 \begin{tabular}{|p{2.5cm}|c|c|c|c|}
 \hline
 \textbf{Project Name}&\textbf{\#Issues}&\textbf{\#Contributors} &\textbf{\#Forks}&\textbf{\#Stars}\\\hline
 % \lipsum[4]\footnote{Foo}\par
Spinnaker& 4595 & 121 & 1.2K & 8.5K \\\hline
 
Cortex & 1120 & 226 & 681 & 4.7K \\\hline
 
Jaeger & 995& 239 & 1.9K & 15.8K \\\hline
 
eShopOnContainers & 986 & 157 & 8.9K & 20.7K \\\hline
 
Goa & 930 & 97 &485 &4.7K \\\hline
 
Light-4j & 584 & 37 & 588 & 3.4K \\\hline
 
Moleculer & 473 & 102 & 497 &5.1K \\\hline
 
Microservices-demo &287 & 55 &2.1K & 3.1K \\ \hline
 
Cliquet & 207 & 19&20&65 \\\hline

Deep-framework &174 &12 &75 &537 \\\hline

Scalecube & 130 &20 &90 &547 \\\hline

Lelylan & 123& 7 &93 & 1.5K \\ \hline

Open-loyalty & 175 &14&80&300 \\ \hline

Spring PetClinic & 69 &32 &1.4K &1.1K\\\hline

Pitstop & 39 &15 &490 &890 \\\hline

\end{tabular}
\end{table*}}

\subsubsection{Step B – Synthesize Issues, Causes, and Solutions}
\label{sec:Ext&Syn}
After the projects were identified, as illustrated in Figure \ref{fig:researchmethod}, extracting and synthesising the issues was divided into the following five parts. 
%\begin{enumerate}
%\item 

\textbf{Raw Data Collection}: We chose 15 microservices projects (see Table \ref{tab:selectedProjects}) as the source for building the dataset to answer our RQs. These 15 projects were chosen because they are significantly larger than other microservices projects hosted on GitHub. Hence, it is highly likely that their contributors had more discussions about the type of issues, causes, and solutions in issue tracking systems. The discussions relating to a software system can usually be captured in issue tracking systems \cite{Viviani21}. We initially extracted 10,222 issue titles, issue links, issue opening and closing dates, and the number of contributors for each issue through our customized Python script (see the Raw Data sheet in \cite{replpack}). We stored this information in MySQL and exported it into MS Excel sheets for further processing. We extracted only closed issues because it could increase the chances of answering all our RQs (e.g., solutions).

%\item 
\textbf{Issues Screening}: The first author further scanned the 10,222 issues to check if an issue has been closed or is still open. All open issues were discarded because an open issue is ongoing with many of its causes unknown and most likely solution(s) not found. Furthermore, after selecting only the closed issues, the first author further eliminated (i) issues without a detailed description, (ii) general questions, opinions, feedback, and ideas, (iii) feature requests (e.g., enhancements, proposals), (iv) announcements (e.g., about new updates), (v) duplicated issues, (vi) issues that had only one participant, and (vii) stale issues. After this step, we got 5,115 issues (see the Selected Issues (Round 1) sheet in \cite{replpack}). We had second round of screening on these 5,115 issues to check whether these issues are related to our RQs or not, and after comprehensively analyzing them we found 2,641 issues that were related to our RQs (see the Selected Issues (Round 2) sheet in \cite{replpack}).

%\item 
\textbf{Pilot Data Extraction}: To gain initial insights into the issues, two authors (i.e., the first and fourth) performed pilot data extraction based on 150 issues (5.67\% of 2,641 issues). The authors focused on issue-specific data, such as issue description (i.e., textual details specified by contributors), type of issue (e.g., testing issue, deployment issue), and frequency of issue (i.e., number of occurrences). Pilot data extraction was checked by the second and third authors to verify and refine the details before final data extraction.

%\item 
%\textbf{Data Extraction}: Issues, causes, solutions, and their corresponding data were extracted based on the guidelines for mining software engineering data from GitHub~\cite{gousios2017mining}. \textcolor{black}{Table \ref{tab:Dataitems} provides a summary of the description for each data item extracted from open-source microservices projects and its relevance to the research questions (RQs). Data items from D1 to D3 capture general information about developer discussions, including issue ID, issue title, and issue link. On the other hand, data items from D4 to D6 are utilized to document the data required to address RQ1, RQ2, and RQ3.}

\textbf{Data Extraction}: Issues, causes, solutions, and their corresponding data were extracted based on the guidelines for mining software engineering data from GitHub~\cite{gousios2017mining}. We extracted issue titles, issue links, the number of participants for each issue, key points of issues, key points of causes, and key points of solutions (see the Raw Data and Initial Codes sheet in \cite{replpack}). Data such as issue title, issue link, and number of participants for each issue were used to capture general information about developer discussions. On the other hand, key points of issues, causes, and solutions were used to document the data for answering RQ1, RQ2, RQ3.
 
%\item 
\textbf{Data Analysis}: To synthesize the issues, we used the thematic analysis approach \cite{cruzes2011recommended} to identify the categories of issues, causes, and solutions. The thematic analysis approach is composed of five steps. (i) Familiarizing with data: The first author repeatedly read the project’s contributor’s discussion and documented all discussed key points about issues, causes, and solutions. (ii) Generating initial codes: after data familiarization, the first author generated an initial list of codes from the extracted data (see the Initial Codes sheet in \cite{replpack}). (iii) Searching for the types of issues, causes, and solutions: The first and second authors analyzed the initially generated codes and brought them under the specific types of issues,causes, and solutions. (iv) Reviewing types of issues, causes, and solutions: All the authors reviewed and refined the coding results with the corresponding types of issues, causes, and solutions. We separated, merged, and dropped several issues, causes, and solutions based on a mutual discussion between all the authors. (v) Defining and naming categories: We defined and further refined all the types of issues, causes, and solutions under precise and clear subcategories and categories (see Figure \ref{fig:Taxonomy}). Throughout this process, we engaged in the collaborative refinement and organization of issues, causes, and solutions into specific subcategories and categories. The naming and assignment of issues, causes, and solutions in microservices systems to the appropriate subcategories were conducted through consensus among all the authors, ensuring the development of a comprehensive and coherent taxonomy. Finally, we introduced three levels of categories for managing the identified issues, causes, and solutions. First, we organized the types of issues, causes, and solutions under a specific subcategory (e.g., \textsc{service dependency} in \textit{Service Design Debt}). Then we arranged the subcategories under a specific category (e.g., \textit{Service Design Debt} in \textbf{Technical Debt}).
%\end{enumerate}

%{\renewcommand{\arraystretch}{1}
%\begin{table*}[t]
%\centering
%\scriptsize
%\caption{Data items to be extracted from open-source microservices projects and their relevant RQs}
 %\label{tab:Dataitems}
 %\begin{tabular}{|p{0.3cm}|p{2.2cm}|p{8.8cm}|p{1.2cm}|}
 %\hline
 %\textbf{\#}&\textbf{Data Item}&\textbf{Description}&\textbf{RQ}\\\hline
 %D1 & Index               & The ID of the issue                                                                & Overview \\\hline
 %D2 & Issue title         & A title of the issue from a contributor that describes what the issue is all about & Overview \\\hline 
 %D3	& Issue link	      & The URL address of the issue	                                                   & Overview \\\hline
 %D4	& Issue key points	  & Key points from the developer discussion for issue identification	               & RQ1\\\hline
 %D5	& Causes key points   & Key points from the developer discussion for cause identification	               & RQ2\\\hline
 %D6	& Solution key points & Key points from the developer discussion for solution identification  	           & RQ3\\\hline
 %\end{tabular}
%\end{table*}}

To minimize any bias during data analysis, each step, i.e., classification, mapping, and documentation, were cross-checked and verified independent of the individual(s) involved in data synthesis. Documentation of the results as answers to RQs is presented as taxonomical classification of issues (Section \ref{sec:results_RQ1}), causes of issues (Section \ref{sec:results_RQ2}), solutions to address the issues (Section \ref{sec:results_RQ3}), and mapping of issues with their causes and solutions (Section \ref{Sec:MappingIssueCausesSolutions}), complemented by details of validity threats (Section \ref{sec:threats}).

\subsection{Phase 2 - Conducting Practitioner Interviews}
\label{sec:RMPhase2}
Since we aim to understand the issues, causes, and solutions of microservices systems from a practitioners' perspective, we opted to conduct interviews to confirm and improve the developed taxonomies. The interview process consists of the following steps.

\subsubsection{Preparing a Protocol}
\label{InterviewsProtocol}
The first author conducted 15 online interviews with microservices practitioners through Zoom, Tencent Meeting, and Microsoft Teams. Before conducting actual interviews, we also conducted two pilot interviews with microservices practitioners to check the understandability and comprehensiveness of interview questions. However, we did not include their answers in our dataset. In total, we conducted 15 actual interviews, and each interview took 35-45 minutes. It is argued that conducting 12 to 15 interviews with homogeneous groups is enough to reach saturation \cite{guest2006many}. In our case, after completing 15 interviews, we noticed a clear saturation in the responses to our interview questions, indicating that we had gathered enough information to address our interview objectives effectively. Therefore, we decided not to conduct any additional interviews. We conducted semi-structured interviews based on an interview guide (see the Interview Questionnaire sheet in \cite{replpack}), which contains closed-ended questions related to participant demographics and open-ended questions related to general categories of microservices issues, causes, and solutions.

The interview process was comprised of three sections. In the first section, we asked 6 demographic questions to understand the interviewee’s background in microservices. We covered various aspects in this section, including the country of the practitioner, major responsibilities, overall experience in the IT industry, experience with implementing microservices systems, the work domain of the organization, and programming languages for developing microservices systems. In the second part, we asked three open-ended questions about the types of issues, causes of issues, and solutions to issues during the development of microservices systems. The purpose of this part was to allow the interviewees to spontaneously express their views about the issues developers face in developing microservices systems, the causes of the issues, and resolution strategies without the interviewer biasing their responses. In the third part, we presented the three taxonomies to the interviewees and asked them to indicate any missing issues, causes, and solutions that have not been explicitly mentioned. All three taxonomies come from identifying, analyzing, and synthesizing the developer discussions from 15 open-source microservices systems. The taxonomy of issues consists of 386 types of issues, 54 issue subcategories, and 18 issue categories. The taxonomy of causes consists of 217 types of causes, 26 cause subcategories, and 8 cause categories. The taxonomy of solutions consists of 171 types of solutions, 33 solution subcategories, and 8 solution categories. At the end of each interview, we thanked the interviewee and briefly informed them of our next plans.

\subsubsection{Conducting Interviews}
To recruit interviewees, we initially reached out to our professional contacts in each country via email. We also requested that they disseminate our interview request among their colleagues who they believed would be suitable candidates for the study. We informed the possible participants that this interview was entirely voluntary with no compensation. With this approach, we recruited 15 microservices practitioners from IT companies in 10 countries: Australia (1), Canada (3), China (1), Chile (1), India (1), Pakistan (2), Sweden (2), Norway (1), the United Kingdom (1), and the United States of America (2). We refer to the interviewees as P1 to P15. Most of the interviewees are mainly software architects and application developers. Their average experience in the IT industry is 10.33 years (Minimum: 5, Maximum: 16, Median: 10, Mode: 9, Standard Deviations: 3.26). The interviewees’ average experience in microservices is 5.33 years (Minimum: 3, Maximum: 8, Mode: 4, Median: 5, Standard Deviations: 1.67).

\subsubsection{Data Analysis}
\label{InterviewsDataAnalysis}
We applied a thematic analysis method \cite{braun2006using} to analyze the recorded interviews. Before applying the thematic analysis method, the first author prepared the text transcripts from audio recordings. The first author read the interview transcripts and coded them using the MAXQDA tool. We dropped several sentences unrelated to ``microservices issues, causes, and solutions''. After removing the extraneous information from the transcribed interviews, the first author coded the interview transcripts’ contents to get the answers to the interview questions. To ensure the quality of the codes, the second author verified the initial codes created by the first author and provided suggestions for improvement. After the initial coding and subsequent enhancements, all the authors collaborated in analyzing the codes. Conflicts in the codes were addressed through collaborative discussions and brainstorming sessions. Finally, we generated a total of 28 types of issues (classified into 15 subcategories and 11 categories), 28 types of causes (classified into 15 subcategories and 7 categories), and 30 types of solutions (classified into 4 subcategories and 3 categories). During the interviews, we got 48 instances of issues, 31 instances of causes, and 36 instances of solutions. Among these instances, we identified 14 types of issues, 21 types of causes, and 23 types of solutions that were not part of the taxonomies we derived from the 15 open-source microservices systems. Later, we exported the analyzed interview data from the MAXQDA tool to an MS Excel sheet (i.e., the Interview Results sheet in \cite{replpack}) to make the part of taxonomies of issues, causes, and solutions in microservices systems from the interview data. 

Except for \textsc{service size}, \textsc{operational and tooling overhead}, and \textsc{team management} issues, all of the issues, causes, and solutions identified through the interviews can be classified under the existing categories (i.e., the output of Phase 1). The instances of issues mentioned by the interviewees include CI/CD Issues (9), Security Issues (6), Service Execution and Communication Issues (5), Database Issues (5), Organizational Issues (5), Testing Issues (5), Monitoring Issues (4), Performance Issues (4), Update and Installation Issues (3), Configuration Issues (1), and Technical Debt (1). The causes mentioned by the interviewees include Service Design and Implementation Anomalies (20), Poor Security Management (2), Legacy Versions, Compatibility, and Dependency Problems (2), Invalid Configuration and Communication Problems (2), General Programming Errors (2), Fragile Code (2), and Insufficient Resources (1). The solutions mentioned by the interviewees include Add Artifacts (32) and Upgrade Tools and Platforms (4).

{\renewcommand{\arraystretch}{1}
\begin{table*}[t]
\centering
\scriptsize
 \caption{Overview of the interviewees and their demographic information}
 \label{tab:IntervieweesDemographics}
 \begin{tabular}{|p{0.25cm}|p{3.8cm}|p{2.4cm}|p{2.8cm}|p{1.4cm}|p{1.4cm}|p{1cm}|} \hline
\textbf{\#} &\textbf{Responsibilities} & \textbf{Languages}& \textbf{Domain}&\textbf{Overall Exp.}	& \textbf{MSA Exp.} &\textbf{Country}\\\hline

P1 &    Software Architect, Developer    &  Python, Java, Node.JS &	E-commerce, Healthcare & 14 Years &	6 Years & Sweden    \\\hline
P2 &	Developer                        &	Java with Spark	      & E-commerce             & 9 Years  &	5 Years	& USA       \\\hline
P3 &	Software Architect               &	Python, Go, Java      &	E-commerce             & 13 Years &	6 Years	& UK        \\\hline
P4 &	Software Architect, Developer    &	Python, Java          &	Educational ERP        & 7 Years  &	3 Years & Pakistan  \\\hline
P5 &	Software Architect               &	Python, Java          & Internet of Things     & 12 Years &	7 Years & Canada    \\\hline
P6 &    Software Architect, Developer    &	Java, Node.JS, Python &	Healthcare             & 15 Years &	8 Years	& Australia \\\hline
P7 &	Software Engineer 	             &  C\#.Net	              & E-commerce, Banking    & 10 Years & 4 Years & Canada    \\\hline
P8 &	DevOps Consultant                &	Kotlin, Python        &	Network applications   & 9 Years  &	6 Years & Canada    \\\hline
P9 &	Software Architect               &	Java                  &	Education, Healthcare  & 5 Years  & 4 Years & Chile     \\\hline
P10 & 	Application Developer, Architect &	Java                  &	Financial (Insurance)  & 10 Years &	8 Years & Sweden    \\\hline
P11 & 	DevOps Consultant                &	Java, Kotlin,Python   & Telecommunication      & 6 Years  & 3 Years & Norway    \\\hline
P12 & 	Application Developer            &	Angular, Puthon       &	Manufacturing          & 8 Years  &	4 Years & China     \\\hline
P13 & 	Principal Consultant             &	Swift                 &	Transportation         & 9 Years  & 5 years & USA       \\\hline
P14 & 	Azure Technical Engineer         &  JavaScript, Golang    & Embedded systems       & 12 Years &	4 Years & Pakistan  \\\hline
P15 &   Software Architect               &   Ruby, UML            &  Payment applications  & 16 Years &	7 Years & India     \\\hline

\end{tabular}

\end{table*}}

\subsection{Phase 3 - Conducting a Survey}
\label{sec:RMPhase3}
A questionnaire-based survey approach is used to evaluate the taxonomies of issues, causes, and solutions based on mining developer discussions and conducting practitioner interviews. We adopted Kitchenham and Pfleeger’s guidelines for conducting surveys \cite{kitchenham2008personal} and used an anonymous survey to increase response rates \cite{tyagi1989effects}.

\subsubsection{Recruitment of Participants and Conducting the Pilot Survey}
\label{PilotSurvey}
After the survey design, we needed to (i) select the survey participants and (ii) conduct a pilot survey for initial assessments (e.g., time taken, clarity of statements, and add, remove, and refine the questions). To select the potential respondents, we used the following contact channels to spread the survey broadly to a wide range of companies from various locations worldwide. The contact channels to recruit the potential participants included (i) professional contacts, researchers of industrial track publications, and authors of web blogs related to microservices, (ii) practitioners and their communities on social coding platforms (e.g., GitHub, Stack Overflow), and (iii) social and professional online networks (LinkedIn, Facebook, Twitter, Google Groups). In the survey invitation email, we also requested the potential participants to share the survey invitation with individuals or groups deemed as relevant participants. Before sending out the invitations, we ensured that we only contact individuals with experience in any aspects of MSA design, development, and/or engineering based on their professional profiles, such as code commits, industrial publications, and professional designations. Based on publicly available email IDs, first, we sent out survey invitations to only a selected set of 30 participants for a pilot survey. Out of the 30 participants contacted for the pilot survey, 10 replied (response rate 33.33\%) from 7 countries. It should be noted that the results of the pilot survey were not incorporated into our primary survey data. However, the pilot survey helped us to refine the survey questionnaire in terms of restructuring the sections and rephrasing some questions for clarity of the survey to ensure that (i) the length of the survey is appropriate, (ii) the terms used in the survey questions are clear and understandable, and (iii) the answers to the survey questions are meaningful.

\begin{figure*}[!htbp]
\centering
\includegraphics[scale=0.35]{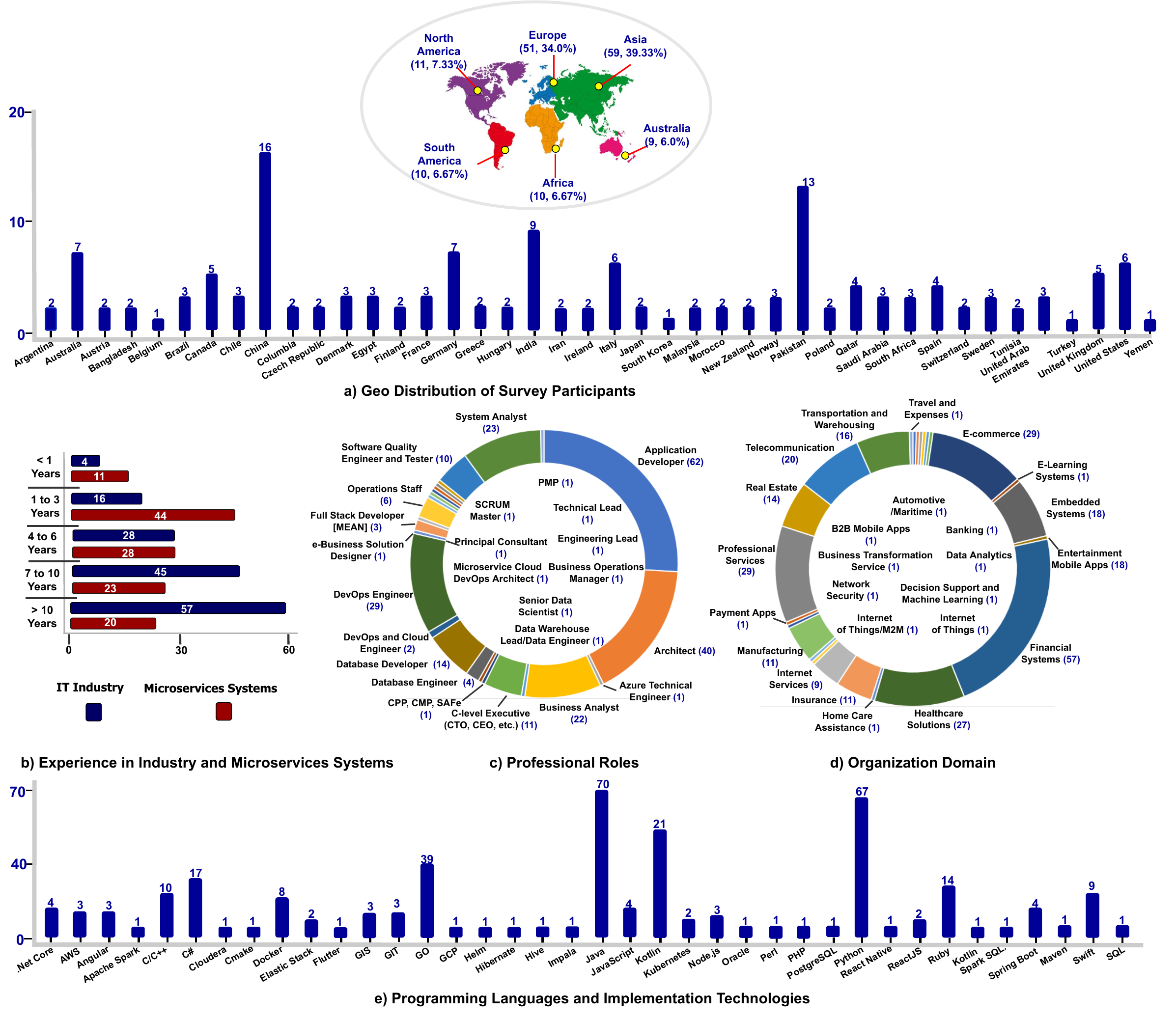}
 \caption{Overview of the demographics of survey participants}
 \label{fig:demography}
\end{figure*}

\subsubsection{Conducting the Web-based Survey}
\label{WebSurvey}
We adopted a cross-sectional survey design, which is appropriate for collecting information at one given point in time across a sample population \cite{kitchenham2008personal}. Surveys can be conducted in many ways, such as Web-based online questionnaires and phone surveys \cite{lethbridge2005studying}. We decided to conduct a Web-based survey because these surveys can help to (i) minimize the time and cost, (ii) collect the responses from geographically distributed respondents, (iii) minimize time zone constraints, and (iv) save the effort of researchers to collect data in a textual, graphical, or structured format \cite{lethbridge2005studying}. To document different types of responses while maintaining the granularity of information, we structured the questionnaire into a total of 12 questions organized under four sections (see the Survey Questionnaire sheet in \cite{replpack}).
%\begin{itemize}
 %\item 
 
\textbf{Demographics}: We asked 6 demographic questions about the background information of the respondents to identify the (i) country or region, (ii) major responsibilities, (iii) overall work experience in the IT industry, (iv) work experience with microservices systems, (v) work domain of the organization, and (vi) programming languages and implementation technologies for developing microservices systems. The demographic information has been collected to (i) identify respondents who do not have sufficient knowledge about microservices, (ii) divide the results into different groups, and (iii) generalize the survey findings for the microservices research and practice community. We received a total of 156 responses, and we excluded 6 responses that were either randomly filled or filled by research students and professors who were not practitioners. In the end, we got 150 valid responses.% It is also important to mention that because the responses to the pilot survey were valid, we also decided to include them in the final survey responses. 
\begin{itemize}
 \item \textbf{Countries}: Respondents came from 42 countries of 6 continents (see Figure \ref{fig:demography}(a)) working in diverse teams and roles to develop microservices systems. The majority of them are from China (16 out of 150, 10.66\%), Pakistan (13 out of 150, 8.66\%), and India (9 out of 150, 6.00\%).
 \item \textbf{Experience}: We asked the participants about their experiences in the IT industry and the development of microservices systems. Figure \ref{fig:demography}(b) shows that the majority of the respondents (57 out of 150, 38.00\%) have worked in the IT industry for more than 10 years and around one third of the respondents (44 out of 150, 29.33\%) have worked with microservices systems for 1 to 3 years. We also received a considerable amount of responses in which practitioners have more than 10 years of experience working with microservices systems (20 out of 150, 13.33\%).
 \item \textbf{Professional Roles}: Figure \ref{fig:demography}(c) shows that the majority of the participants were application developer (62 out of 150, 41.33\%), architect (40 out of 150, 26.66\%), and DevOps engineer (29 out of 150, 19.8\%). Note that one participant may have multiple major responsibilities in the company, and consequently, the sum of the percentages exceeds 100\%.
 \item \textbf{Application Domains}: Figure \ref{fig:demography}(d) shows the domains of the participants' organizations where the microservice practitioners worked. Financial Systems (57 out of 150, 38.00\%), E-commerce (29 out of 150, 19.33\%), and Professional Services (29 out of 150, 19.33\%) are the dominant domains. Note that one organization where a practitioner worked may have one or more application domains.
 \item \textbf{Programming Languages and Implementation Technologies}: Figure \ref{fig:demography}(e) shows that 38 programming languages and technologies were used to develop microservices systems, in which “Java" (70 out of 150, 46.66\%), “Python" (67 out of 150, 44.66\%), and “GO" (39 out of 150, 26.00\%) are the most frequently used languages for developing microservices systems. 
\end{itemize}
 
\textbf{Microservices Practitioners’ Perspective}: To evaluate the taxonomies of issues, causes, and solutions, we asked six survey questions (both Likert scale and open-ended, see the Survey Questionnaire sheet in \cite{replpack}) from microservices practitioners. We provided a list of 19 issue categories and asked survey participants to respond to each category on a 5-point Likert scale (Very Often, Often, Sometimes, Rarely, Never). Similarly, regarding causes, we provided 8 categories and asked practitioners to respond to each category on a 5-point Likert scale (Strongly Agree, Agree, Neutral, Disagree, Strongly Disagree). We also provided 8 categories of solutions and asked practitioners to respond to each category on a 5-point Likert scale. Along with these 5-point Likert scales, we added one option to know the familiarity of the survey respondents with the listed categories. We also asked three open-ended questions to identify the missing issue, causes, and solutions in the provided categories. All the open-ended responses were further analyzed through thematic classification and adjusted in the taxonomies.
%\end{itemize}

\subsubsection{Data Analysis}
We used descriptive statistics and constant comparison techniques \cite{hoda2017becoming} to analyze the quantitative (i.e., closed-ended questions) and qualitative (i.e., open-ended questions) responses to the survey questions, respectively. To better understand practitioners’ perspectives through Likert answers on issues, causes, and solutions in microservices systems, we used the Wilcoxon rank-sum test, i.e., participants who have $\le$ 6 years of experience (49 responses) versus participants who have $\ge$ 6 years of experience with microservices systems (101 responses). We used 6 years as a breaking point for separating the groups because 6 years is almost in the middle of practitioners' experience with microservices systems. We used the~\faBalanceScale{} symbol to indicate a significant difference between participant groups who have \textbf{Experience $\le$ 6 years} vs. \textbf{Experience $\ge$ 6 years}. To analyze the results of open-ended questions, we employed the constant comparison method from grounded theory \cite{hoda2017becoming}, which allowed us to systematically identify and understand various issues, causes, and solutions from practitioners' perspective. We began the process by coding the responses and assigning relevant labels. Throughout the analysis, we persistently compared new responses with existing codes (i.e., types of issues, causes, and solutions from the taxonomies) and made refinements as needed.

\section{Results}
\label{sec:results}

This section presents the analyzed results of this study, addressing the four RQs outlined in Section \ref{RQs}. The analyzed results are further organized as categories (e.g., \textbf{Technical Debt}), subcategories (e.g., \textit{Code Debt}), and types (e.g., \textsc{code smell}). We present categories in \textbf{boldface}, subcategories in \textit{italic}, and types in \textsc{small capitals}. The relevant examples are provided as quoted messages along with their issue ID numbers to facilitate the traceability to our dataset (see the Initial Codes sheet in \cite{replpack}). We report the types of issues in Section \ref{sec:results_RQ1} (see Figure \ref{fig:Taxonomy}), the types of causes in Section \ref{sec:results_RQ2} (see Table \ref{tab:CausesTaxnomey}), the types of solutions in Section \ref{sec:results_RQ3} (see Table \ref{tab:SolutionsTaxnomey}), and practitioners' perspective on these taxonomies in Section \ref{sec:results_RQ4} (see Table \ref{tab:RQ4}).

\subsection{Types of Issues (RQ1)}
\label{sec:results_RQ1}

\begin{landscape}
\begin{figure}
\includegraphics[width=\linewidth, height=0.68\linewidth]{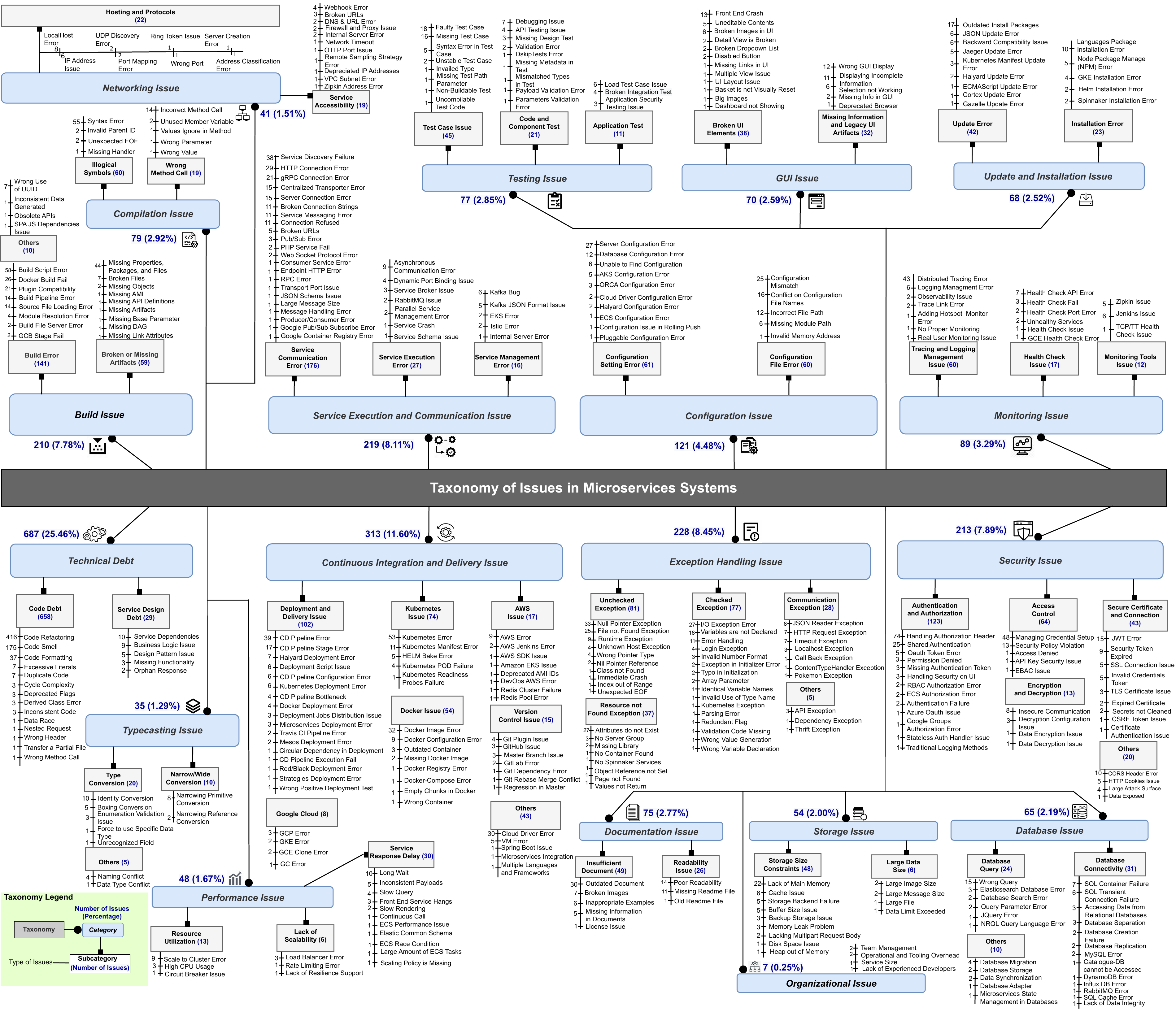}
 \caption{A taxonomy of issues in microservices systems}
 \label{fig:Taxonomy}
 \end{figure}
\end{landscape}

The taxonomy of issues in microservices systems is provided in Figure \ref{fig:Taxonomy}. The taxonomy of issues is derived by mining developer discussions (i.e., 2,641 instances of issues, see Section \ref{sec:Ext&Syn}), conducting practitioner interviews (i.e., 48 instances of issues, see Section \ref{InterviewsDataAnalysis}), and conducting a survey (9 instances of issues, see Section \ref{sec:results_RQ4}). Therefore, we got a total of 2,698 instances of issues. The results show that Technical Debt (687 out of 2698), Continues Integration and Delivery (313 out of 2698), and Service Execution and Communication (219 out of 2698) issues are most frequently discussed. It is worth noting that one issue may cover more than one topic. However, we consider the main topic of an issue where there is a high probability of identifying the causes or solutions. The number of issues in each issue type, subcategory, and category are also shown in Figure \ref{fig:Taxonomy}.

\textbf{1. Technical Debt} (687/2698, 25.46\%): Technical Debt (TD) is “\textit{a metaphor reflecting technical compromises that can yield short-term benefit but may hurt the long-term health of a software system”} \cite{li2015systematic}. This is the largest category in the taxonomy of microservices issues and includes a wide range of issues related to \textit{Code Debt} and \textit{Service Design Debt}. The interviewees also mentioned several issues regarding this category, and one representative quotation is depicted below. 

\faHandORight{} “\textit{The complexity introduces several types of technical debt both in the design and development phases of microservices systems}” \textbf{Developer, P2}.

We identified and classified 19 types of TD issues in 2 subcategories (see Figure \ref{fig:Taxonomy} and the Issue Taxonomy sheet in \cite{replpack}). Each of them is briefly described below.

\begin{itemize}
\item \textit{Code Debt} (658, 24.47\%) refers to the issues of source code, which could adversely impact the code quality. This subcategory mainly gathers issues related to \textsc{code refactoring}, \textsc{code smell}, \textsc{code formatting}, \textsc{excessive literals}, and \textsc{duplicate code}. For instance, developers refactored the code of the Spinnaker project by ``\textit{adding hal command for tweaking the component sizings, \#11}''. In addition, several other types of \textit{Code Debt}, such as \textsc{code formatting}, \textsc{excessive literals}, and \textsc{duplicate code} also negatively affect the code legibility of microservices systems.
 
\item \textit{Service Design Debt} (29, 1.07\%) refers to the violation of adopting successful practices (e.g., MSA patterns) for designing open-source microservices systems. The issues in this subcategory are mainly related to \textsc{service dependency}, \textsc{business logic issue}, and \textsc{design pattern issue}. For instance, the developers of the moleculer project reported the issue of service design debt in which “\textit{Service A requires module A. If service A changed, the runner reloads, but if module A changed, the runner does not reload, \#1873}”.
\end{itemize}

\textbf{2. Continuous Integration and Delivery (CI/CD) Issue} (313/2698, 11.60\%): CI/CD refers to the automation process that enables development teams to frequently develop, test, deploy, and modify software systems (e.g., microservices systems). Usually, a variety of tools and technologies are used to implement the CI/CD process. A wide range of CI/CD issues has been identified by mining microservices systems. However, we also found a few issues that the interviewees mentioned, and one representative quotation is depicted below.

\faHandORight{} “\textit{The key issues of CI/CD for practitioners are many small independent code bases, multiple languages, frameworks, microservices integration, load testing, managing releases, and continued service updates}”, \textbf{DevOps Consultant, P8}. 

We identified and classified 55 types of CI/CD issues in 7 subcategories (see Figure \ref{fig:Taxonomy} and the Issue Taxonomy sheet in \cite{replpack}). Each of them is briefly described below. 

\begin{itemize}
\item \textit{Deployment and Delivery Issue} (105, 3.89\%) reports the problems that occur during the deployment and delivery of microservices systems. We identified 17 types of issues in this subcategory, which are mainly related to \textsc{cd pipeline error}, \textsc{cd pipeline stage}, \textsc{halyard deployment}, and \textsc{deployment script} errors. For example, regarding \textsc{cd pipeline error}, one contributor of the eShopOnContainers project highlighted that “\textit{Jenkins pipeline is failing with error 403, \#990}”. 

\item \textit{Kubernetes Issue} (74, 2.74\%) reports the CI/CD issues specific to Kubernetes which is an open-source system for automatic deployment, scaling, and management of containerized applications (e.g., microservices systems). Most problems of the subcategory are related to general \textsc{kubernetes}, \textsc{helm bake}, \textsc{kubernetes manifest} errors. For example, the contributors of the eShopOnContainers project were “\textit{unable to list Kubernetes resources using default ASK to create a script, \#688}”). 

\item \textit{Docker Issue} (50, 1.85\%): Docker is an open-source platform that helps practitioners with continuous testing, deployment, executing, and delivering applications (e.g., microservices systems). Our results indicate that most of the issues related to Dockers are \textsc{docker image error}, \textsc{docker configuration error}, and \textsc{outdated container}. For instance, the contributors of the eShopOnContainers project pointed out that the “\textit{building of the solution using docker-compose was failing, \#2464}” due to a docker configuration error.

\item \textit{Amazon Web Services (AWS) Issue} (17, 0.63\%): AWS provides an on-demand cloud computing platform for creating, testing, delivering, and managing applications. The most frequent types of this subcategory are general \textsc{aws error} and \textsc{aws jenkins error}. As an example, some developers of the Spinnaker project faced a situation in which “\textit{AWS Jenkins multi Debian package jobs fail to bake, \#2429}”.

\item \textit{Version Control Issue} (16, 0.69\%) is related to version control and management systems. In general, the major issues in this subcategory are related to Git, such as \textsc{git plugin}, \textsc{master branch}, and \textsc{gitlab} issues.
 
\item \textit{Google Cloud Issue} (8, 0.29\%): Google Cloud provides infrastructure services for creating and managing projects (e.g., microservices systems) and resources. This subcategory contains issues related to the Google Cloud platform for microservices systems. The most reported issues are \textsc{gcp error} \textsc{gke error}, and \textsc{gce clone error}. For instance, in the Spinnaker project, practitioners identified that “\textit{GCP: instance group port name Mapping is not working properly, \#2496}”.

 \item \textit{Others} (38, 1.40\%): This subcategory gathers issues related to \textsc{cloud driver error}, \textsc{virtual machine error}, and \textsc{spring boot error}.

\end{itemize}

\textbf{3. Exception Handling Issue} (228/2698, 8.45\%): Exception handling is used to respond the unexpected errors during the running state of software systems (e.g., microservices systems), and it helps to avoid the software system being crashed unexpectedly. This category represents the issues practitioners face when handling various kinds of exceptions in microservices systems. We identified and classified 44 types of Exception Handling issues in 5 subcategories (see Figure \ref{fig:Taxonomy} and the Issue Taxonomy sheet in \cite{replpack}). Each of them is briefly described below.

\begin{itemize}
\item\textit{Unchecked Exception} (81, 3.00\%): These exceptions cannot be checked on the program’s compile time and throw the errors while executing the program’s instructions.  We identified 10 types of issues in this subcategory, in which the top three types are \textsc{null pointer exception}, \textsc{file not found exception}, and \textsc{runtime exception}.
 
\item\textit{Checked Exception} (77, 2.85\%): These exceptions can be checked on the program’s compile time. Checked exceptions could be fully or partially checked exceptions.  We identified 16 types of issues in this subcategory, mainly related to \textsc{io exception}, \textsc{variables are not declared}, and \textsc{error handling}.
 
\item\textit{Resource not Found Exception} (37, 1.33\%): These exceptions occur when some services cannot find the required resources for executing operations. We identified 8 types of issues in this subcategory. Most of them are related to \textsc{attributes do not exist}, \textsc{no server group}, and \textsc{missing library}.

\item\textit{Communication Exception} (28, 1.03\%): These are exceptions thrown when the client services cannot communicate with the producer services. We identified 7 types of issues in this subcategory, mainly related to \textsc{http request exception} and \textsc{timeout exception}.
 
\item\textit{Others} (5, 0.18\%): This subcategory gathers issues related to \textsc{api exception}, \textsc{dependency exception}, and \textsc{thrift exception}.
\end{itemize}

\textbf{4. Service Execution and Communication Issue} (219/2698, 8.11\%): Communication problems are obvious when microservices communicate across multiple servers and hosts in a distributed environment. Services interact using e.g., HTTP, AMQP, and TCP protocols depending on the nature of services. The interviewees also mentioned a few issues regarding this category, and one representative quotation is depicted below.

\faHandORight{} “\textit{The poor implementation of microservices communication is also a source of insecure communication, latency, lack of scalability, and errors and fault identification on runtime}” (\textbf{P7, Software Engineer}).

We identified and classified 34 types of service execution and communication issues in 3 subcategories (see Figure \ref{fig:Taxonomy} and the Issue Taxonomy sheet in \cite{replpack}). Each of them is briefly described below

\begin{itemize}
\item \textit{Service Communication} (166, 6.15\%): There are different ways of communication (e.g., synchronous communication, asynchronous message passing) between microservices. This subcategory covers the issues of service communication in which the majority of the issues are related to \textsc{service discovery failure}, \textsc{http connection error}, and \textsc{grpc connection error}.
 
 \item \textit{Service Execution} (27, 1.00\%): This subcategory contains the issues regarding \textsc{asynchronous communication}, \textsc{dynamic port binding}, \textsc{rabbitmq messaging}, and \textsc{service broker} during service execution. These issues occur due to various reasons. For example, a dependency issue between microservices occurred when “\textit{integration commands were sent asynchronously, \#1710}”. We also found several issues regarding \textsc{dynamic port binding}. For instance, a server module of the light-4j project could not dynamically allocate a “\textit{port on the same host with a given range, \#1742}”. 
 
 \item \textit{Service Management} (17, 0.63\%): This subcategory covers the issues that occur in the distributed event store platform, service management platform, and service networking layer. The majority of the problems that happen in this subcategory are \textsc{kafka bug}, \textsc{kafka josn format issue}, and \textsc{eks (elastic kubernetes service) error}.
\end{itemize}

\textbf{5. Security Issue} (213/2698, 7.89\%): Microservices provide public interfaces, use network-exposed APIs for communicating with other services, and are developed by using polyglot technologies and toolsets that may be insecure. This makes microservices a potential target for cyber-attacks; therefore, security in microservices systems demands serious attention. Mining microservices systems have identified a wide range of security issues. Among those issues, microservices practitioners also mentioned several other security issues during the interviews. One representative quotation is depicted below. 

\faHandORight{} “\textit{The other problem is securing microservices at different levels. Specifically, we deal with microservices-based IOT applications that have more insecure points than traditional ones}” (\textbf{P5, Solution Architect}).

We identified and classified 37 types of security issues in 4 subcategories (see Figure \ref{fig:Taxonomy} and the Issue Taxonomy sheet in \cite{replpack}). Each of them is briefly described below.

\begin{itemize}
\item \textit{Authentication and Authorization} (123, 4.55\%): Authentication is the process of identifying a user, whereas authorization determines the access rights of a specific user to system resources. We found that the majority of the authentication and authorization issues are related to \textsc{handling authorization header} (e.g., Basic Auth, OAuth, OAuth 2.0) and \textsc{shared authentication} issues. \textsc{handling authorization header} is generally used to implement authorization mechanisms. Our study found several issues related to Basic Auth, OAuth, and OAuth 2.0 header failure and the non-availability of these security headers. For example, the developers of the eShopOnContainers project reported the issue about “\textit{invalid\textunderscore request on auth from Swagger for Location API, \#1990}”. Moreover, we found several issues regarding improper implementation of \textsc{shared authentication} methods in microservices systems.

\item \textit{Access Control} (64, 2.37\%) is a fundamental element in securing the infrastructure of microservices or any software systems. Access control could be role- or attribute-based in a microservices system. The major types of issues in this subcategory are \textsc{managing credential setup} and \textsc{security policy violation}.

\item \textit{Secure Certificate and Connection} (43, 1.59\%): Our study reports several issues regarding implementing security certificates and standards, such as SSL, TSL, and JWT, which are used to secure communication between client-server, service-to-service, or between microservices. For example, we found a \textsc{jwt error} and an \textsc{ssl connection issue} in the Goa and Spinnaker project respectively. We also found several other types of secure certificate and connection issues, such as \textsc{security token expired}, \textsc{tls certificate issue}, and \textsc{expired certificate}.

\item \textit{Encryption and Decryption} (13, 0.48\%) is used to convert plain text into ciphertext and ciphertext into plain text to secure the information. We identified three types of issues related to this subcategory are \textsc{data encryption}, \textsc{data decryption}, and \textsc{configuration decryption}.
\end{itemize}

\textbf{6. Build Issue} (210/2698, 7.78\%): Build is a process of preparing an application program for software release by collecting and compiling all required source files. The outcome of this process could be several types of artifacts, such as binaries and executable programs. We identified and classified 20 types of build issues in 3 subcategories (see Figure \ref{fig:Taxonomy} and the Issue Taxonomy sheet in \cite{replpack}). Each of them is briefly described below.

\begin{itemize}
\item \textit{Build Error} (141, 5.22\%): We identified several types of build errors which can interrupt the build process of microservices systems. We found that the majority of build errors are related to \textsc{build script}, \textsc{plugin compatibility}, \textsc{docker build fail}, \textsc{build pipeline error}, \textsc{build file server error}, \textsc{source file loading}, and \textsc{module resolution}.

\item \textit{Broken and Missing Artifacts} (59, 2.18\%): These issues occur during the build process's parsing stage when the build systems verify the required information (e.g., files, packages, designated locations) in the build script files before executing the build tasks. This subcategory mainly covers the issues related to \textsc{missing properties, packages, and files}, \textsc{broken files}, and \textsc{missing objects}. We also identified several other types of broken and missing artifacts which include \textsc{missing ami (amazon machine image)}, \textsc{missing base parameter in the client}, and \textsc{missing link attributes}.

\item \textit{Others} (10, 0.37\%): This subcategory mainly gathers issues related to \textsc{wrong use of universally unique identifier}, and \textsc{inconsistent data generated}. 
\end{itemize}

\textbf{7. Configuration Issue} (121/2698, 4.48\%):  Configuration is a process of controlling and tracking and making consistent all the required instances for software systems(e.g., microservices systems). Microservices systems have multiple instances and third party applications to configure. This category gathers configuration issues during microservices system development, implementation, and deployment phases. Besides identifying the configuration issues from the OSS microservices system, microservices practitioners also indicated configuration issues during the interviews. One representative quotation is depicted in the following.

\faHandORight{} “\textit{The major challenge for me is the poor microservices’ configuration, which grows as the application size grows—the configuration effect on implementation and deployment phases of the microservices systems. The poor configuration of microservices may lead to increased latency and decrease the speed of microservices calls between different services}” (\textbf{P12, Application Developer}).

We identified and classified 16 types of configuration issues in 2 subcategories (see Figure \ref{fig:Taxonomy} and the Issue Taxonomy sheet in \cite{replpack}). Each of them is briefly reported below.

\begin{itemize}
\item \textit{Configuration Setting Error} (61, 2.26\%): This subcategory contains issues associated with the setup configuration of different types of servers, databases, and cloud infrastructure platforms. The main types of issues in this subcategory are \textsc{server configuration error}, \textsc{database configuration error}, and \textsc{aks (azure kubernetes service) configuration error}.

\item \textit{Configuration File Error} (60, 2.22\%): This subcategory covers the issues that mainly occur due to providing incorrect values in environment setting variables. The major types of issues in this subcategory are \textsc{configuration mismatch}, \textsc{conflict in configuration file names}, and \textsc{incorrect file path}.
\end{itemize}

\textbf{8. Monitoring Issue} (89/2698, 3.29\%): The dynamic nature of microservices systems needs monitoring infrastructures to diagnose and report errors, faults, failure, and performance issues. This category reports issues related to monitoring microservices systems. Several interviewees also mentioned monitoring issues for microservices systems. One representative quotation is depicted in the following.

\faHandORight{} “\textit{Microservices systems host containerized or virtualized across distributed private, public, hybrid, and multi-cloud environments. Monitoring highly distributed systems like microservices systems through traditional monitoring tools is a challenging experience because these tools only focus on a specific component or the overall operational health of the system}” (\textbf{P14, Azure Technical Engineer}).

We identified and classified 17 types of monitoring issues in 3 subcategories (see Figure \ref{fig:Taxonomy} and the Issue Taxonomy sheet in \cite{replpack}). Each of them is briefly described below.

\begin{itemize}
\item \textit{Tracing and Logging Management Issue} (60, 2.22\%): One prominent challenge of monitoring microservices systems is the collection of logs from containers and distributed tracing. We identified 7 types of issues in this subcategory, mainly related to \textsc{distributed tracing error}, \textsc{logging management error}, and \textsc{observability issue}.

\item \textit{Health Check Issue} (17, 0.63\%): This subcategory deals with the problems related to the health monitoring of microservices systems. We identified six types of issues, mainly related to \textsc{health check API error}, \textsc{health check fail}, and \textsc{health check port error}. 

\item\textit{Monitoring Tool Issue} (12, 0.44\%): We also identified several issue discussions, in which microservices practitioners discussed the problems of three monitoring tools, including \textsc{zipkin issue}, \textsc{jenkins issue}, and \textsc{tcp/tt health check issue}.
\end{itemize}

\textbf{9. Compilation Issue} (79/2698, 2.92\%): This category reports compilation errors, which mainly occur when the compiler cannot compile source code due to errors in the code or errors with the compiler itself. We identified and classified 9 types of compilation issues in 2 subcategories (see Figure \ref{fig:Taxonomy} and the Issue Taxonomy sheet in \cite{replpack}). Each of them is briefly described below.

\begin{itemize}
 \item \textit{Illogically Symbols} (60/2698, 2.22\%): These issues occur when developers use illegal characters or incorrect syntax during the coding, for instance, \textsc{syntax error}, \textsc{invalid parent id}, and \textsc{unexpected end of file}.
 
 \item \textit{Wrong Method Call} (19/2698, 0.70\%): These issues occur when the compiler tries to search for definitions of methods by invoking them through method calls and finds \textsc{wrong parameter}, \textsc{wrong method call}, and \textsc{incorrect values}.
\end{itemize}

\textbf{10. Testing Issue} (77/2698, 2.85\%): Microservices systems pose significant challenges for testing because of many services, inter-communication processes, dependencies, network communication, and other factors. Among those issues, microservices practitioners also mentioned several other testing issues during the interviews. One representative quotation is depicted in the following. 

\faHandORight{} “\textit{Testing is another issue that I think is more challenging in microservices systems. I also think deploying each microservice as a singular entity and testing them is tedious and brings several problems. For example, testing coordination among multiple microservices when deploying one service as a singular entity}” (\textbf{P2, Application Developer}).

We identified and classified 20 types of testing issues in 3 subcategories (see Figure \ref{fig:Taxonomy} and the Issue Taxonomy sheet in \cite{replpack}). Each of them is briefly described below.

\begin{itemize}
\item \textit{Test Case Issue} (45, 1,66\%): This subcategory covers the problems with test cases written to evaluate the expected output compliance with specific requirements for the microservices systems. Most of them are related to \textsc{faulty test case}, \textsc{missing test case}, and \textsc{syntax error in test case}.
 
\item \textit{Code and Component Test} (21, 0.77\%): This subcategory deals with the issues mainly related to \textsc{debugging}, \textsc{api testing}, and \textsc{missing design test} of microservices systems.
 
\item \textit{Application Test} (11, 0.40\%): This subcategory gathers the issues related to overall microservices application testing, which include \textsc{load test case}, \textsc{broken integration test}, and \textsc{application security testing} issues.
\end{itemize}

\textbf{11. Documentation Issue} (75/2698, 2.77\%): Documentation for microservices systems may suffer from several problems. We identified and classified 8 types of documentation issues in two subcategories (see Figure \ref{fig:Taxonomy} and the Issue Taxonomy sheet in \cite{replpack}). Each of them is briefly described below.

\begin{itemize}

\item \textit{Insufficient Document} (49, 1.81\%): This subcategory covers the problems related to \textsc{outdated document}, \textsc{broken images}, and \textsc{inappropriate examples}. For instance, one contributor of the eShopOnContainers project mentioned the issue of “\textit{Out of date wiki guide for vs2015, \#2030}”.

\item \textit{Readability Issue} (26, 0.95\%): This subcategory is related to readability problems with provided documentation. The leading types of readability issues are \textsc{poor readability}, \textsc{missing readme file}, and \textsc{old readme file}.
\end{itemize}

\textbf{12. Graphical User Interface (GUI) Issue} (70/2698, 2.59\%): This category reports the problems that can wreck the GUI of microservices systems. We identified and classified 68 types of GUI issues in 3 subcategories (see Figure \ref{fig:Taxonomy} and the Issue Taxonomy sheet in \cite{replpack}). Each of them is briefly described below.

\begin{itemize}
\item \textit{Broken User Interface Elements} (38, 1.41\%) are dysfunctional User Interface (UI) elements (e.g., buttons or text fields) that can become the reason for inconsistencies in the page layout across different devices (e.g., mobile and desktop browsers). This subcategory represents the faults mainly related to \textsc{front end crash}, \textsc{uneditable contents}, and \textsc{broken images in ui}.

\item \textit{Missing Information and Legacy UI Artifacts} (32, 1.19\%): This subcategory contains the problems of wrong and incomplete information along with outdated UI artifacts, mainly related to \textsc{wrong gui display}, \textsc{displaying incomplete information}, and \textsc{selection not working}.
\end{itemize}

\textbf{13. Update and Installation Issue} (68/2698, 2.52\%): This category gathers the identified problems related to update and installation of the packages, libraries, tools, and containerization platforms required to develop and manage microservices systems. We identified and classified 16 types of update and installation issues in 2 subcategories (see Figure \ref{fig:Taxonomy} and the Issue Taxonomy sheet in \cite{replpack}). Each of them is briefly described below.

\begin{itemize}

\item \textit{Update Error} (45, 1.67\%): This subcategory represents the errors in which developers face the issues of outdated packages, platforms, technologies, and backward compatibility. The leading types of issues are \textsc{outdated install packages} ,\textsc{backward compatibility issue}, and \textsc{json update error}.

\item \textit{Installation Error} (23, 0.85\%): The development of microservices systems can be interrupted because of failure to install required languages, packages, and platforms. The top three types of issues are \textsc{language package installation error}, \textsc{npm (node package manager) error}, and \textsc{gke (google kubernetes engine) installation error}.
\end{itemize}

\textbf{14. Database Issue} (65/2698, 2.40\%): The ownership of the microservices system database is usually distributed, and most of the microservices are autonomous and have a private data store relevant to their functionality. The distributed nature of microservices systems brings challenges like database implementation, data accessibility, and database connectivity. The microservices practitioners also mentioned several other database issues during the interviews. One representative quotation is depicted below. 

\faHandORight{} “\textit{Relational database for microservices systems. This issue occurs when we migrate from monolithic applications to microservices systems. It was mainly because of the missing transaction management system for getting data from the database of the old application (that was a relational database) through the microservices application}” (\textbf{P6, Software Architect, Developer}).

We identified and classified 24 types of database issues in 3 subcategories (see Figure \ref{fig:Taxonomy} and the Issue Taxonomy sheet in \cite{replpack}). Each of them is briefly described below
\begin{itemize}

\item \textit{Database Connectivity (26, 0.96\%)}: This subcategory refers to the issues that occur while establishing a database connection. The leading types of issues in this subcategory are \textsc{sql container failure} , \textsc{sql transient connection failure} , and \textsc{database creation failure}.

\item \textit{Database Query} (31, 1.14\%): The errors in this subcategory cover search query issues during the development of microservices systems. Such issues may cause a database performance bottleneck. The leading types of issues in this subcategory are \textsc{wrong query}, \textsc{elasticsearch database error}, and \textsc{database search error}.

\item \textit{Others} (8, 0.29\%): Other types of database issues that cannot be classified into the above subcategories are included in this subcategory, which are mainly related to \textsc{database migration}, \textsc{database storage}, and \textsc{database adapter}. textcolor{red}{For example, one developer of the Spinnaker project mentioned that “\textit{DB adapter issue - can not perform a textual search with a list, \#1374}”}.
\end{itemize}

\textbf{15. Storage Issue} (54/2698, 2.00\%): This category reports storage space problems during the development, execution, and management of microservices systems. We identified and classified 13 types of storage issues in 2 subcategories (see Figure \ref{fig:Taxonomy} and the Issue Taxonomy sheet in \cite{replpack}), and each subcategory is further described below.

\begin{itemize}
\item \textit{Storage Size Constraints} (48, 1.78\%): Different microservices have different data storage requirements. This subcategory covers the storage size constraints, mainly related to \textsc{lack of main memory}, \textsc{cache issue}, and \textsc{storage backend failure}. 
 
\item \textit{Large Data Size} (6, 0.21\%): This subcategory gathers the issues related to large data size, including \textsc{large image size}, \textsc{large message size}, and \textsc{large file}.
\end{itemize}

\textbf{16. Performance Issue} (45/2698, 1.67\%): Microservices systems offer various advantages over monolithic systems. However, several types of issues wreak the performance of microservices systems. The interviewees also mentioned performance overhead as an issue, and one representative quotation is depicted below.

\faHandORight{} “\textit{Unlike a monolithic application whose deployment and management are seemingly easier due to centralized control and monitoring, a microservices-based application has numerous independent services that may be deployed on different infrastructures and platforms. Such an aspect increases its performance overhead.  I also think that microservices systems consume more resources, creating a heavy burden for servers. In the response achieving the performance goal become questionable}” (\textbf{P15, Software Architect}).

We identified and classified 16 types of performance issues in three subcategories (see Figure \ref{fig:Taxonomy} and the Issue Taxonomy sheet in \cite{replpack}). Each of them is briefly described below.

\begin{itemize}
\item \textit{Service Response Delay} (28, 1.03\%): This subcategory gathers the types of issues regarding delay in service response that ruins the performance of microservices systems, such as \textsc{long wait}, \textsc{inconsistent payloads}, and \textsc{slow query}. 

\item \textit{Resource Utilisation} (13, 0.48\%): In this subcategory, we collected the issues that degrade the performance of microservices systems due to resource utilisation. These issues are mainly related to \textsc{load balancer error}, \textsc{high cpu usage}, and \textsc{rate limiting error}. One example of the load balancer issue mentioned by a Spinnaker project contributor is “\textit{Failed to create a Load Balancer when creating a new application, \#1583}”.

\item \textit{Lack of Scalability} (4, 0.14\%): Scalability is the system's ability to respond the user demands and change workload by adding or removing resources. This subcategory gathers the issues related to scalability that can hinder microservices system growth, for instance, \textsc{scale to cluster error} and \textsc{circuit breaker issue}.
\end{itemize}

\textbf{17. Networking Issue} (41/2698, 1.51\%): Deploying, executing, and communicating in microservices systems over the network is complex. It is observed that many problems may disrupt the network. We identified and classified 20 types of networking issues in 2 subcategories (see Figure \ref{fig:Taxonomy} and the Issue Taxonomy sheet in \cite{replpack}). Each of them is briefly described below.

\begin{itemize}
\item \textit{Hosting and Protocols} (22, 0.81\%): This subcategory represents the issues related to hosting protocols, ports, and topologies for microservices systems. The leading types of issues in this subcategory are \textsc{localhost error}, \textsc{ip address issue}, and \textsc{udp (user datagram protocol) discovery error}.  
\item \textit{Service Accessibility} (19, 0.70\%): This subcategory of issues represent the cases where microservices practitioners face service accessibility problems. The leading types of issues are \textsc{webhook error}, \textsc{broken urls}, and \textsc{dns (domain name system) error}.
\end{itemize}

\textbf{18. Typecasting Issue} (35/2698, 1.29\%): This category is related to the typecasting issues that occur when assigning a value of one primitive data type to another type. We identified and classified 9 types of typecasting issues in 2 subcategories (see Figure \ref{fig:Taxonomy} and the Issue Taxonomy sheet in \cite{replpack}). Each of them is briefly described below.

\begin{itemize}
\item \textit{Type Conversion} (20, 0.74\%): This subcategory deals with the issues when variables are not correctly converted from one type to another. The top three types of conversion issues are \textsc{identity conversion}, \textsc{boxing conversion}, and \textsc{enumeration validation issue}.

\item \textit{Narrow/Wide Conversion} (15, 0.55\%): These issues occur when the compiler converts variables of a larger type into a smaller type (e.g., double to float) or a smaller type into a larger type (e.g., float to double). The top two types of narrow/wide conversion issues are \textsc{narrowing primitive conversion} and \textsc{narrowing reference conversion}. 
\end{itemize} 

\textbf{19. Organizational Issue} (7/2698, 0.25\%): We derived this category based on the interviewees’ feedback on the taxonomy of the issues (see Figure \ref{fig:Taxonomy} and the Issue Taxonomy sheet in \cite{replpack}). The interviewees only mentioned three types of issues in this category, including \textsc{team management}, \textsc{operational and tooling overhead}, and \textsc{service size}. We depicted two representative quotations below.

\faHandORight{} “\textit{One of the critical challenges in organizations is team management according to available people, their expertise, and their working habits}” (\textbf{P1, Software Architect, Developer}). 

\faHandORight{} “\textit{Creating a reasonable size for each microservices' (I mean, each microservice should have sufficient responsibilities). This is a bit tricky because most of the issues are rooted here}” (\textbf{P6, Software Architect, Developer}). 

\begin{tcolorbox} [sharp corners, boxrule=0.1mm,]
\footnotesize
\textbf{Key Findings of RQ1}: We identified 2,641 instances of issues by mining developer discussions in 15 open-source microservices systems with 48 instances of issues mentioned by the interviewees and 9 instances of issues mentioned by the survey participants, which are 2,698 issues in total. The issue taxonomy consists of 19 categories, 54 subcategories, and 402 types, indicating the diversity of the issues in microservices systems. The majority of issues are related to Technical Debt (25.46\%), CI/CD (11.60\%), and Exception Handling (8.45\%).
\end{tcolorbox}

\subsection{Causes of Issues (RQ2)}
\label{sec:results_RQ2}

The taxonomy of causes of microservices issues is provided in Table \ref{tab:CausesTaxnomey}. It is worth mentioning that not all the issue discussions provide the information about their causes. Therefore, we identified 2,225 issue discussions containing information about the causes. The taxonomy of causes is derived by mining developer discussions (i.e., 2,225 instances of causes), conducting practitioner interviews (i.e., 31 instances of causes, see Section \ref{InterviewsDataAnalysis}), and conducting a survey (i.e., 11 instances of causes, see Section \ref{sec:results_RQ4}). Hence, we got a total of 2,267 instances of causes.  We identified a total of 228 types of causes that can be classified into 8 categories and 26 subcategories. Due to space limitations, we only list the top two types of causes for each subcategory in Table \ref{tab:CausesTaxnomey}. The detail of the types of causes can be found in the dataset \cite{replpack}. The results show that General Programming Error (860 out of 2267), Missing Features and Artifacts (386 out of 2267), and Invalid Configuration \& Communication Problems (382 out of 2267) are the top three categories of causes. Each cause category is briefly discussed below.

\textbf{1. General Programming Error (GPE)} (860/2267, 37.93\%): This category captures the causes that are based on a broad range of errors occurred in different phases of microservices system development, such as coding (e.g., syntax errors), testing (e.g., incorrect test cases), and maintenance (e.g., wrong examples in documentation) phases. We identified and classified 78 types of GPE causes in six subcategories (see Table \ref{tab:CausesTaxnomey} and the Cause Taxonomy sheet in \cite{replpack}). Each of them is briefly described below.

\begin{itemize}
\item \textit{Compile Time Error} (377, 16.62\%): This subcategory gathers the causes of issues in which microservices practitioners violate the rules of writing syntax for microservices systems codes. These causes must be addressed before the program can be compiled. We identified 16 types of causes in this subcategory in which the top three are \textsc{semantic error}, \textsc{syntax error in code}, and \textsc{variable mutations}. 
 
\item \textit{Erroneous Method Definition and Execution} (262, 11.55\%): The causes in this subcategory are related to incorrect or partly correct definitions and executions of methods associated with object messages. Generally, methods are referred as class building blocks linked together for sharing and processing data to produce the desired results. We identified 27 types of causes in this subcategory in which the top three are \textsc{lack of cohesion in methods}, \textsc{long message chain}, and \textsc{wrong parameterization}.
 
\item \textit{Incorrect Naming and Data Type} (157, 6.92\%): This subcategory covers the causes related to choosing incorrect names and data types for identifiers, methods, packages, and other entities in the source code. Our taxonomy contains 18 types of causes in this subcategory, and among them \textsc{wrong data conversion}, \textsc{wrong data type}, and \textsc{wrong use of data types} are the top three types of causes. 
 
\item \textit{Testing Error} (25, 1.10\%): This subcategory covers the causes behind testing issues in microservices systems. In this subcategory, we identified 6 types of causes in which the top two types of causes are \textsc{incorrect test case} and \textsc{incorrect syntax in test cases}.
 
\item \textit{Poor Documentation} (22, 0.97\%): The documentation of software systems may contain critical information that describes the software product capabilities for system stakeholders. We identified 7 types of causes in this subcategory, in which the top three are \textsc{typo in documents}, and \textsc{wrong example in documents}. 
\end{itemize}

\begin{itemize}
\item \textit{Query and Database Issue} (11, 0.48\%): This subcategory contains the causes behind database issues in microservices systems. We identified 4 types of causes in this subcategory, and the top three types of causes are \textsc{wrong query parameters}, \textsc{missing query parameters}, and \textsc{incorrect querying range}.
\end{itemize}

\textbf{2. Missing Features and Artifacts (MFA)} (386/2267, 17.02\%): This category represents the causes behind microservices systems issues that occur due to missing required features, packages, files, variables, and documentation and tool support. We identified and classified 27 types of MFA causes in four subcategories (see Table \ref{tab:CausesTaxnomey} and the Cause Taxonomy sheet in \cite{replpack}). Each of them is briefly described below.

\begin{itemize}
\item \textit{Missing Features} (186, 8.20\%) denotes the nonexistence of system functionality in microservices systems. In this subcategory, we identified 14 types of causes, in which the top two types of causes are \textsc{missing required system features} and \textsc{missing security features}.
 
\item \textit{Missing Documentation and Tool Support} (104,4.58\%): Proper documentation and tool support is vital to keep the record of and track changes between system requirements, architecture, and source code. It also guide various system stakeholders (e.g., architects, developers, end-users) regarding design, architecture, and coding standards use in microservices systems. Moreover, several types of tool support are also necessary for different phases of microservices system development (e.g., development and deployment). The absence of proper documentation and tool support can bring several types of issues to microservices systems. This subcategory contains 4 types of missing documentation and tool support causes, and mong them \textsc{missing readme file} and \textsc{missing development and deployment tool support} are identified as the leading causes.

\item \textit{Missing Packages and Files} (68, 2.99\%): This subcategory groups the causes related to absence of required resources, packages, and files for developing, deploying, and executing microservices systems. We collected 9 types of missing packages and files causes, in which the leading three types of causes are \textsc{missing resource}, \textsc{missing required package}, and \textsc{missing api}.
 
\item \textit{Missing Variables} (28, 1.23\%): A few missing variables are also identified as the causes for several microservices issues. Compilers throw the error messages of missing variables if variables are set to a nonexistent directory or have the wrong names. This subcategory contains 7 types of missing variables causes, and among them \textsc{missing environment variable} and \textsc{missing properties} are identified as the leading causes.
\end{itemize}

\textbf{3. Invalid Configuration and Communication (ICC) Problem} (382/2267, 16.85\%): Considering a large number of microservices, their distributed nature, and third-party plugins, microservices systems need to be configured for complete business operations properly. Each microservice has its instances and process, and services interact with each other using several inter-service communication protocols (e.g., HTTP, gRPC, message brokers AMQP). One of the interviewees also mentioned the following cause regarding invalid configuration and communication.  

\faHandORight{} “\textit{Microservices systems typically use one or more infrastructure and 3rd party services. Examples of infrastructure services include a service registry, a message broker, and a database server. During the configuration of microservices, a service must be provided with configuration data that tells it how to connect to the external or 3rd party services — for example, the database network location and credentials}” (\textbf{P1, Software Architect, Developer}).

We identified and classified 29 types of causes in two subcategories (see Table \ref{tab:CausesTaxnomey} and the Cause Taxonomy sheet in \cite{replpack}). Each of them is briefly described below.

\begin{itemize}
\item \textit{Incorrect Configuration} (290, 12.79\%): Configuration management in microservices systems is a hefty task because microservices are scattered across multiple servers, containers, databases, and storage units. Each microservices may have multiple instances. Therefore, an incorrect configuration may lead to several types of errors. This subcategory contains 16 causes for different types of microservices issues, in which the leading two causes are \textsc{incorrect configuration setting} and \textsc{wrong connection closure}.
 
\item \textit{Server and Access Problem} (92, 4.05\%): Each microservices acts as a miniature application that communicates with each other. We need to configure the infrastructure layers of the microservice system for sharing different types of resources. A poor configuration may lead to problems of accessibility for servers and other resources that bring multiple issues in microservices systems. In this subcategory, we identified 13 types of causes, in which the leading three types of causes are \textsc{transient failure}, \textsc{service registry error}, and \textsc{wrong communication protocol}.
\end{itemize}

\textbf{4. Legacy Versions, Compatibility, and Dependency (LC\&D) Problem} (222/2267, 9.79\%): This category represents a broad range of causes arising from outdated repositories, applications, documentation versions, development and deployment platforms, APIs, libraries, and packages. One of the interviewees also mentioned several causes regarding microservices issues, especially compatibility and dependency. One representative quotation is depicted below. 

\faHandORight{} “\textit{Along with the legacy code version, some of the critical reasons for the microservices issues are i) no clear strategy for code repository and branching, the mix of technologies (each team uses its way of development), dependencies on other services which are not released yet but going for integration and load testing, development pace varies from team to team, lack of centralized release manager, incompatibility of a new version of the service with previous services.}” (\textbf{P8, DevOps Consultant}).

We identified and classified 28 types of causes in five subcategories (see Table \ref{tab:CausesTaxnomey} and the Cause Taxonomy sheet in \cite{replpack}). Each of them is briefly described below. 

\begin{itemize}
\item \textit{Compatibility and Dependency} (59, 2.60\%): A typical microservices system consists of several independent services running on multiple servers or hosts \cite{waseem2022}. However, some microservices also depend on other microservices to complete business operations. Usually, practitioners ensure the compatibility (e.g., backward compatibility) of each microservice with previous versions of the microservices systems during the upgrading. In this subcategory, we identified 5 types of causes, in which the leading three types of causes are \textsc{compatibility error}, and \textsc{outdated dependency}.
 
\item \textit{Outdated and Inconsistent Repositories} (53, 2.33\%): This subcategory gathers the causes which are the source of the issues that occur when the online code repository of version control systems has updated the local files repository. The most frequent causes in this subcategory are \textsc{old repository version}, \textsc{old dev branch}, and \textsc{version conflicts}.
 
\item \textit{Outdated Application and Documentation Version} (45, 1.98\%): We identified several causes arising from using outdated applications and documentation versions in the selected systems, which are mainly related to \textsc{document not updated}, and \textsc{deprecated software version}. 
 
\item \textit{Outdated Development and Deployment Platforms} (38, 1.67\%): Development and deployment platforms help developers build, test, and deploy microservices systems efficiently. We identified 10 types of causes in this subcategory, and among them \textsc{old Kubernetes version} and \textsc{old visual studio version} are the top two leading type of causes. 
 
\item \textit{Outdated APIs, Libraries, and Packages} (27, 1.09\%): This subcategory covers the causes arising from the usage of outdated APIs, libraries, and packages. The most frequent types of causes in this subcategory are \textsc{outdated version of library} and \textsc{old version of package}.
\end{itemize}

\textbf{5. Service Design and Implementation Anomalies (SD\&IA)} (174/2267, 7.67\% ): This category covers the causes of issues when microservices practitioners cannot address the associated complexity of distributed systems at the design level. The interviewees also mentioned several causes related to service design and implementation anomalies, and one representative quotation is depicted below.

\faHandORight{} “\textit{According to my experience, the primary reason behind the design issues is the distributed nature of microservices systems, which is becoming increasingly complicated with growing systems—especially where multiple subsystems also need to be integrated}” (\textbf{P7, Software Engineer}).

We identified and classified 17 types of SD\&IA causes in 2 subcategories (see Table \ref{tab:CausesTaxnomey} and the Cause Taxonomy sheet in \cite{replpack}). Each of them is briefly described below.
 
\begin{itemize}
\item \textit{Code Design Anomaly} (130, 5.73\%): Code design anomalies refer to poorly written code that may lead to several problems (e.g., difficulties in maintenance or future enhancements) in microservices systems. Our taxonomy gets 12 types of causes in this subcategory and among them \textsc{poor code readability}, \textsc{messy code}, and \textsc{data clumps} are the top three types of causes.

\item \textit{System Design Anomaly} (44, 1.94\%): System design anomalies refer to poorly designed microservices architecture that may lead to maintainability, scalability, and performance issues. This subcategory covers 5 types of causes. Among them, the top three types of causes are \textsc{wrong dependencies chain}, \textsc{lack of asrs}, and \textsc{wrong application decomposition}.
\end{itemize}

\textbf{6. Poor Security Management (PSM)} (126/2267, 5.55\%): Microservices systems are distributed over data centres, cloud providers, and host machines. The security of microservices systems is a multi-faceted problem that requires a layered solution to cope with various types of vulnerabilities. The interview participants also mentioned several causes for security issues, and one representative quotation is depicted below.

\faHandORight{} “\textit{I think the basic reasons behind the security issues are poor understanding of microservices architecture, large attack surfaces (many distributed points),  error-prone encryption techniques while services are communicating, and insecure physical devices}” (\textbf{P5, Solution Architect}).

 We identified and classified 21 types of PSM causes in 3 subcategories (see Table \ref{tab:CausesTaxnomey} and the Cause Taxonomy sheet in \cite{replpack}). Each of them is briefly described below.

\begin{itemize}
\item \textit{Coding Level} (55, 2.42\%): This subcategory collects the causes where strict security principles and practices are not followed to prevent potential vulnerabilities while writing code of microservices systems. This subcategory covers 7 types of coding level causes, and among them, the top three types of causes are \textsc{unsafe code}, \textsc{malformed input}, and \textsc{wrong implementation of security api}.
 
\item \textit{Communication Level} (40, 1.76\%): Given the polyglot and distributed nature of microservices systems, practitioners need to secure the inter-microservices communication. This subcategory covers 10 types of communication level causes, and among them, the top two types of causes are \textsc{security dependencies} and \textsc{wrong access control}.
 
\item \textit{Application Level} (29, 1.27\%): The application level of security refers to security practices implemented at the interface between an application and various components (e.g., databases, containers) of microservices systems. This subcategory only contains 3 types of causes that are \textsc{insecure configuration management}, \textsc{missing security features}, and \textsc{violation of the security policies}.
\end{itemize}

\textbf{7. Insufficient Resources (IR)} (93/2267, 4.10\% ): Microservices systems are at risk of delivering the required outcome without sufficient resources. We identified and classified 4 types of IR causes in two subcategories (see Table \ref{tab:CausesTaxnomey} and the Cause Taxonomy sheet in \cite{replpack}). Each of them is briefly described below.

\begin{itemize}
\item \textit{Memory Issue} (68, 2.99\%): Microservices systems are developed by using multiple languages (e.g., Java, Python, C++) and platforms (e.g., containers, virtual machines). Some languages and platforms consume more memory than others. For instance, C/C++ consumes less memory than Java, and Python and Perl consume less memory than C/C++. This subcategory covers 2 types of causes that are \textsc{limited memory for process execution}, and \textsc{ide problem with memory}. 

\item \textit{Lack of Human Resources, Tools and Platforms } (25, 1.10\%): This subcategory deals with the causes related to tools and platforms support for microservices systems. Only 3 types of causes related to this subcategory are identified, including \textsc{lack of tool support}, and\textsc{deployment platform problem}.
\end{itemize}

\textbf{8. Fragile Code (FC)} (24/2267, 1.05\%): Fragile code refers to code that is difficult to change, and a minor modification in fragile code may break the service or module. We identified and classified 6 types of FC causes in two subcategories (see Table \ref{tab:CausesTaxnomey} and the Cause Taxonomy sheet in \cite{replpack}). Each of them is briefly described below.
\begin{itemize}
\item \textit{Poor Implementation of Code} (20, 0.88\%): This subcategory gathers the causes of poor code quality, which exhibits the buggy behaviour of microservices systems. We identified 3 types of causes in this subcategory, including \textsc{poor object-oriented design}, \textsc{poor code reusability}, and \textsc{unnecessary code}.
 
\item \textit{Poor Code Flexibility} (11, 0.48\%): Code flexibility is important for long-lived microservices project code bases, however, we found a few issues that occurred due to poor code flexibility. The 3 types of causes in this subcategory are \textsc{divergent changes}, \textsc{delayed refactoring}, and \textsc{poorly organized code}.
\end{itemize}

{\renewcommand{\arraystretch}{1}
\begin{table*}[!h]
\centering
\scriptsize

\caption{Taxonomy of causes of issues in microservices systems}
\label{tab:CausesTaxnomey}
\begin{tabular}{|c|l|l|}
\hline
\multicolumn{1}{|l|}{\textbf{Category of Causes}} &
 \textbf{Subcategory of Causes} &
 \textbf{Type of Causes} \\ \hline

 \multirow{12}{*}{\begin{tabular}[c]{@{}c@{}}General Programming \\Error (GPE) (860)\end{tabular}} &
 \multirow{2}{*}{Compile Time Error (377)} &
 Semantic Error (205) \\ \cline{3-3} 
 &
 &
 Syntax Error in Code (112) \\ \cline{2-3} 
 &
 \multirow{2}{*}{\begin{tabular}[c]{@{}l@{}}Erroneous Method Definition and Execution (262)\end{tabular}} &
 Lack of Cohesion in Methods (38) \\ \cline{3-3} 
 &
 &
 Long Message Chain (36) \\ \cline{2-3} 
 &
 \multirow{2}{*}{\begin{tabular}[c]{@{}l@{}}Incorrect Naming and Data Type (157)\end{tabular}} &
 Wrong Data Conversion (39) \\ \cline{3-3} 
 &
 &
 Wrong Data Type (20) \\ \cline{2-3} 
 &
 \multirow{2}{*}{Testing Error (25)} &
 Incorrect Test Case (20) \\ \cline{3-3} 
 &
 &
 Wrong Implementation of Security Pattern (5) \\ \cline{2-3} 
 &
 \multirow{2}{*}{Poor Documentation (22)} &
 Typo in Documents (9) \\ \cline{3-3} 
 &
 &
 Wrong Example in Documents (7) \\ \cline{2-3} 
 &
 \multirow{2}{*}{Query and Database Issue (13)} &
 Wrong Query Parameters (6) \\ \cline{3-3} 
 &
 &
 Missing Query Parameters (3) \\ \hline

 \multirow{8}{*}{\begin{tabular}[c]{@{}c@{}}Missing Features \\and\\ 
 Artifacts (MFA) (386)\end{tabular}} &
 \multirow{2}{*}{Missing Features (186)} &
 Missing Required System Features (111) \\ \cline{3-3} 
 &
 &
 Missing Security Features (37) \\ \cline{2-3} 
 &
 \multirow{2}{*}{Missing Documentation and Tool Support (104)} &
 Missing Readme File (88) \\ \cline{3-3} 
 &
 &
 Missing Tool Support (14) \\ \cline{2-3} 
 &
 \multirow{2}{*}{Missing Packages and Files (68)} &
 Missing Resource (41) \\ \cline{3-3} 
 &
 &
 Missing Required Package (15) \\ \cline{2-3} 
 &
 \multirow{2}{*}{Missing Variables (28)} &
 Missing Environment Variable (14) \\ \cline{3-3} 
 &
 &
 Missing Properties (8) \\ \hline

 \multirow{4}{*}{\begin{tabular}[c]{@{}c@{}}Invalid Configuration \\and\\ Communication (ICC) Problem (382)\end{tabular}} &
 \multirow{2}{*}{Incorrect Configuration (285)} &
 Incorrect Configuration Setting (227) \\ \cline{3-3} 
 &
 &
 Wrong Connection Closure (24) \\ \cline{2-3} 
 &
 \multirow{2}{*}{Server and Access Problem (92)} &
 Transient Failure (21) \\ \cline{3-3} 
 &
 &
 Service Registry Error (24) \\ \hline

 \multirow{10}{*}{\begin{tabular}[c]{@{}c@{}}Legacy Versions, Compatibility,\\ and\\ Dependency (LC\&D) Problem (222)\end{tabular}} &
 \multirow{2}{*}{Compatibility and Dependency (59)} &
 Compatibility Error (42) \\ \cline{3-3} 
 &
 &
 Outdated Dependency (9) \\ \cline{2-3} 
 &
 \multirow{2}{*}{Outdated and Inconsistent Repositories (53)} &
 Old Repository Version (38) \\ \cline{3-3} 
 &
 &
 Old DEV Branch (7) \\ \cline{2-3} 
 &
 \multirow{2}{*}{Outdated Application and Documentation Version (45)} &
 Document not Updated (33) \\ \cline{3-3} 
 &
 &
 Deprecated Software Version (6) \\ \cline{2-3} 
 &
 \multirow{2}{*}{Outdated Development and Deployment Platforms (38)} &
 Old Kubernetes Version (7) \\ \cline{3-3} 
 &
 &
 Old IDE Version (6) \\ \cline{2-3} 
 &
 \multirow{2}{*}{Outdated APIs, Libraries, and Packages (27)} &
 Outdated Version of Library (15) \\ \cline{3-3} 
 &
 &
 Old Version of Required Package (11) \\ \hline

 \multirow{4}{*}{\begin{tabular}[c]{@{}c@{}}Service Design\\ and\\ Implementation Anomalies (SD\&IA) (174)\end{tabular}} &
 \multirow{2}{*}{Code Design Anomaly (125)} &
 Poor Code Readability (54) \\ \cline{3-3} 
 &
 &
 Messy Code (35) \\ \cline{2-3} 
 &
 \multirow{2}{*}{System Design Anomaly (44)} &
 Wrong Dependencies Chain (16) \\ \cline{3-3} 
 &
 &
 Lack of ASRs (5) \\ \hline

 \multirow{6}{*}{\begin{tabular}[c]{@{}c@{}}Poor Security\\ Management (PSM) (126)\end{tabular}} &
 \multirow{2}{*}{Coding Level (55)} &
 Unsafe Code (23) \\ \cline{3-3} 
 &
 &
 Malformed Input (12) \\ \cline{2-3} 
 &
 \multirow{2}{*}{Communication Level (42)} &
 Security Dependencies (8) \\ \cline{3-3} 
 &
 &
 Wrong Access Control (5) \\ \cline{2-3} 
 &
 \multirow{2}{*}{Application Level (29)} &
 Insecure Configuration Management (17) \\ \cline{3-3} 
 &
 &
 Missing Security Features (7) \\ \hline
 \multirow{4}{*}{\begin{tabular}[c]{@{}c@{}}Insufficient Resources\\ (IR) (93)\end{tabular}} &
 \multirow{2}{*}{Memory Issue (68)} &
 Limited Memory for Process Execution (66) \\ \cline{3-3} 
 &
 &
 IDE Problem with Memory (2) \\ \cline{2-3} 
 &
 \multirow{2}{*}{Lack of Human Resources,Tools and Platforms (25)} &
 Lack of Tool Support (7) \\ \cline{3-3} 
 &
 &
 Deployment Platform Problem (5) \\ \hline

 \multirow{4}{*}{\begin{tabular}[c]{@{}c@{}}Fragile Code\\ (FC) (24)\end{tabular}} &
 \multirow{2}{*}{Poor Implementation of Code (20)} &
 Poor Object Oriented Design (6) \\ \cline{3-3} 
 &
 &
 Poor Code Reusability (5) \\ \cline{2-3} 
 &
 \multirow{2}{*}{Poor Code Flexibility (11)} &
 Tightly Services Components (7) \\ \cline{3-3} 
 &
 &
 Divergent Changes (9) \\ \hline
\end{tabular}
\end{table*}
}

\begin{tcolorbox} [sharp corners, boxrule=0.1mm,]
\footnotesize
\textbf{Key Findings of RQ2}: We found that not all the issue discussions provide the information about their causes, and finally we identified 2,225 issue discussions containing information about the causes with 31 causes mentioned by the interviewees and 11 causes indicated by the survey participants, which are 2,267 causes instances in total. The cause taxonomy of microservices issues consists of 8 categories, 26 subcategories, and 228 types. The majority of causes are related to General Programming Errors (37.93\%), Missing Features and Artifacts (17.02\%), and Invalid Configuration and Communication (16.85\%).
\end{tcolorbox}

\subsection{Solutions of Issues (RQ3)}
\label{sec:results_RQ3}
The taxonomy of solutions for microservices issues is provided in Table \ref{tab:SolutionsTaxnomey}. It is worth mentioning that not all the issue discussions provide the information about their solutions. Therefore, we identified 1,899 issue discussions containing the information about the solutions. The taxonomy of solutions is derived by mining developer discussions (i.e., 1,899 solutions), conducting practitioner interviews (i.e., 36 solutions, see Section \ref{sec:Ext&Syn}), and conducting a survey (i.e., one instance of solution, see Section \ref{sec:results_RQ4}). Hence, we got a total of 1,936 solutions. We identified a total of 196 types of solutions that can be classified in 8 categories and 35 subcategories. Due to space limitations, we only list the top two types of solutions for each subcategory in Table \ref{tab:SolutionsTaxnomey}. The detail of the types of solutions can be found in the dataset \cite{replpack}. The results show that Fix Artifacts (1056 out of 1936), Add Artifacts (360 out of 1936), and Modify Artifacts (210 out of 1936) are the top three categories of solutions. Each solution category is briefly discussed below.

\textbf{1. Fix Artifacts} (1056/1936, 54.54\%): During the analysis of developer discussions about microservices issues, we identified that most developers did not explicitly mention any solution to the problems. They fixed the issue in the local repository and send the fixed code (e.g., through a pull request) to the maintainer of the public repository. In this case, from the developer discussions, we could not find exactly what they added, removed, or modified in the project to fix a specific issue. Therefore, we named this category Fix Artifacts. We identified and classified 25 types of fix artifacts solutions in 4 subcategories (see Table \ref{tab:SolutionsTaxnomey} and the Solution Taxonomy sheet in \cite{replpack}). Each of them is briefly described below.

\begin{itemize}
\item \textit{Fix Code Issue} (912, 47.10\%): The solutions in this subcategory are related to the direct repair of the source code by developers. We identified 14 types of solutions in this subcategory and among them \textsc{fix source code of issues}, \textsc{fix illegal symbols (syntax) in code}, and \textsc{clean code} are the top three types of solutions.

\item \textit{Fix Testing Issue} (107, 5.52\%): This subcategory covers the types of solutions with which testing issues have been fixed. We collected 2 types of solutions in this subcategory that are \textsc{debug code} and \textsc{add test services in containers}.

\item \textit{Fix Build Issue} (33, 1.70\%): Build systems are essential for developing, deploying, and maintaining microservices systems. In contrast, build failures frequently occur across the development life cycle, bringing non-negligible costs in microservices system development. We collected 3 types of solutions, including \textsc{fix errors in build files}, \textsc{correct build type}, and \textsc{hide fail status}.

\item \textit{Fix GUI Issue} (4, 0.20\%): We collected two types of solutions related to repairing the graphical user interface of microservices systems, and they are \textsc{fix backwards incompatible ui} and \textsc{fix broken screenshot}. 
\end{itemize}

\textbf{2. Add Artifacts} (360/1936, 18.59\%): This category covers the types of solutions for addressing missing features, packages, files, variables, and documentation and tool support issues. The interview participants also mentioned that they adopted solutions for addressing several types (e.g., CI/CD, Security) of issues that have occurred because of missing features, and one representative quotation is depicted below.

\faHandORight{} “\textit{To address these issues, we recently adopted the `service mesh' approach. This approach helps address security, latency and scalability, fault identification, and runtime error detection in microservices systems. A service mesh pattern can also provide features for a service health check with the lowest latency to address errors and fault identification issues}” \textbf{(P7, Software Engineer)}.

We identified and classified 33 types of solutions in 10 subcategories (see Table \ref{tab:SolutionsTaxnomey} and the Solution Taxonomy sheet in \cite{replpack}). Each of them is briefly described below.

\begin{itemize}
\item \textit{Add Features and Services} (131, 6.76\%): Developers added required features and services to microservices systems. System features and services refer to a process that accepts one or more inputs and returns outputs for particular system functionality \cite{paolucci2002semantic}. We identified 5 types of solutions in this subcategory, and among them, \textsc{add missing features}, \textsc{add security features}, and \textsc{add communication protocols} are the top three types of solutions.
 
\item \textit{Add Files, Templates, and Interfaces} (39, 2.01\%): This subcategory covers the solutions for adding missing files, templates, and interfaces in microservices systems. We identified 3 types of solutions, including \textsc{add files}, \textsc{add templates}, and \textsc{add interfaces}.
 
\item \textit{Add Methods and Modules} (33, 1.70\%): We identified the solutions in which developers mainly added or repaired the missing methods and modules to address several types (e.g., Compilation, Service Execution and Communication, Build) of issues in microservices systems. We identified 3 types of solutions, including \textsc{add constructor}, \textsc{add parameters in methods}, and \textsc{add security certificates}.

\item \textit{Add Test Cases} (28, 1.44\%): We identified the solutions in which developers added test cases to address testing issues of microservices systems. We identified 4 types of solutions in this subcategory in which the top three are \textsc{add test cases to validate service}, \textsc{generate correct test cases}, and \textsc{add test cases to validate data type}.

\item \textit{Add Classes and Packages} (27, 1.39\%): This subcategory covers the solutions in which developers added classes and packages to address the microservices issues. We identified 3 types of solutions, including \textsc{add packages}, \textsc{add properties}, and \textsc{add objects}.

\item \textit{Implement Patterns and Strategies (23, 1.18\%)}: The solutions in this subcategory are based on interviewees’ feedback in which they mentioned 19 MSA patterns and strategies for addressing microservices design issues. The most frequently mentioned patterns and strategies are \textsc{service mesh architecture}, \textsc{service instance per container}, and \textsc{serverless deployment}.
 
\item \textit{Add Data Types, Identifiers, and Loops} (21, 1.08\%): This subcategory contains the solutions for addressing errors at the program initialization level. We identified 5 types of solutions in which the top three are \textsc{add identifiers}, \textsc{add data types}, and \textsc{add query parameters}. 

\item \textit{Add Dependencies and Metrics} (7, 0.36\%): The issues related to missing metrics (e.g., monitoring) and dependencies can be resolved by adding the required metrics and dependencies in the microservices systems. We identified 3 types of solutions, including \textsc{add monitoring metrics}, \textsc{add dependencies}, and \textsc{add stack trace}.

\item \textit{Add APIs, Namespaces, and Plugins} (7, 0.36\%): APIs, namespaces, and plugins are the essential part of any microservices systems. This subcategory covers the solutions for addressing missing APIs, namespaces, and plugins. We identified three types of solutions in this subcategory, including \textsc{add apis}, \textsc{add namespaces}, and \textsc{add plugins}.

\item \textit{Add Logs} (4, 0.20\%): Logging activity is specifically related to monitoring microservices systems. We identified 4 types of solutions that have been used to address monitoring issues (e.g., missing logs), in which the top three are \textsc{add trace id}, \textsc{add trace logging}, and \textsc{add header logging}.
\end{itemize}

\textbf{3. Modify Artifacts} (212/1936, 10.95\%): Besides adding new artifacts to the existing system to address the microservices issues, we also identified a large number of solutions in which the developers explicitly mentioned how they modified modules, services, packages, APIs, scripts, methods, objects, data types, identifiers, databases, and documentation to address the microservices issues. We identified and classified 31 types of solutions in 5 subcategories (see Table \ref{tab:SolutionsTaxnomey} and the Solution Taxonomy sheet in \cite{replpack}). Each of them is briefly described below.

\begin{itemize}
\item \textit{Modify Methods and Objects} (78, 4.03\%): We found that most developers corrected the proprieties (e.g., method calls, operations, parameters) associated with methods and objects of classes to address various types of issues (e.g., code smells, excessive literals) in microservices systems. We identified 13 types of solutions in this subcategory, among them \textsc{correct method definition}, \textsc{redefine method operations}, and \textsc{correct method parameters} are the top three types of solutions.
 
\item \textit{Modify Packages, Modules, and Documentation} (64, 3.30\%): This subcategory covers the packages, modules, and documentation that are modified to address the microservices issues. We identified 4 types of solutions in this subcategory, and among them \textsc{improve documentation} (“\textit{Resolve 'Create Stack' issue and update docs, \#336}”) and \textsc{update packages} are the two leading types of solutions.

\item \textit{Modify Data Types and Identifiers} (51, 2.63\%): This subcategory covers the data types and identifiers that are modified to address the microservices issues. We collected 5 types of solutions, and among them \textsc{correct naming}, \textsc{correct data types}, and \textsc{correct nil value} are the top three types of solutions. 
 
\item \textit{Modify APIs, Services, and Scripts} (13, 0.67\%): This subcategory covers the solutions in which APIs, services, and scripts are modified to address the microservices issues. We identified 6 types of solutions in this subcategory, and among them \textsc{update scripts}, \textsc{update syntax}, and \textsc{modify services} are the top three types of solutions.
 
\item \textit{Modify Databases} (6, 0.31\%): This subcategory covers the solutions in which database query strings and tables are modified to address the microservices issues. We identified 2 types of solutions related to this subcategory, including \textsc{modify query strings} and \textsc{modify database tables}. 
\end{itemize}

\textbf{4. Remove Artifacts} (39/1936, 2.01\%): This category covers the solutions in which artifacts are removed to address several types of microservices issues. We identified and classified 14 types of solutions in 5 subcategories (see Table \ref{tab:SolutionsTaxnomey} and the Solution Taxonomy sheet in \cite{replpack}). Each of them is briefly described below.

\begin{itemize}
\item \textit{Remove Data Types, Methods, Objects, and Plugins} (20, 1.03\%): This subcategory covers the solutions in which data types, methods, objects, and plugins are removed to address the microservices issues. We identified 4 types of solutions in this subcategory, including \textsc{remove data types}, \textsc{remove methods}, \textsc{remove conflicting plugins}, and \textsc{remove objects}. 
 
\item \textit{Remove Dependencies and Databases} (8, 0.41\%): This subcategory covers the solutions in which dependencies and database images are removed to address the microservices issues. We identified 2 types of solutions in this subcategory, including \textsc{remove conflicting dependencies} and \textsc{remove database images}.
 
\item \textit{Remove Logs} (5, 0.25\%): Logging is required to track the communication and identify the failure in microservices systems. This subcategory covers the solutions in which log messages and transaction IDs are removed to address the microservices issues. We identified 3 types of solutions in this subcategory to address the microservices issues, including \textsc{eliminate log messages}, \textsc{remove transaction id for logging}, and \textsc{unregister from registry}.
 
\item \textit{Remove Documentation} (5, 0.25\%): Several microservices issues were addressed by removing unnecessary or wrong information from project documentation. We identified 2 types of solutions in this subcategory, including \textsc{remove unnecessary information} and \textsc{remove empty tags}.
 \end{itemize}

\textbf{5. Manage Infrastructure} (160/1936, 8.26\%): This category captures the solutions that are based on efficient resource utilization for addressing the microservices issues. We identified and classified 20 types of solutions in 2 subcategories (see Table \ref{tab:SolutionsTaxnomey} and the Solution Taxonomy sheet in \cite{replpack}). Each of them is briefly described below.

\begin{itemize}
\item \textit{Manage Storage} (136, 7.02\%): Typical microservices systems store their data in dedicated databases for each service. We found several microservices issues that occurred due to the lack of data storage. We identified 9 types of solutions in this subcategory, in which the top three are \textsc{allocate storage} , \textsc{clean cache}, and \textsc{extend memory}.

\item \textit{Manage Networking} (24, 1.23\%): Networking is complicated in microservices systems due to managing an explosion of service connections over the distributed network. We collected 10 types of solutions in which the top three are \textsc{change proxy settings}, \textsc{server resource management}, and \textsc{disable server groups}.
\end{itemize}

\textbf{6. Manage Configuration and Execution} (50/1936, 2.58\%): Managing configuration and execution enables developers to track changes in microservices systems and their consuming applications over time, for example, the ability to track the version history of configuration changes for multiple instances of microservices systems. We identified and classified 13 types of solutions in 2 subcategories (see Table \ref{tab:SolutionsTaxnomey} and the Solution Taxonomy sheet in \cite{replpack}). Each of them is briefly described below.

\begin{itemize}
\item \textit{Manage Execution} (29, 1.49\%): This subcategory collects the solutions for issues related to managing commands for executing and configuring microservices systems. We identified 5 types of solutions in this subcategory, and among them \textsc{execution and configuration management} and \textsc{execute multiple commands} are the top two types of solutions.

\item \textit{Manage Configuration} (21, 1.08\%): Managing configurations of each microservice and their instances separately is a tedious and time-consuming task. In this subcategory, we identified 8 types of solutions, and the top three types of solutions are \textsc{documentation for configuration management}, \textsc{change configuration files}, and \textsc{correct uuid}. 
\end{itemize}

\textbf{7. Upgrade Tools and Platforms} (47/1936, 2.42\%): The updates of used tools and platforms to develop and manage microservices systems help developers address several issues due to old or legacy versions and protect the microservices systems from security breaches. The interviewees also mentioned a few solutions to address the microservices issues with the help of tools, and one representative quotation is mentioned below.

\faHandORight{} “\textit{Normally, we adopt the solutions according to the type of issues. It could include adding several tools, importing different packages, and adopting successful practices. For example, we improve security by using several open-source API gateways such as OKTA, Spring Cloud gateway, JWT token, and Spring Security. To address service communication issues, we mainly use different ways of communication according to the needs of projects, such as Kafka, RabbitMQ, and Service Mesh. In addition, we automated our continuous integration and delivery process using AWS Code Pipeline. AWS Code Pipeline automates the project release process’s build, test, and deployment phases}” \textbf{(P1, Software Architect, Developer)}.

We identified and classified 21 types of solutions in two subcategories (see Table \ref{tab:SolutionsTaxnomey} and the Solution Taxonomy sheet in \cite{replpack}). Each of them is briefly described below.

\begin{itemize}
\item \textit{Upgrade Deployment, Scaling, and Management Platforms} (39, 2.01\%): This subcategory gathers the solutions for addressing the microservices issues related to the deployment, scaling, and management of CI/CD tools and platforms. We identified 16 types of solutions, and the top 3 types of solutions are \textsc{upgrade container logging}, \textsc{upgrade load balancer}, and \textsc{upgrade docker files}. 

\item \textit{Upgrade Development and Monitoring Tool Support} (8, 0.41\%): Development and monitoring are crucial tasks for developers due to the distributed nature of microservices systems. We identified 5 types of solutions in this subcategory, and the top three types of solutions are \textsc{upgrade kafka flags}, \textsc{upgrade zipkin thrift}, and \textsc{disable tracking}.
\end{itemize}

\textbf{8. Import/Export Artifacts} (12/1936, 0.61\%): We found several issues that can be fixed by importing and exporting various artifacts. We identified and classified 3 types of solutions in 2 subcategories (see Table \ref{tab:SolutionsTaxnomey} and the Solution Taxonomy sheet in \cite{replpack}). Each of them is briefly described below.

\begin{itemize}
\item \textit{Import Artifacts} (9, 0.46\%): This subcategory gathers the solutions related to importing packages and libraries. We identified 2 types of solutions in this subcategory, which are \textsc{import packages}, and \textsc{import libraries} to address the scalability issues in microservices systems. 
 
\item \textit{Export Artifacts} (3, 0.15\%): In this subcategory, we identified only one type of solutions \textsc{export packages} to fix the Configuration issues in microservices systems (e.g., \textsc{conflict on configuration file names}).
\end{itemize}

{\renewcommand{\arraystretch}{1}
\begin{table*}[!htb]
\centering
\scriptsize
\caption{Taxonomy of solutions of issues in microservices systems}
\label{tab:SolutionsTaxnomey}
\begin{tabular}{|l|l|l|}
\hline
\textbf{Category of Solutions} &
 \textbf{Subcategory of Solutions} &
 \textbf{Types of Solutions} \\ \hline
 
 \multirow{10}{*}{Fix Artifacts (1056)} &
 \multirow{2}{*}{Fix Code Issue (912)} &
 Fix Source Code of Issues (791) \\ \cline{3-3} 
 &
 &
 Fix Illegal Symbols (Syntax) in Code (79) \\ \cline{2-3} 
 &
 \multirow{2}{*}{Fix Testing Issue (107)} &
 Debug Code (105) \\ \cline{3-3} 
 &
 &
 Add Test Services in Containers (2) \\ \cline{2-3} 
 &
 \multirow{2}{*}{Fix Build Issue (33)} &
 Fix Errors in Build File (17) \\ \cline{3-3} 
 &
 &
 Correct Build Type (8) \\ \cline{2-3} 
% &
 %\multirow{2}{*}{\begin{tabular}[c]{@{}l@{}}Fix Exception Handling issue (13)\end{tabular}} &
 %Exception Handling (8) \\ \cline{3-3} 
% &
 %&
% Add Error Handling (4) \\ \cline{2-3} 
 &
 \multirow{2}{*}{Fix GUI Issue (4)} &
 Fix Backwards Incompatible UI (3) \\ \cline{3-3} 
 &
 &
 Fix Broken Screenshot (1) \\ \hline

\multirow{19}{*}{Add Artifacts (360)} &
 \multirow{2}{*}{\begin{tabular}[c]{@{}l@{}}Add Features and Services (144)\end{tabular}} &
 Add Missing Feature (114) \\ \cline{3-3} 
 &
 &
 Add Security Feature (26) \\ \cline{2-3} 
 &
 \multirow{2}{*}{\begin{tabular}[c]{@{}l@{}}Add Interfaces, Templates, and Files (39)\end{tabular}} &
 Add File (37) \\ \cline{3-3} 
 &
 &
 Add Templates (1) \\ \cline{2-3} 
 &

 \multirow{3}{*}{\begin{tabular}[c]{@{}l@{}}Add Methods and Modules (33)\end{tabular}} &
 Add constructor (24) \\ \cline{3-3} 
 &
 &
Add Parameters in Method (8) \\ \cline{3-3} 
 &
 &
 Add Security Certificates (1) \\ \cline{2-3} 
 &

 \multirow{2}{*}{Add Classes and Packages (27)} &
 Add Packages (15) \\ \cline{3-3} 
 &
 &
 Add Properties(11) \\ \cline{2-3} 
 &

 \multirow{2}{*}{Add Test Cases (28)} &
 Add Test Cases to Validate Service (21) \\ \cline{3-3}
 &
 &
 Generate Correct Test Case (3) \\ \cline{2-3} 
 &
 
\multirow{2}{*}{Implement Patterns and Strategies (23)} &
Service Mesh Architecture (2)\\ \cline{3-3}
 &
 &
Serverless Deployment (2)\\ \cline{2-3} 
 &
 
 \multirow{2}{*}{\begin{tabular}[c]{@{}l@{}}Add Data Types, Identifiers, and Loops (21)\end{tabular}} &
Add Identifiers (11) \\ \cline{3-3} 
 &
 &
 Add Data Types (5) \\ \cline{2-3} 
 &
 \multirow{2}{*}{\begin{tabular}[c]{@{}l@{}}Add Dependencies and Metrics (7)\end{tabular}} &
 Add Monitoring Metrics (3) \\ \cline{3-3} 
 &
 &
 Add Dependencies (2) \\ \cline{2-3} 
 &
 \multirow{2}{*}{\begin{tabular}[c]{@{}l@{}}Add APIs, Namespaces, and Plugins (7)\end{tabular}} &
 Add APIs (4) \\ \cline{3-3} 
 &
 &
 Add Namespaces (2) \\ \cline{2-3} 
 &

% \multirow{2}{*}{\begin{tabular}[c]{@{}l@{}}Add Tool Support (5)\end{tabular}} &
% Intelligent Monitoring Tools (3)\\ \cline{3-3} 
%  &
%  &
% Kafka (1)\\ \cline{2-3} 
%  &
 \multirow{2}{*}{Add Logs (4)} &
 Add Trace ID (1) \\ \cline{3-3} 
 &
 &
 Add General Purpose Logger (1) \\ \hline

\multirow{10}{*}{Modify Artifacts (212)} &
 \multirow{2}{*}{\begin{tabular}[c]{@{}l@{}}Modify Methods and Objects (78)\end{tabular}} &
 Correct Method Definition (51) \\ \cline{3-3} 
 &
 &
 Redefine Method Operations (8) \\ \cline{2-3} 
 &
 \multirow{2}{*}{\begin{tabular}[c]{@{}l@{}}Modify Package, Module, and Documentation (64)\end{tabular}} &
 Improve Documentation (56) \\ \cline{3-3} 
 &
 &
 Update Packages (6) \\ \cline{2-3} 
 &
 \multirow{2}{*}{\begin{tabular}[c]{@{}l@{}}Modify Data Types and Identifiers (51)\end{tabular}} &
 Correct Naming (29) \\ \cline{3-3} 
 &
 &
 Correct Data Types (8) \\ \cline{2-3} 
 &
 \multirow{2}{*}{\begin{tabular}[c]{@{}l@{}}Modify APIs, Services, and Scripts(13)\end{tabular}} &
 Update Scripts (5) \\ \cline{3-3} 
 &
 &
 Update Syntax (3) \\ \cline{2-3} 
 &
 \multirow{2}{*}{Modify Database (4)} &
 Modify Database Tables (2) \\ \cline{3-3} 
 &
 &
 Modify Query Strings (3) \\ \hline

\multirow{10}{*}{Remove Artifacts (38)} &
 \multirow{2}{*}{\begin{tabular}[c]{@{}l@{}}Remove Data Types, Methods, Objects, and Plugins (20)\end{tabular}} &
 Remove Conflicting Dependencies (17) \\ \cline{3-3} 
 &
 &
Remove Database Images (1) \\ \cline{2-3} 
 &
 \multirow{2}{*}{\begin{tabular}[c]{@{}l@{}}Remove Dependencies and Databases (8)\end{tabular}} &
 Remove Conflicting Dependencies (7) \\ \cline{3-3} 
 &
 &
 Remove Code Dependencies (1) \\ \cline{2-3} 
 &
 \multirow{2}{*}{Remove Logs (5)} &
 Eliminate Log Messages(3) \\ \cline{3-3} 
 &
 &
 Remove Transaction ID for Logging(1) \\ \cline{2-3} 
 &

%\multirow{2}{*}{\begin{tabular}[c]{@{}l@{}}Remove Data Type, Methods and objects (3)\end{tabular}} &
 %Remove Data Type (1) \\ \cline{3-3} 
 %&
 %&
 %Remove Method (1) 
\multirow{2}{*}{Remove Documentation (5)} &
 Remove Unnecessary Information (3) \\ \cline{3-3} 
 &
 &
 Remove Empty Tags (1) \\ \hline

\multirow{6}{*}{Manage Infrastructure (160)} &
 \multirow{2}{*}{Manage Storage (82)} &
 Allocate Storage (77) \\ \cline{3-3} 
 &
 &
 Clean Cache (5) \\ \cline{2-3} 
 &

 %\multirow{2}{*}{Manage Devices (44)} &
 %Add I/O Resources (42) \\ \cline{3-3} 
 %&
 %&
 %Disable WIFI (2) \\ \cline{2-3} 
 %&
 
 \multirow{2}{*}{Manage Networking (24)} &
 Change Proxy Setting (8) \\ \cline{3-3} 
 &
 &
 Server Resource Management (5) \\ \hline
\multirow{4}{*}{\begin{tabular}[c]{@{}l@{}}Manage Configuration \\ and Execution (50)\end{tabular}} &
 \multirow{2}{*}{Manage Execution (29)} &
 Manage Configuration Commands (24) \\ \cline{3-3} 
 &
 &
 Execute Multiple Commands (2) \\ \cline{2-3} 
 &
 \multirow{2}{*}{Manage Configuration (21)} &
 Documentation for Configuration Management (7) \\ \cline{3-3} 
 &
 &
 Change Configuration Files (4) \\ \hline

\multirow{4}{*}{\begin{tabular}[c]{@{}l@{}}Upgrade Tools\\ and Platforms (43)\end{tabular}} &
 \multirow{2}{*}{\begin{tabular}[c]{@{}l@{}}Upgrade Deployment, Scaling, and Management Platforms (35)\end{tabular}} &
 Upgrade Container Logging (6) \\ \cline{3-3} 
 &
 &
Upgrade Load Balancer (4) \\ \cline{2-3} 
 &
 \multirow{2}{*}{\begin{tabular}[c]{@{}l@{}}Upgrade Development and Monitoring Tool Support (8)\end{tabular}} &
 Upgrade Kafka Flags (2) \\ \cline{3-3} 
 &
 &
Upgrade Zipkin Thrift (2) \\ \hline

\multirow{3}{*}{Import/Export Artifacts (12)} &
 \multirow{2}{*}{Import Artifacts (9)} &
 Import Packages (5) \\ \cline{3-3} 
 &
 &
 Import Libraries (4) \\ \cline{2-3} 
 &
 Export Artifacts (3) &
 Export Packages (3) \\ \hline
\end{tabular}
\end{table*}}

\begin{tcolorbox} [sharp corners, boxrule=0.1mm,]
\footnotesize
\textbf{Key Findings of RQ3}: We found that not all the issue discussions provide the information about their solutions, and finally we identified 1,899 issue discussions containing information about the solutions with 36 solutions mentioned by the interviewees and 1 solution mentioned by the survey participants, which are 1,936 solutions in total. The solution taxonomy consists of 8 categories, 32 subcategories, and 177 types of solutions. The majority of solutions are related to Fix Artifacts (54.54\%), Add Artifacts (18.59\%), and Modify Artifacts (10.95\%).

\end{tcolorbox}

\subsection{Practitioners' Perspective (RQ4)}
\label{sec:results_RQ4}
We conducted a cross-sectional survey to evaluate the taxonomies of issues, causes, and solutions in microservices systems built in RQ1, RQ2, and RQ3. We provided a list of 19 issue categories and asked survey participants to respond to each category on a 5-point Likert scale (Very Often, Often, Sometimes, Rarely, Never). Similarly, regarding causes and solutions, we provided 8 cause categories and 8 solution categories, and asked practitioners to respond to each category on a 5-point Likert scale (Strongly Agree, Agree, Neutral, Disagree, Strongly Disagree). We also asked three open-ended questions to identify the missing issues, causes, and solutions in the provided categories. We received 150 valid responses completed by microservices practitioners from 42 countries across 6 continents. The results of RQ4 are summarized in two tables (i.e., Table \ref{tab:RQ4} and Table \ref{tab:sig_value_group}). This section also provides representative quotations from practitioners for answering open-ended questions with the \faHandORight{} sign. The practitioners’ perspectives on the issues, causes, and solutions categories in microservices systems are presented in Table \ref{tab:RQ4}. Due to the limited space, we only presented the ‘percentage’ and ‘mean’ values of the practitioners’ responses to each category of issues, causes, and solutions. The survey results about the practitioners' perspectives on microservices issues, causes, and solutions are briefly reported below.

\textbf{Practitioners' Perspective on Issues}: We asked the microservices participants which issue they faced while developing microservices systems. The majority of the respondents mentioned that they face all of the issues while developing microservices systems. The practitioners' responses presented in Table \ref{tab:RQ4} show that IC1: Technical Debt (46.67\% Very Often, 24.00\% Often), IC2:  Continuous Integration and Delivery Issue (26.67\% Very Often, 42.67\% Often), and IC5: Security Issue (18.67\% Very Often, 64.00\% Often) occur most frequently than other categories of issues. Some practitioners also suggested 6 types of issues (with 9 instances of issues) that were not part of the initial taxonomy of issues in microservices systems. We added microservices practitioners' suggested types of issues in Figure \ref{fig:Taxonomy} and the Issue Taxonomy sheet in \cite{replpack}. One representative quotation about other types of issues is depicted below.

\faHandORight{} “\textit{In my understanding, there should be some other issues such as lack of (i) understanding of the implementation domain, (ii) defined process for designing, developing, and deploying the projects, and (iii) expertise in programming languages for developing microservices systems}” \textbf{(Application Developer, SP2)}.

\textbf{Practitioners' Perspective on Causes}: Our survey participants also confirmed the identified causes that could lead to issues in microservices systems. The practitioners' responses presented in Table \ref{tab:RQ4} show that CC1: General Programming Error (41.33\% Strongly Agree, 38.67\% Agree), CC8: Fragile Code (23.33\% Strongly Agree, 46.67\% Agree), and CC2: Missing Features and Artifacts (12.67\% Strongly Agree, 55.33\% Agree) are the top three cause categories. Some practitioners also suggested 5 types of causes (with 11 instances of causes) that were not part of the initial taxonomy of causes in microservices systems. We added microservices practitioners' suggested types of causes in Table \ref{tab:RQ4} and the Cause Taxonomy sheet in \cite{replpack}. One representative quotation about suggested types of causes is depicted below. 

\faHandORight{} “\textit{Several microservices issues can be occurred because of (i) separate physical database, (ii) lack of resilience support for the whole application, (iii) excessive tooling, (iv) lack of CI/CD (e.g., DevOps) culture in organizations, and (v) lack of practitioners in teams who have multiple skills}” \textbf{(Business Analyst, SP120)}

\textbf{Practitioners' Perspective on Solutions}: We also recorded the practitioners' perspective about the solutions for the issues occurring during microservices system development. The practitioners' responses presented in Table \ref{tab:RQ4} show that SC1: Add Artifacts (42.00\% Strongly Agree, 39.33\% Agree), SC3: Modify Artifact (27.33\% Strongly Agree, 37.33\% Agree), and SC6: Manage Configuration and Execution (18.00\% Strongly Agree, 48.67\% Agree) and are the top three solution categories that have been used to address the microservices issues. One practitioner also suggested one type of solution (with one instance of solution) that was not part of the initial taxonomy of solutions in microservices systems. We added microservices practitioners' suggested types of solutions in Table \ref{tab:RQ4} and the Solution Taxonomy sheet in \cite{replpack}. One representative quotation about the suggested solutions is depicted below.

\faHandORight{} “\textit{Regular training of the employees on the latest technologies and cloud platforms for developing and managing microservices systems can address several types of security, communication, and deployment issues}” (\textbf{DevOps \& Cloud Engineer, SP92}).

{\renewcommand{\arraystretch}{1}
\begin{table*}[!htb]
\centering
\scriptsize
\caption{Practitioners’ perspective (in \%) on the issue, cause, and solution categories in microservices systems}
\label{tab:RQ4}
\begin{tabular}{|lllllll|}
%\begin{tabular}{|>{\color{black}}l|>{\color{black}}l|>{\color{black}}l|>{\color{black}}l|>{\color{black}}l|>{\color{black}}l|>{\color{black}}l|}
\hline
\multicolumn{7}{|l|}{\textbf{Microservices Issues (VO-Very   Often, O-Often, S-Sometimes, R-Rarely, N-Never)}} \\ \hline
\multicolumn{1}{|l|}{\textbf{Categories}} & \multicolumn{1}{l|}{\textbf{VO}} & \multicolumn{1}{l|}{\textbf{O}} & \multicolumn{1}{l|}{\textbf{S}} & \multicolumn{1}{l|}{\textbf{R}} & \multicolumn{1}{l|}{\textbf{N}} & \textbf{Mean} \\ \hline

\multicolumn{1}{|l|}{Technical Debt} & \multicolumn{1}{l|}{\cellcolor[HTML]{70AD47}46.67} & \multicolumn{1}{l|}{\cellcolor[HTML]{A9D08E}24.00} & \multicolumn{1}{l|}{\cellcolor[HTML]{E2EFDA}12.67} & \multicolumn{1}{l|}{\cellcolor[HTML]{E2EFDA}5.33} & \multicolumn{1}{l|}{\cellcolor[HTML]{E2EFDA}1.33} & \cellcolor[HTML]{E2EFDA}3.93 \\ \hline
\multicolumn{1}{|l|}{Continuous Integration and Delivery (CI/CD) Issue} & \multicolumn{1}{l|}{\cellcolor[HTML]{A9D08E}26.67} & \multicolumn{1}{l|}{\cellcolor[HTML]{70AD47}42.67} & \multicolumn{1}{l|}{\cellcolor[HTML]{E2EFDA}15.33} & \multicolumn{1}{l|}{\cellcolor[HTML]{E2EFDA}7.33} & \multicolumn{1}{l|}{\cellcolor[HTML]{E2EFDA}8.00} & \cellcolor[HTML]{E2EFDA}3.73 \\ \hline
\multicolumn{1}{|l|}{Exception Handling Issue} & \multicolumn{1}{l|}{\cellcolor[HTML]{E2EFDA}14.67} & \multicolumn{1}{l|}{\cellcolor[HTML]{A9D08E}29.33} & \multicolumn{1}{l|}{\cellcolor[HTML]{A9D08E}27.33} & \multicolumn{1}{l|}{\cellcolor[HTML]{E2EFDA}13.33} & \multicolumn{1}{l|}{\cellcolor[HTML]{E2EFDA}14.67} & \cellcolor[HTML]{E2EFDA}3.14 \\ \hline
\multicolumn{1}{|l|}{Service Execution and Communication Issue} & \multicolumn{1}{l|}{\cellcolor[HTML]{E2EFDA}15.33} & \multicolumn{1}{l|}{\cellcolor[HTML]{70AD47}44.67} & \multicolumn{1}{l|}{\cellcolor[HTML]{A9D08E}22.67} & \multicolumn{1}{l|}{\cellcolor[HTML]{E2EFDA}10.00} & \multicolumn{1}{l|}{\cellcolor[HTML]{E2EFDA}7.33} & \cellcolor[HTML]{E2EFDA}3.51 \\ \hline
\multicolumn{1}{|l|}{Security Issue} & \multicolumn{1}{l|}{\cellcolor[HTML]{E2EFDA}18.67} & \multicolumn{1}{l|}{\cellcolor[HTML]{70AD47}64.00} & \multicolumn{1}{l|}{\cellcolor[HTML]{E2EFDA}8.67} & \multicolumn{1}{l|}{\cellcolor[HTML]{E2EFDA}6.00} & \multicolumn{1}{l|}{\cellcolor[HTML]{E2EFDA}2.67} & \cellcolor[HTML]{E2EFDA}3.90 \\ \hline

\multicolumn{1}{|l|}{Build Issue} & \multicolumn{1}{l|}{\cellcolor[HTML]{E2EFDA}15.33} & \multicolumn{1}{l|}{\cellcolor[HTML]{70AD47}43.33} & \multicolumn{1}{l|}{\cellcolor[HTML]{A9D08E}23.33} & \multicolumn{1}{l|}{\cellcolor[HTML]{E2EFDA}10.00} & \multicolumn{1}{l|}{\cellcolor[HTML]{E2EFDA}8.00} & \cellcolor[HTML]{E2EFDA}3.48 \\ \hline
\multicolumn{1}{|l|}{Configuration Issue} & \multicolumn{1}{l|}{\cellcolor[HTML]{E2EFDA}14.00} & \multicolumn{1}{l|}{\cellcolor[HTML]{A9D08E}38.67} & \multicolumn{1}{l|}{\cellcolor[HTML]{A9D08E}25.33} & \multicolumn{1}{l|}{\cellcolor[HTML]{E2EFDA}10.67} & \multicolumn{1}{l|}{\cellcolor[HTML]{E2EFDA}11.33} & \cellcolor[HTML]{E2EFDA}3.33\\ \hline
\multicolumn{1}{|l|}{Monitoring Issue} & \multicolumn{1}{l|}{\cellcolor[HTML]{E2EFDA}12.00} & \multicolumn{1}{l|}{\cellcolor[HTML]{70AD47}40.67} & \multicolumn{1}{l|}{\cellcolor[HTML]{A9D08E}22.67} & \multicolumn{1}{l|}{\cellcolor[HTML]{E2EFDA}16.67} & \multicolumn{1}{l|}{\cellcolor[HTML]{E2EFDA}6.67} & \cellcolor[HTML]{E2EFDA}3.31 \\ \hline
\multicolumn{1}{|l|}{Compilation Issue} & \multicolumn{1}{l|}{\cellcolor[HTML]{E2EFDA}16.00} & \multicolumn{1}{l|}{\cellcolor[HTML]{A9D08E}34.67} & \multicolumn{1}{l|}{\cellcolor[HTML]{E2EFDA}15.33} & \multicolumn{1}{l|}{\cellcolor[HTML]{E2EFDA}14.00} & \multicolumn{1}{l|}{\cellcolor[HTML]{A9D08E}20.00} & \cellcolor[HTML]{E2EFDA}3.13 \\ \hline
\multicolumn{1}{|l|}{Testing Issue} & \multicolumn{1}{l|}{\cellcolor[HTML]{A9D08E}26.67} & \multicolumn{1}{l|}{\cellcolor[HTML]{A9D08E}32.67} & \multicolumn{1}{l|}{\cellcolor[HTML]{A9D08E}20.00} & \multicolumn{1}{l|}{\cellcolor[HTML]{E2EFDA}8.00} & \multicolumn{1}{l|}{\cellcolor[HTML]{E2EFDA}11.33} & \cellcolor[HTML]{E2EFDA}3.51 \\ \hline
\multicolumn{1}{|l|}{Documentation Issue} & \multicolumn{1}{l|}{\cellcolor[HTML]{E2EFDA}13.33} & \multicolumn{1}{l|}{\cellcolor[HTML]{70AD47}42.00} & \multicolumn{1}{l|}{\cellcolor[HTML]{E2EFDA}14.67} & \multicolumn{1}{l|}{\cellcolor[HTML]{E2EFDA}11.33} & \multicolumn{1}{l|}{\cellcolor[HTML]{E2EFDA}18.67} &\cellcolor[HTML]{E2EFDA}3.20  \\ \hline
\multicolumn{1}{|l|}{Graphical User Interface (GUI) Issue} & \multicolumn{1}{l|}{\cellcolor[HTML]{E2EFDA} 7.33} & \multicolumn{1}{l|}{\cellcolor[HTML]{A9D08E}30.67} & \multicolumn{1}{l|}{\cellcolor[HTML]{A9D08E}22.00} & \multicolumn{1}{l|}{\cellcolor[HTML]{E2EFDA}16.67} & \multicolumn{1}{l|}{\cellcolor[HTML]{A9D08E}20.67} &\cellcolor[HTML]{E2EFDA}2.79 \\ \hline
\multicolumn{1}{|l|}{Update and Installation Issue} & \multicolumn{1}{l|}{\cellcolor[HTML]{E2EFDA} 10.67} & \multicolumn{1}{l|}{\cellcolor[HTML]{70AD47}44.00} & \multicolumn{1}{l|}{\cellcolor[HTML]{A9D08E}20.67} & \multicolumn{1}{l|}{\cellcolor[HTML]{E2EFDA}12.00} & \multicolumn{1}{l|}{\cellcolor[HTML]{E2EFDA}12.00} & \cellcolor[HTML]{E2EFDA}3.27 \\ \hline
\multicolumn{1}{|l|}{Database Issue} & \multicolumn{1}{l|}{\cellcolor[HTML]{E2EFDA}18.00} & \multicolumn{1}{l|}{\cellcolor[HTML]{70AD47}40.67} & \multicolumn{1}{l|}{\cellcolor[HTML]{E2EFDA}18.00} & \multicolumn{1}{l|}{\cellcolor[HTML]{E2EFDA}16.00} & \multicolumn{1}{l|}{\cellcolor[HTML]{E2EFDA}7.33} &\cellcolor[HTML]{E2EFDA}3.46  \\ \hline
\multicolumn{1}{|l|}{Storage Issue} & \multicolumn{1}{l|}{\cellcolor[HTML]{E2EFDA} 12.00} & \multicolumn{1}{l|}{\cellcolor[HTML]{A9D08E}38.67} & \multicolumn{1}{l|}{\cellcolor[HTML]{A9D08E}22.00} & \multicolumn{1}{l|}{\cellcolor[HTML]{E2EFDA}17.33} & \multicolumn{1}{l|}{\cellcolor[HTML]{E2EFDA}10.00} &\cellcolor[HTML]{E2EFDA}3.25  \\ \hline
\multicolumn{1}{|l|}{Performance Issue} & \multicolumn{1}{l|}{\cellcolor[HTML]{E2EFDA}17.33} & \multicolumn{1}{l|}{\cellcolor[HTML]{70AD47}46.00} & \multicolumn{1}{l|}{\cellcolor[HTML]{E2EFDA}14.67} & \multicolumn{1}{l|}{\cellcolor[HTML]{E2EFDA}14.67} & \multicolumn{1}{l|}{\cellcolor[HTML]{E2EFDA}6.67} &\cellcolor[HTML]{E2EFDA}3.51  \\ \hline
\multicolumn{1}{|l|}{Networking Issue} & \multicolumn{1}{l|}{\cellcolor[HTML]{E2EFDA}12.00} & \multicolumn{1}{l|}{\cellcolor[HTML]{A9D08E}34.67} & \multicolumn{1}{l|}{\cellcolor[HTML]{A9D08E}22.00} & \multicolumn{1}{l|}{\cellcolor[HTML]{A9D08E}20.00} & \multicolumn{1}{l|}{\cellcolor[HTML]{E2EFDA}10.67} &\cellcolor[HTML]{E2EFDA}3.15   \\ \hline
\multicolumn{1}{|l|}{Typecasting Issue} & \multicolumn{1}{l|}{\cellcolor[HTML]{E2EFDA} 12.67} & \multicolumn{1}{l|}{\cellcolor[HTML]{A9D08E}34.00} & \multicolumn{1}{l|}{\cellcolor[HTML]{A9D08E}21.33} & \multicolumn{1}{l|}{\cellcolor[HTML]{E2EFDA}10.67} & \multicolumn{1}{l|}{\cellcolor[HTML]{E2EFDA}18.00} & \cellcolor[HTML]{E2EFDA}3.03 \\ \hline
\multicolumn{1}{|l|}{Organizational Issue} & \multicolumn{1}{l|}{\cellcolor[HTML]{E2EFDA} 18.67} & \multicolumn{1}{l|}{\cellcolor[HTML]{70AD47}64.00} & \multicolumn{1}{l|}{\cellcolor[HTML]{E2EFDA}8.67} & \multicolumn{1}{l|}{\cellcolor[HTML]{E2EFDA}6.00} & \multicolumn{1}{l|}{\cellcolor[HTML]{E2EFDA}2.67} &\cellcolor[HTML]{E2EFDA}3.90   \\ \hline

\multicolumn{7}{|l|}{\textbf{Causes of Issues (SA-Strongly Agree, A-Agree, UD-Undecided, D-Disagree, SD-Strongly Disagree)}} \\ \hline
\multicolumn{1}{|l|}{\textbf{Categories}} & \multicolumn{1}{l|}{\textbf{SA}} & \multicolumn{1}{l|}{\textbf{A}} & \multicolumn{1}{l|}{\textbf{U}} & \multicolumn{1}{l|}{\textbf{D}} & \multicolumn{1}{l|}{\textbf{SD}} & \textbf{Mean} \\ \hline
\multicolumn{1}{|l|}{General Programming Error} & \multicolumn{1}{l|}{\cellcolor[HTML]{70AD47}41.33} & \multicolumn{1}{l|}{\cellcolor[HTML]{A9D08E}38.67} & \multicolumn{1}{l|}{\cellcolor[HTML]{E2EFDA}12.00} & \multicolumn{1}{l|}{\cellcolor[HTML]{E2EFDA}6.00} & \multicolumn{1}{l|}{\cellcolor[HTML]{E2EFDA}2.00} & \cellcolor[HTML]{E2EFDA}4.11\\ \hline
\multicolumn{1}{|l|}{Missing Features and Artifacts} & \multicolumn{1}{l|}{\cellcolor[HTML]{E2EFDA}12.67} & \multicolumn{1}{l|}{\cellcolor[HTML]{70AD47}55.33} & \multicolumn{1}{l|}{\cellcolor[HTML]{A9D08E}25.33} & \multicolumn{1}{l|}{\cellcolor[HTML]{E2EFDA}4.67} & \multicolumn{1}{l|}{\cellcolor[HTML]{E2EFDA}2.67} & \cellcolor[HTML]{E2EFDA}3.70 \\ \hline
\multicolumn{1}{|l|}{Invalid Configuration and Communication Problem} & \multicolumn{1}{l|}{\cellcolor[HTML]{E2EFDA}18.00} & \multicolumn{1}{l|}{\cellcolor[HTML]{70AD47}47.33} & \multicolumn{1}{l|}{\cellcolor[HTML]{A9D08E}20.00} & \multicolumn{1}{l|}{\cellcolor[HTML]{E2EFDA}10.00} & \multicolumn{1}{l|}{\cellcolor[HTML]{E2EFDA}4.67} & \cellcolor[HTML]{E2EFDA}3.64 \\ \hline
\multicolumn{1}{|l|}{Legacy Versions, Compatibility, and Dependency Problem} & \multicolumn{1}{l|}{\cellcolor[HTML]{E2EFDA}12.67} & \multicolumn{1}{l|}{\cellcolor[HTML]{70AD47}41.33} & \multicolumn{1}{l|}{\cellcolor[HTML]{A9D08E}27.33} & \multicolumn{1}{l|}{\cellcolor[HTML]{E2EFDA}13.33} & \multicolumn{1}{l|}{\cellcolor[HTML]{E2EFDA}5.33} &\cellcolor[HTML]{E2EFDA}3.43 \\ \hline
\multicolumn{1}{|l|}{Service Design and Implementation Anomaly} & \multicolumn{1}{l|}{\cellcolor[HTML]{E2EFDA}19.33} & \multicolumn{1}{l|}{\cellcolor[HTML]{70AD47}48.00} & \multicolumn{1}{l|}{\cellcolor[HTML]{E2EFDA}18.67} & \multicolumn{1}{l|}{\cellcolor[HTML]{E2EFDA}8.67} & \multicolumn{1}{l|}{\cellcolor[HTML]{E2EFDA}5.33} &\cellcolor[HTML]{E2EFDA}3.67  \\ \hline
\multicolumn{1}{|l|}{Poor Security Management} & \multicolumn{1}{l|}{\cellcolor[HTML]{E2EFDA}18.67} & \multicolumn{1}{l|}{\cellcolor[HTML]{70AD47}42.67} & \multicolumn{1}{l|}{\cellcolor[HTML]{A9D08E}21.33} & \multicolumn{1}{l|}{\cellcolor[HTML]{E2EFDA}10.00} & \multicolumn{1}{l|}{\cellcolor[HTML]{E2EFDA}7.33} & \cellcolor[HTML]{E2EFDA}3.55 \\ \hline
\multicolumn{1}{|l|}{Insufficient Resources} & \multicolumn{1}{l|}{\cellcolor[HTML]{E2EFDA}16.00} & \multicolumn{1}{l|}{\cellcolor[HTML]{70AD47}46.00} & \multicolumn{1}{l|}{\cellcolor[HTML]{A9D08E}22.67} & \multicolumn{1}{l|}{\cellcolor[HTML]{E2EFDA}10.00} & \multicolumn{1}{l|}{\cellcolor[HTML]{E2EFDA}5.33} & \cellcolor[HTML]{E2EFDA}3.57 \\ \hline
\multicolumn{1}{|l|}{Fragile Code} & \multicolumn{1}{l|}{\cellcolor[HTML]{A9D08E}23.33} & \multicolumn{1}{l|}{\cellcolor[HTML]{70AD47}46.67} & \multicolumn{1}{l|}{\cellcolor[HTML]{E2EFDA}12.67} & \multicolumn{1}{l|}{\cellcolor[HTML]{E2EFDA}12.67} & \multicolumn{1}{l|}{\cellcolor[HTML]{E2EFDA}4.67} & \cellcolor[HTML]{E2EFDA}3.71 \\ \hline

\multicolumn{7}{|l|}{\textbf{Solutions for Issues (SA-Strongly Agree, A-Agree, UD-Undecided, D-Disagree, SD-Strongly Disagree)}} \\ \hline
\multicolumn{1}{|l|}{\textbf{Categories}} & \multicolumn{1}{l|}{\textbf{SA}} & \multicolumn{1}{l|}{\textbf{A}} & \multicolumn{1}{l|}{\textbf{U}} & \multicolumn{1}{l|}{\textbf{D}} & \multicolumn{1}{l|}{\textbf{SD}} & \textbf{Mean} \\ \hline

\multicolumn{1}{|l|}{Add Artifacts} & \multicolumn{1}{l|}{\cellcolor[HTML]{70AD47}42.00} & \multicolumn{1}{l|}{\cellcolor[HTML]{A9D08E}39.33} & \multicolumn{1}{l|}{\cellcolor[HTML]{E2EFDA}8.67} & \multicolumn{1}{l|}{\cellcolor[HTML]{E2EFDA}6.00} & \multicolumn{1}{l|}{\cellcolor[HTML]{E2EFDA}3.33} &\cellcolor[HTML]{E2EFDA}4.11\\ \hline
\multicolumn{1}{|l|}{Remove Artifacts} & \multicolumn{1}{l|}{\cellcolor[HTML]{E2EFDA}16.00} & \multicolumn{1}{l|}{\cellcolor[HTML]{70AD47}60.67} & \multicolumn{1}{l|}{\cellcolor[HTML]{E2EFDA}14.00} & \multicolumn{1}{l|}{\cellcolor[HTML]{E2EFDA}6.67} & \multicolumn{1}{l|}{\cellcolor[HTML]{E2EFDA}2.00} & \cellcolor[HTML]{E2EFDA}3.83 \\ \hline
\multicolumn{1}{|l|}{Modify Artifact} & \multicolumn{1}{l|}{\cellcolor[HTML]{A9D08E}27.33} & \multicolumn{1}{l|}{\cellcolor[HTML]{A9D08E}37.33} & \multicolumn{1}{l|}{\cellcolor[HTML]{A9D08E}25.33} & \multicolumn{1}{l|}{\cellcolor[HTML]{E2EFDA}6.00} & \multicolumn{1}{l|}{\cellcolor[HTML]{E2EFDA}3.33} &\cellcolor[HTML]{E2EFDA}3.80  \\ \hline
\multicolumn{1}{|l|}{Manage Infrastructure} & \multicolumn{1}{l|}{\cellcolor[HTML]{E2EFDA}14.00} & \multicolumn{1}{l|}{\cellcolor[HTML]{70AD47}48.00} & \multicolumn{1}{l|}{\cellcolor[HTML]{A9D08E}23.33} & \multicolumn{1}{l|}{\cellcolor[HTML]{E2EFDA}10.00} & \multicolumn{1}{l|}{\cellcolor[HTML]{E2EFDA}4.00} &\cellcolor[HTML]{E2EFDA}3.58 \\ \hline

\multicolumn{1}{|l|}{Fix Artifacts} & \multicolumn{1}{l|}{\cellcolor[HTML]{E2EFDA}18.00	} & \multicolumn{1}{l|}{\cellcolor[HTML]{70AD47}46.67} & \multicolumn{1}{l|}{\cellcolor[HTML]{A9D08E}24.00} & \multicolumn{1}{l|}{\cellcolor[HTML]{E2EFDA}6.00} & \multicolumn{1}{l|}{\cellcolor[HTML]{E2EFDA}4.67} &\cellcolor[HTML]{E2EFDA}3.68 \\ \hline

\multicolumn{1}{|l|}{Manage Configuration and Execution} & \multicolumn{1}{l|}{\cellcolor[HTML]{E2EFDA}18.00} & \multicolumn{1}{l|}{\cellcolor[HTML]{70AD47}48.67} & \multicolumn{1}{l|}{\cellcolor[HTML]{E2EFDA}18.67} & \multicolumn{1}{l|}{\cellcolor[HTML]{E2EFDA}9.33} & \multicolumn{1}{l|}{\cellcolor[HTML]{E2EFDA}4.67} &\cellcolor[HTML]{E2EFDA}3.66\\ \hline

\multicolumn{1}{|l|}{Upgrade Tools and Platforms} & \multicolumn{1}{l|}{\cellcolor[HTML]{E2EFDA}15.33} & \multicolumn{1}{l|}{\cellcolor[HTML]{70AD47}52.00} & \multicolumn{1}{l|}{\cellcolor[HTML]{A9D08E}22.00} & \multicolumn{1}{l|}{\cellcolor[HTML]{E2EFDA}8.00} & \multicolumn{1}{l|}{\cellcolor[HTML]{E2EFDA}2.00} & \cellcolor[HTML]{E2EFDA}3.71 \\ \hline

\multicolumn{1}{|l|}{Import/Export Artifacts} & \multicolumn{1}{l|}{\cellcolor[HTML]{E2EFDA}14.67} & \multicolumn{1}{l|}{\cellcolor[HTML]{70AD47}50.00} & \multicolumn{1}{l|}{\cellcolor[HTML]{A9D08E}28.67} & \multicolumn{1}{l|}{\cellcolor[HTML]{E2EFDA}3.33} & \multicolumn{1}{l|}{\cellcolor[HTML]{E2EFDA}2.67} &\cellcolor[HTML]{E2EFDA}3.71 \\ \hline
\end{tabular}
\end{table*} }

\textbf{Statistical significance on the issue, cause, and solution categories in microservices systems}: We analyzed the practitioners' responses across one pair of demographic groups in Table \ref{tab:sig_value_group}. The first column of Table \ref{tab:sig_value_group} lists the categories of issues, causes, and solutions presented to the survey participants. The subsequent columns of Table \ref{tab:sig_value_group} show the \textit{Likert Distribution}, \textit{Mean}, \textit{p-value}, and \textit{Effect Size}. Analyzing survey data using the \textit{Likert Distribution}, \textit{Mean}, \textit{p-value}, and \textit{Effect Size} provides insights into response patterns, central tendency, statistical significance, and practical significance, respectively. The practitioners' responses are grouped into Experience $\le$ 6 Years vs. Experience > 6 Years group to check the test whether there are statistical differences between the two groups on the same variable or not. 

The \textit{Likert Distribution} shows the level of agreement and importance for each issue, cause, and solution category. In contrast, the \textit{Mean} indicates the average of the Likert distribution for the issue, cause, and solution categories. The \textit{p-value} indicates statistical differences between Experience $\le$ 6 Years and Experience > 6 Years in the Experience Based Grouping column. We used the non-parametric Mann–Whitney U test to test the null hypothesis (i.e., there is no significant difference between the responses in both groups). We describe the impact of the groups on the responses as significant if the \textit{p-value} is less than 0.05 (see \faBalanceScale{} in Table \ref{tab:sig_value_group}). The \textit{Effect Size} is measured by taking the mean difference of Experience > 6 Years and Experience $\le$ 6 Years.

\textbf{Observations}: We made the following observations based on the practitioners' responses.
\begin{itemize}
 \item There are no major statistically significant differences in practitioners' responses on the issue, cause, and solution categories in microservices systems.
 
 \item The observed statistically significant differences between experienced-based grouping indicate that the experience of microservices practitioners does not affect the survey responses.
 
 \item The survey findings indicate that most issues are related to Technical Debt, CI/CD, and Security; most causes are associated with General Programming Errors, Fragile Code, and Missing Features and Artifacts; and most solutions are associated with Add Artifacts, Modify Artifacts, and Manage Configuration and Execution.

\item More than 50\% of the respondents indicated that they frequently (Very Often and Often) encountered different issues belonging to the given list of issue categories.

\item Our results indicate that many practitioners rarely or never face GUI (16.67\% Rarely, Never 20.67\%), Compilation (14.00\% Rarely, 20.00\% Never), and Networking issues (20.00\% Rarely, 10.66\% Never) in microservices system development. 
\end{itemize}

\begin{tcolorbox} [sharp corners, boxrule=0.1mm,]
\footnotesize
\textbf{Key Findings of RQ4}: (i) There are no major statistically significant differences in the issue, cause, and solution categories in practitioners' responses, (ii) practitioners frequently face Technical Debt, CI/CD, and Security issues, (iii) most causes are associated with General Programming Errors, Fragile Code, and Missing Features and Artifacts, and (iv) most solutions are associated with Add Artifacts, Modify Artifact, and Manage Configuration and Execution. The survey participants generally confirmed the categories of issues, causes, and solutions derived from mining developer discussions in open-source microservices systems. However, the survey participants also indicated several other types of issues, causes, and solutions that help improve our taxonomies.
\end{tcolorbox}

{\renewcommand{\arraystretch}{1}
\begin{table*}[!htb]
\centering
\scriptsize
%\footnotesize
\caption{Statistical significance on the issue, cause, and solution categories in microservices systems}
\label{tab:sig_value_group}
\begin{tabular}{|rccccc|}
\hline
\multicolumn{1}{|r|}{\multirow{2}{*}{\textbf{Categories}}} &
 \multicolumn{1}{c|}{\multirow{2}{*}{\textbf{\#}}} &
 \multicolumn{2}{c|}{\textbf{Likert Distributions}} &
 \multicolumn{2}{c|}{\textbf{Experience Based Grouping}} \\ \cline{3-6}
 %\multicolumn{2}{c|}{\textbf{Countries Wise Grouping}} \\ \cline{3-8} 
\multicolumn{1}{|r|}{} &
 \multicolumn{1}{c|}{} &
 \multicolumn{1}{c|}{\textbf{In Total}} &
 \multicolumn{1}{c|}{\textbf{Mean}} &
 \multicolumn{1}{c|}{\textbf{P-value}} &
 \multicolumn{1}{c|}{\textbf{Effect Size}} \\ \hline 
 %\multicolumn{1}{c|}{\textbf{P-value}} &
 %\textbf{Effect Size} \\ \hline
 
\multicolumn{6}{|l|}{\textbf{Microservices Issues}} \\ \hline
\multicolumn{1}{|r|}{Technical Debt} &
 \multicolumn{1}{c|}{IC1} &
 \multicolumn{1}{c|}{\includegraphics[width = 0.6cm, height = 0.19cm]{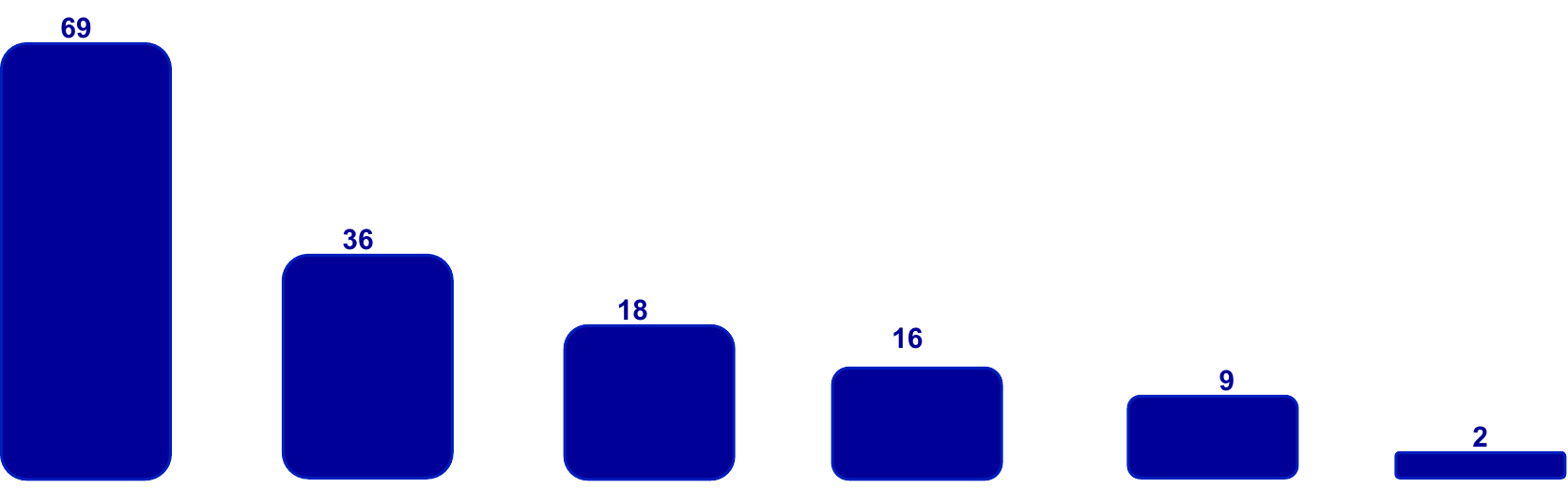}} &
 \multicolumn{1}{c|}{3.89} &
 \multicolumn{1}{c|}{0.47} &
 \multicolumn{1}{c|}{0.45}\\\hline
% \multicolumn{1}{c|}{0.60} &
 %0.40 \\ \hline

\multicolumn{1}{|r|}{Continuous Integration and Delivery (CI/CD) Issue} &
 \multicolumn{1}{c|}{IC2} &
 \multicolumn{1}{c|}{\includegraphics[width = 0.6cm, height = 0.19cm]{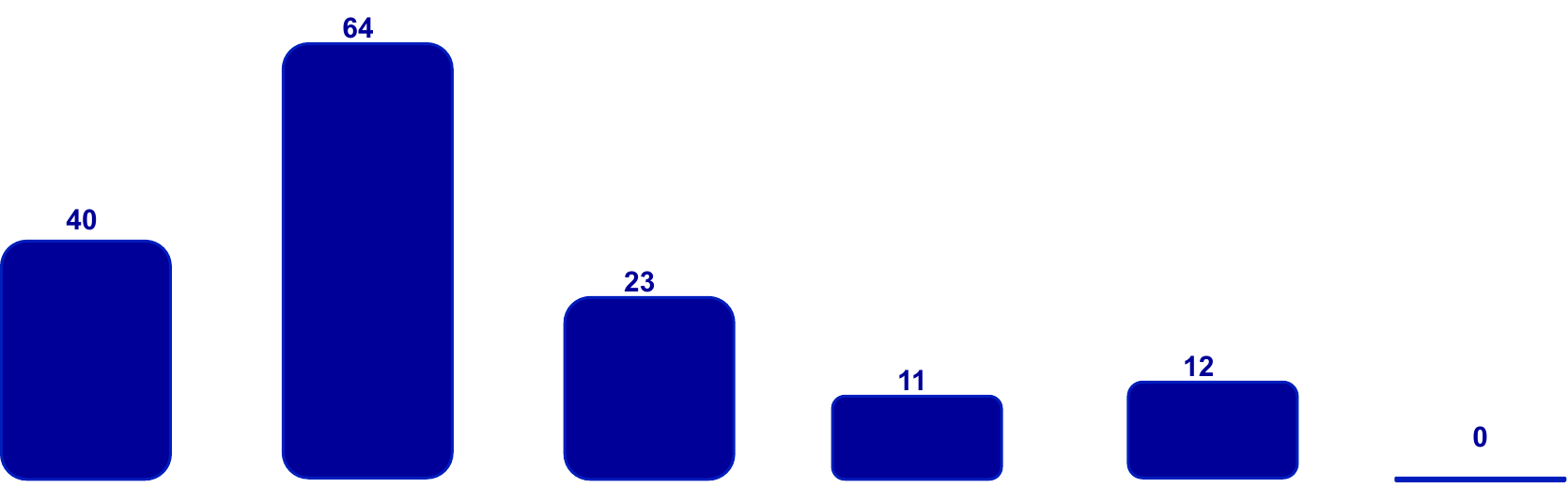}} &
 \multicolumn{1}{c|}{3.73} &
 \multicolumn{1}{c|}{0.35} &
 \multicolumn{1}{c|}{0.30} \\\hline
 %\multicolumn{1}{c|}{0.76} &
 %0.15 \\ \hline

\multicolumn{1}{|r|}{Exception Handling Issue} &
 \multicolumn{1}{c|}{IC3} &
 \multicolumn{1}{c|}{\includegraphics[width = 0.6cm, height = 0.19cm]{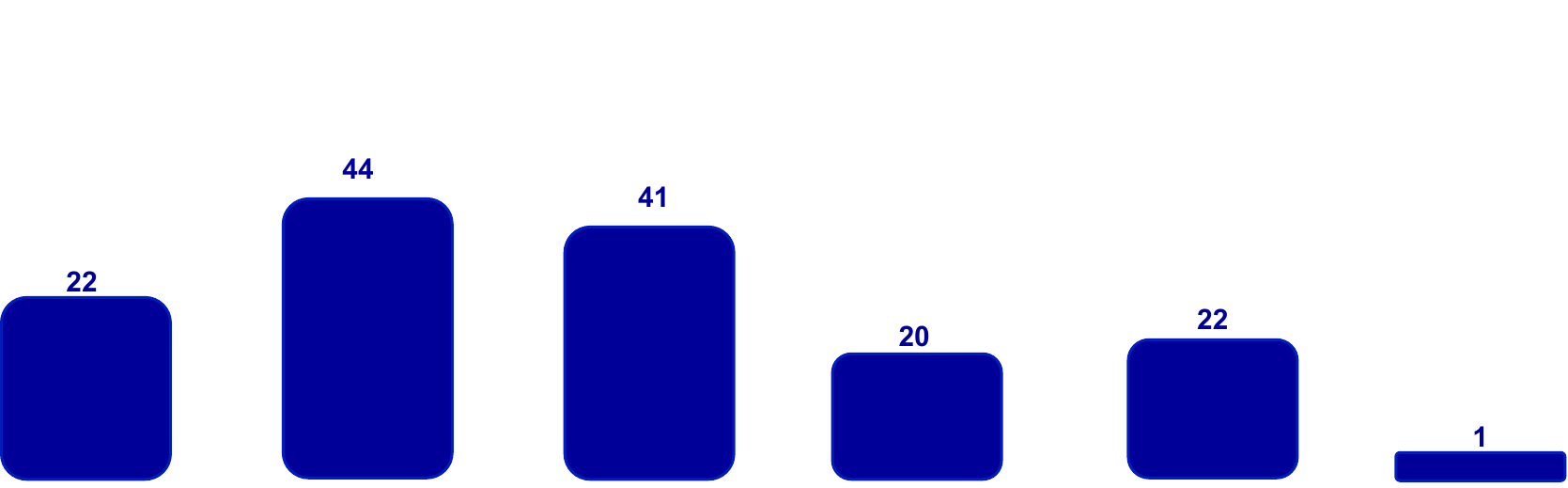}} &
 \multicolumn{1}{c|}{3.14} &
 \multicolumn{1}{c|}{\faBalanceScale{}  0.02} &
 \multicolumn{1}{c|}{0.05} \\\hline
% \multicolumn{1}{c|}{0.92} &
% -0.13 \\ \hline

\multicolumn{1}{|r|}{Service Execution and Communication Issue} &
 \multicolumn{1}{c|}{IC4} &
 \multicolumn{1}{c|}{\includegraphics[width = 0.6cm, height = 0.19cm]{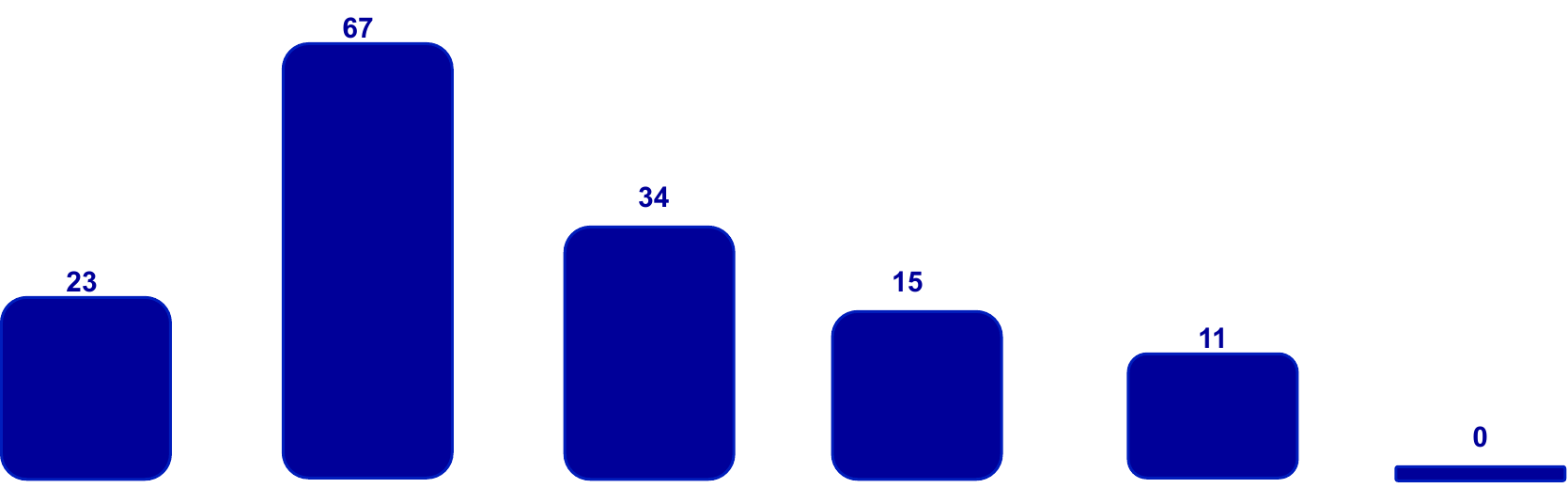}} &
 \multicolumn{1}{c|}{3.51} &
 \multicolumn{1}{c|}{0.17} &
 \multicolumn{1}{c|}{0.35}\\\hline
 %\multicolumn{1}{c|}{0.53} &
 %0.20 \\ \hline

\multicolumn{1}{|r|}{Security Issue} &
 \multicolumn{1}{c|}{IC5} &
 \multicolumn{1}{c|}{\includegraphics[width = 0.6cm, height = 0.19cm]{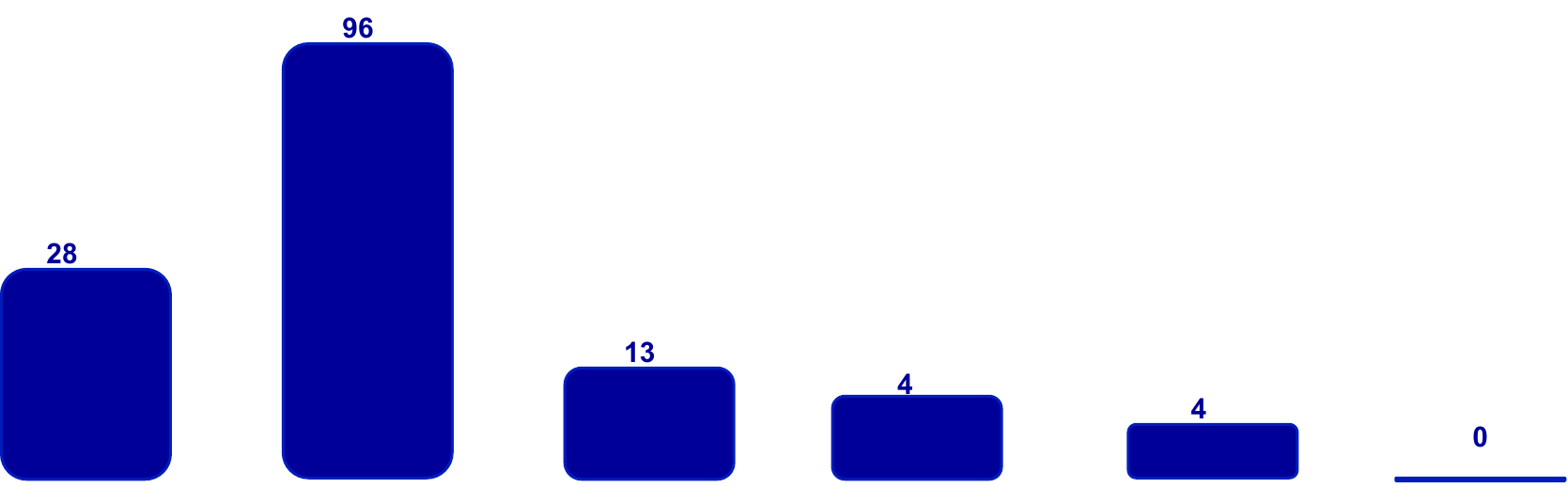}} &
 \multicolumn{1}{c|}{3.90} &
 \multicolumn{1}{c|}{0.67} &
 \multicolumn{1}{c|}{0.16} \\\hline
 %\multicolumn{1}{c|}{0.53} &
 %0.26 \\ \hline

\multicolumn{1}{|r|}{Build Issue} &
 \multicolumn{1}{c|}{IC6} &
 \multicolumn{1}{c|}{\includegraphics[width = 0.6cm, height = 0.19cm]{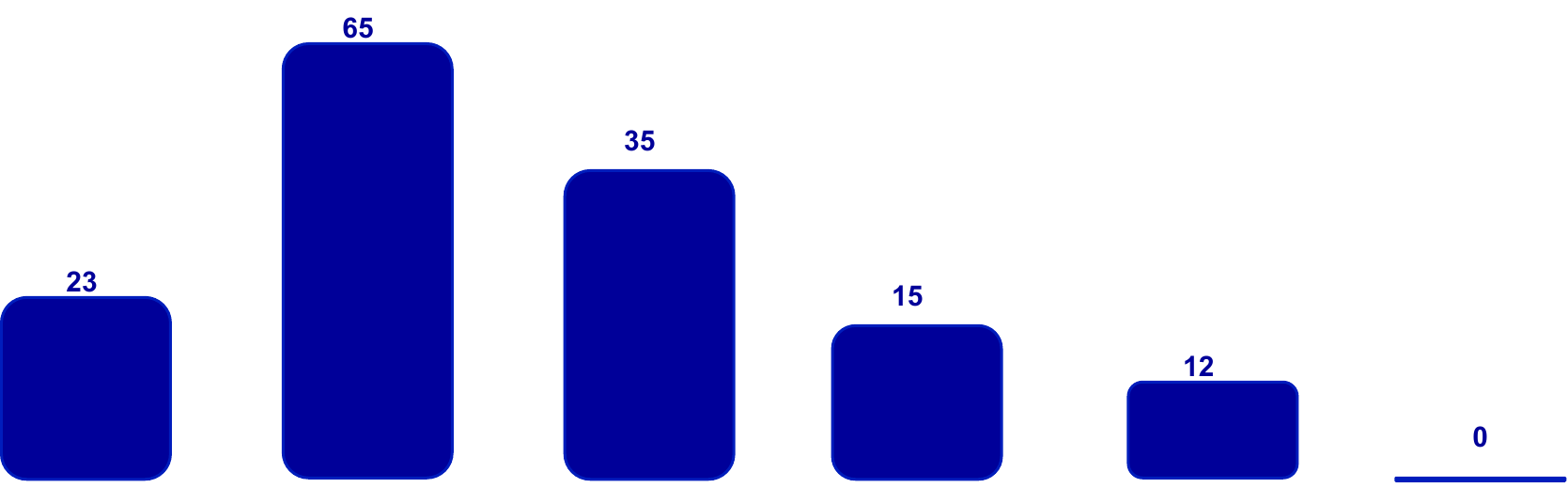}} &
 \multicolumn{1}{c|}{3.48} &
 \multicolumn{1}{c|}{0.14} &
 \multicolumn{1}{c|}{0.12} \\\hline
% \multicolumn{1}{c|}{0.67} &
% 0.41 \\ \hline

\multicolumn{1}{|r|}{Configuration Issue} &
 \multicolumn{1}{c|}{IC7} &
 \multicolumn{1}{c|}{\includegraphics[width = 0.6cm, height = 0.19cm]{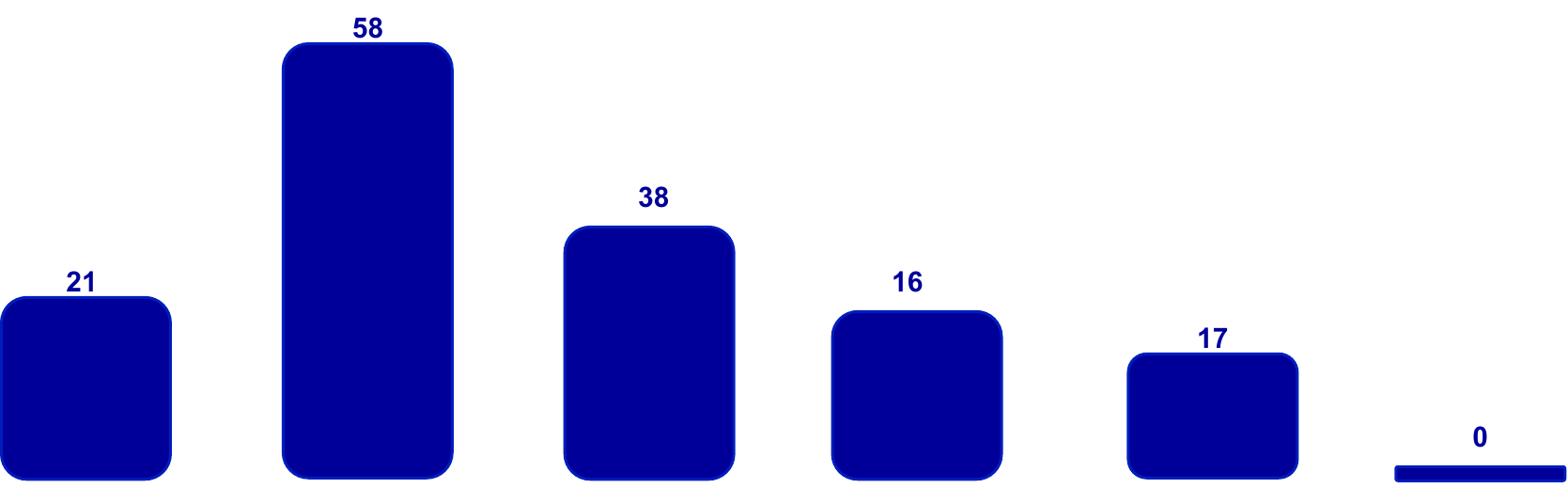}} &
 \multicolumn{1}{c|}{3.33} &
 \multicolumn{1}{c|}{0.14} &
 \multicolumn{1}{c|}{-0.06}\\\hline
% \multicolumn{1}{c|}{0.76} &
% 0.36 \\ \hline

\multicolumn{1}{|r|}{Monitoring Issue} &
 \multicolumn{1}{c|}{IC8} &
 \multicolumn{1}{c|}{\includegraphics[width = 0.6cm, height = 0.19cm]{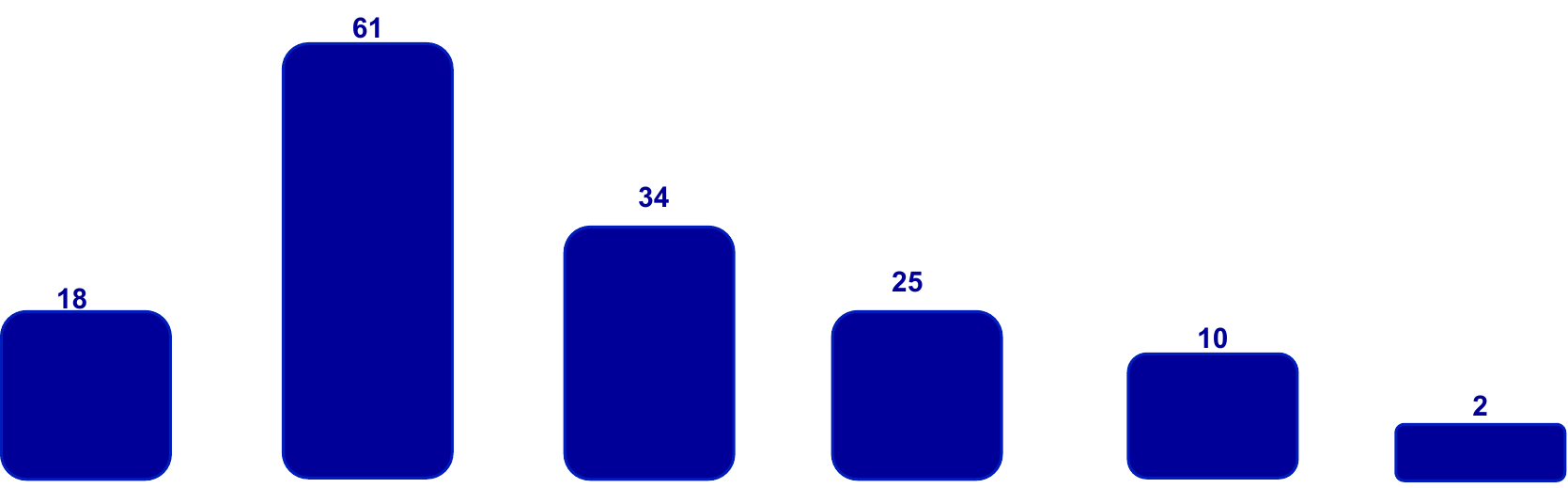}} &
 \multicolumn{1}{c|}{3.31} &
 \multicolumn{1}{c|}{0.14} &
 \multicolumn{1}{c|}{0.23}\\\hline
% \multicolumn{1}{c|}{1.00} &
% -0.36 \\ \hline

\multicolumn{1}{|r|}{Compilation Issue} &
 \multicolumn{1}{c|}{IC9} &
 \multicolumn{1}{c|}{\includegraphics[width = 0.6cm, height = 0.19cm]{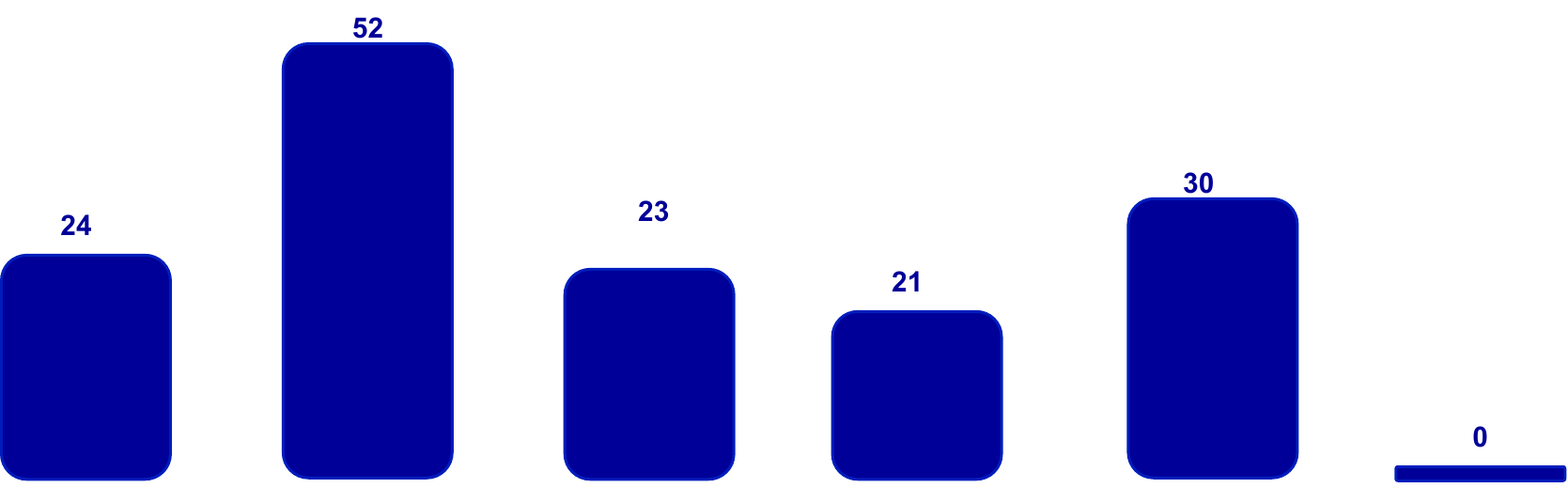}} &
 \multicolumn{1}{c|}{3.13} &
 \multicolumn{1}{c|}{\faBalanceScale{}  0.03} &
 \multicolumn{1}{c|}{-0.06}\\\hline
% \multicolumn{1}{c|}{0.47} &
% 0.49 \\ \hline

\multicolumn{1}{|r|}{Testing Issue} &
 \multicolumn{1}{c|}{IC10} &
 \multicolumn{1}{c|}{\includegraphics[width = 0.6cm, height = 0.19cm]{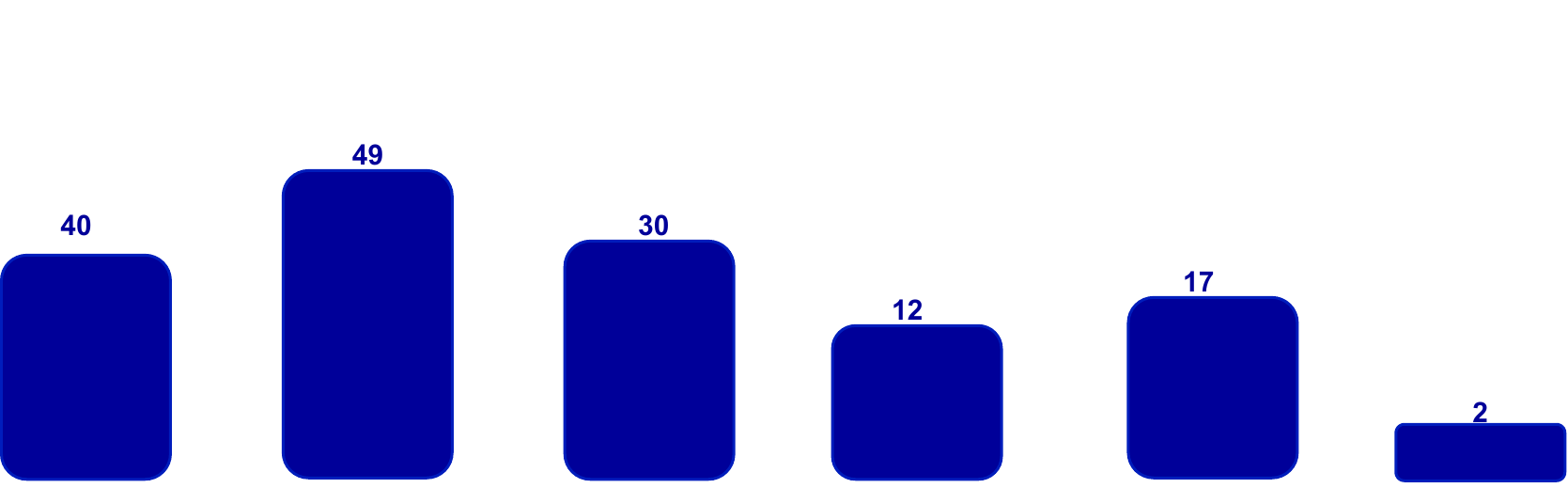}} &
 \multicolumn{1}{c|}{3.51} &
 \multicolumn{1}{c|}{0.09} &
 \multicolumn{1}{c|}{-0.36}\\\hline
 %\multicolumn{1}{c|}{0.83} &
% 0.24 \\ \hline

\multicolumn{1}{|r|}{Documentation Issue} &
 \multicolumn{1}{c|}{IC11} &
 \multicolumn{1}{c|}{\includegraphics[width = 0.6cm, height = 0.19cm]{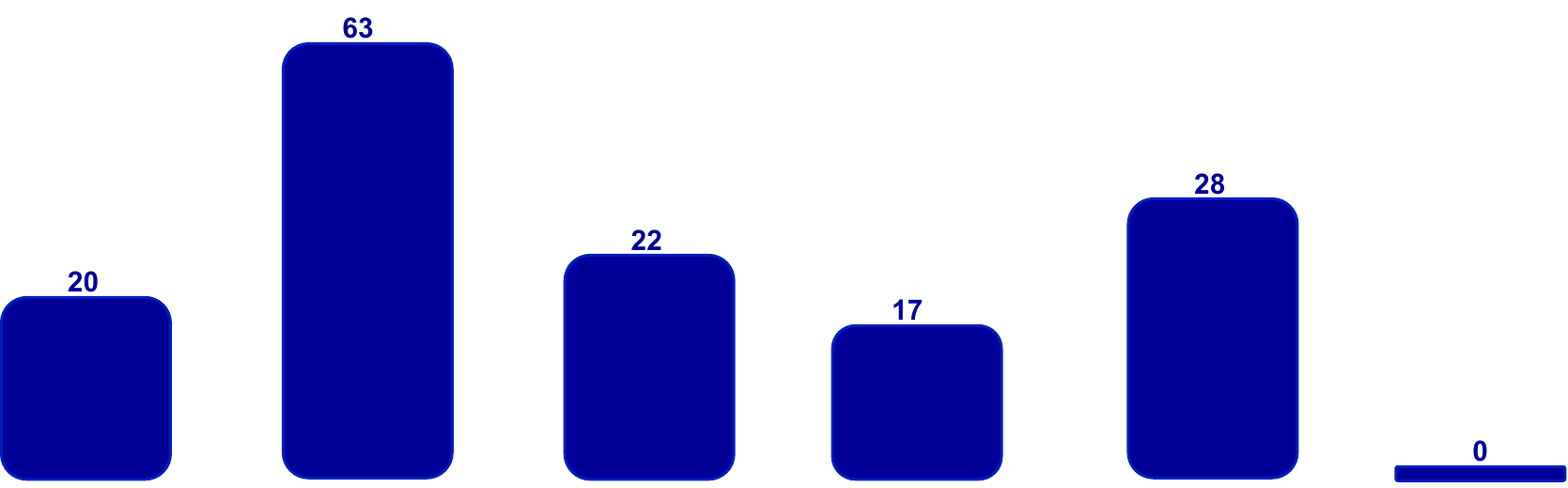}} &
 \multicolumn{1}{c|}{3.20} &
 \multicolumn{1}{c|}{0.02} &
 \multicolumn{1}{c|}{0.31}\\\hline
% \multicolumn{1}{c|}{0.40} &
% 0.25 \\ \hline

\multicolumn{1}{|r|}{Graphical User Interface (GUI) Issue} &
 \multicolumn{1}{c|}{IC12} &
 \multicolumn{1}{c|}{\includegraphics[width = 0.6cm, height = 0.19cm]{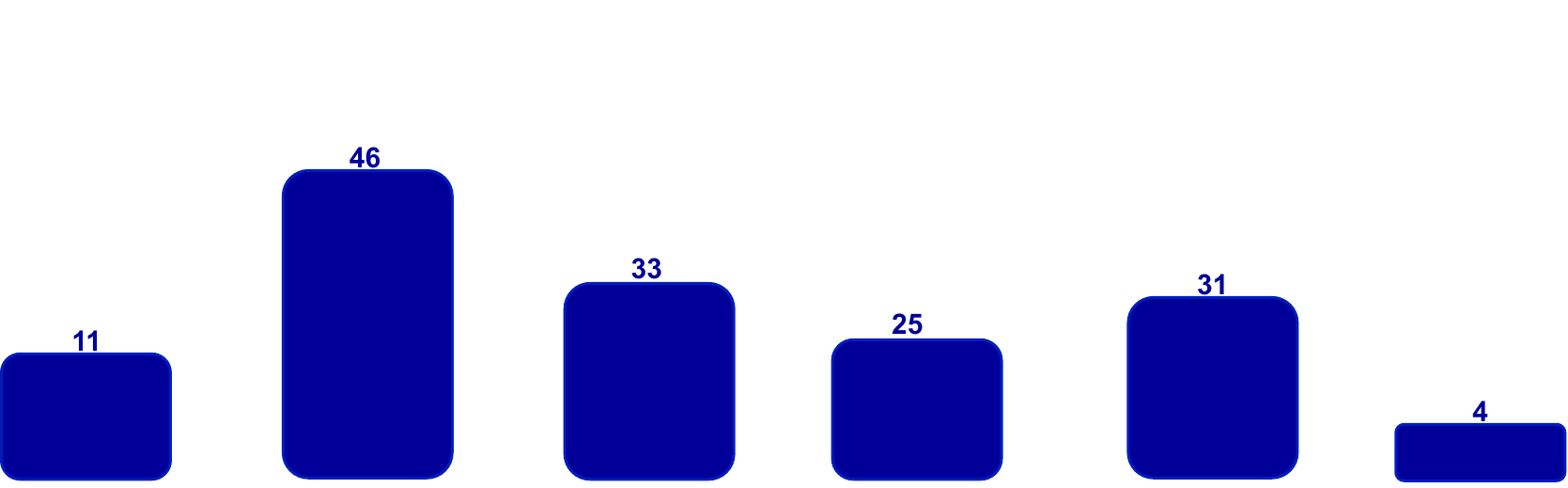}} &
 \multicolumn{1}{c|}{2.79} &
 \multicolumn{1}{c|}{0.08} &
 \multicolumn{1}{c|}{0.60}\\\hline
 %\multicolumn{1}{c|}{0.40} &
 %0.16 \\ \hline
 
\multicolumn{1}{|r|}{Update and Installation Issue} &
 \multicolumn{1}{c|}{IC13} &
 \multicolumn{1}{c|}{\includegraphics[width = 0.6cm, height = 0.19cm]{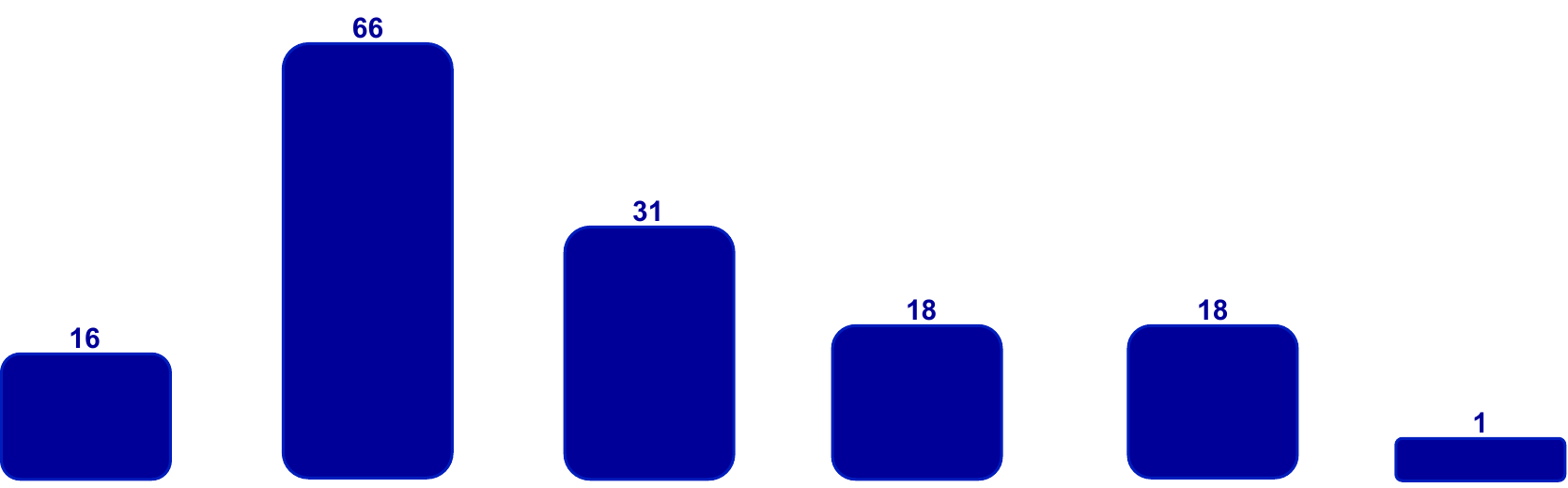}} &
 \multicolumn{1}{c|}{3.27} &
 \multicolumn{1}{c|}{\faBalanceScale{}  0.04} &
 \multicolumn{1}{c|}{-0.28}\\\hline
 %\multicolumn{1}{c|}{0.67} &
 %0.36 \\ \hline

\multicolumn{1}{|r|}{Database Issue} &
 \multicolumn{1}{c|}{IC14} &
 \multicolumn{1}{c|}{\includegraphics[width = 0.6cm, height = 0.19cm]{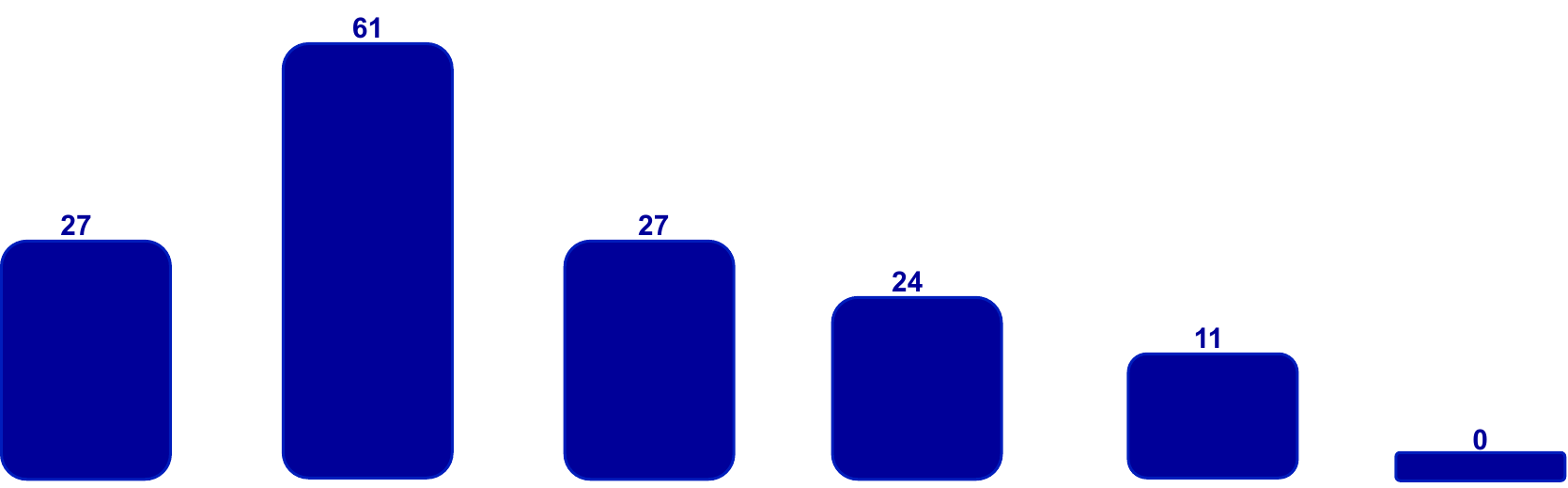}} &
 \multicolumn{1}{c|}{3.46} &
 \multicolumn{1}{c|}{0.25} &
 \multicolumn{1}{c|}{0.03}\\\hline
 %\multicolumn{1}{c|}{0.67} &
 %0.66 \\ \hline

\multicolumn{1}{|r|}{Storage Issue} &
 \multicolumn{1}{c|}{IC15} &
 \multicolumn{1}{c|}{\includegraphics[width = 0.6cm, height = 0.19cm]{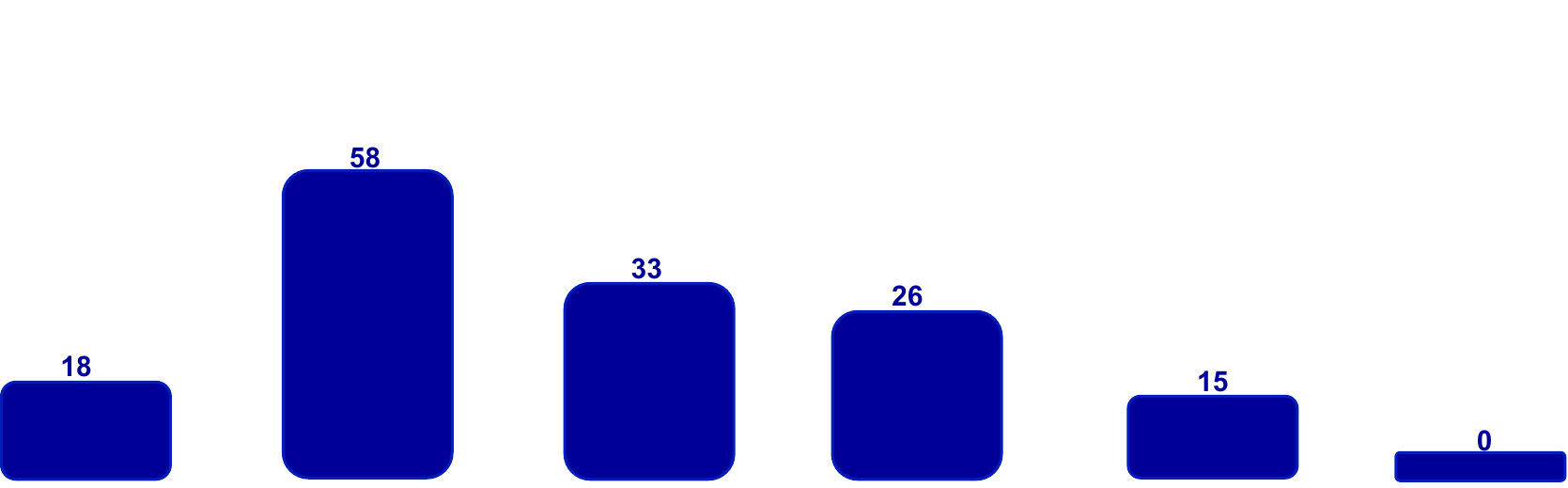}} &
 \multicolumn{1}{c|}{3.25} &
 \multicolumn{1}{c|}{0.12} &
 \multicolumn{1}{c|}{0.31}\\\hline
 %\multicolumn{1}{c|}{1.00} &
 %0.40 \\ \hline

\multicolumn{1}{|r|}{Performance Issue} &
 \multicolumn{1}{c|}{IC16} &
 \multicolumn{1}{c|}{\includegraphics[width = 0.6cm, height = 0.19cm]{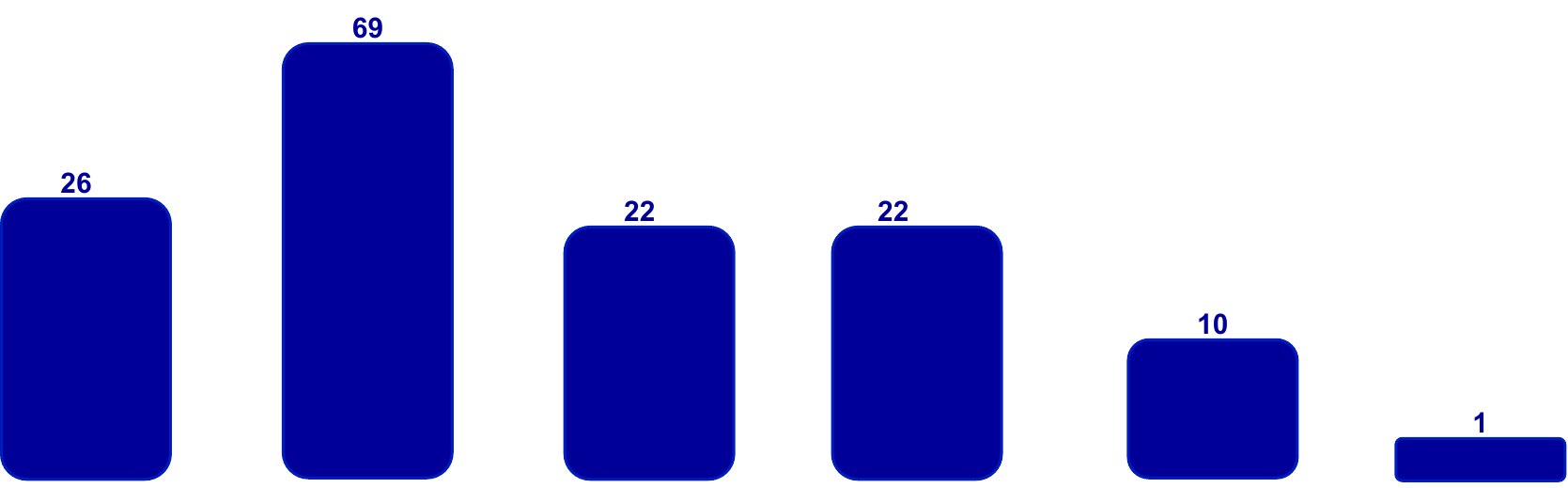}} &
 \multicolumn{1}{c|}{3.51} &
 \multicolumn{1}{c|}{0.21} &
 \multicolumn{1}{c|}{-0.02}\\\hline
 %\multicolumn{1}{c|}{0.47} &
 %0.24 \\ \hline

\multicolumn{1}{|r|}{Networking Issue} &
 \multicolumn{1}{c|}{IC17} &
 \multicolumn{1}{c|}{\includegraphics[width = 0.6cm, height = 0.19cm]{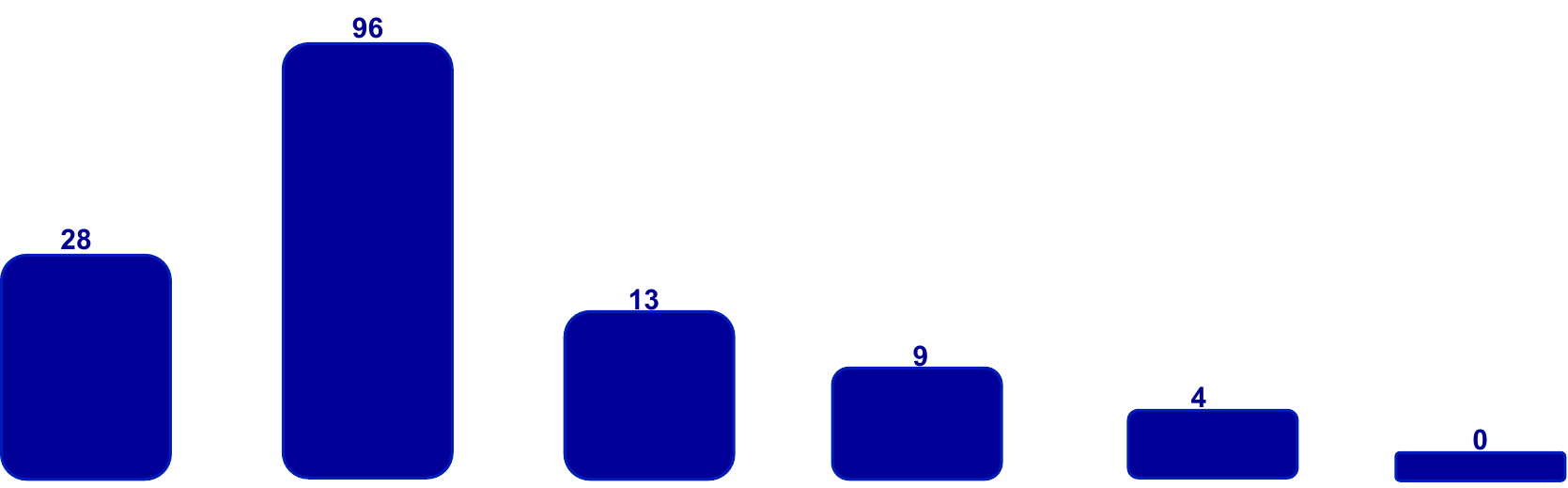}} &
 \multicolumn{1}{c|}{3.15} &
 \multicolumn{1}{c|}{\faBalanceScale{}  0.04} &
 \multicolumn{1}{c|}{-0.07}\\\hline
 %\multicolumn{1}{c|}{0.92} &
 %0.11 \\ \hline

\multicolumn{1}{|r|}{Typecasting Issue} &
 \multicolumn{1}{c|}{IC18} &
 \multicolumn{1}{c|}{\includegraphics[width = 0.6cm, height = 0.19cm]{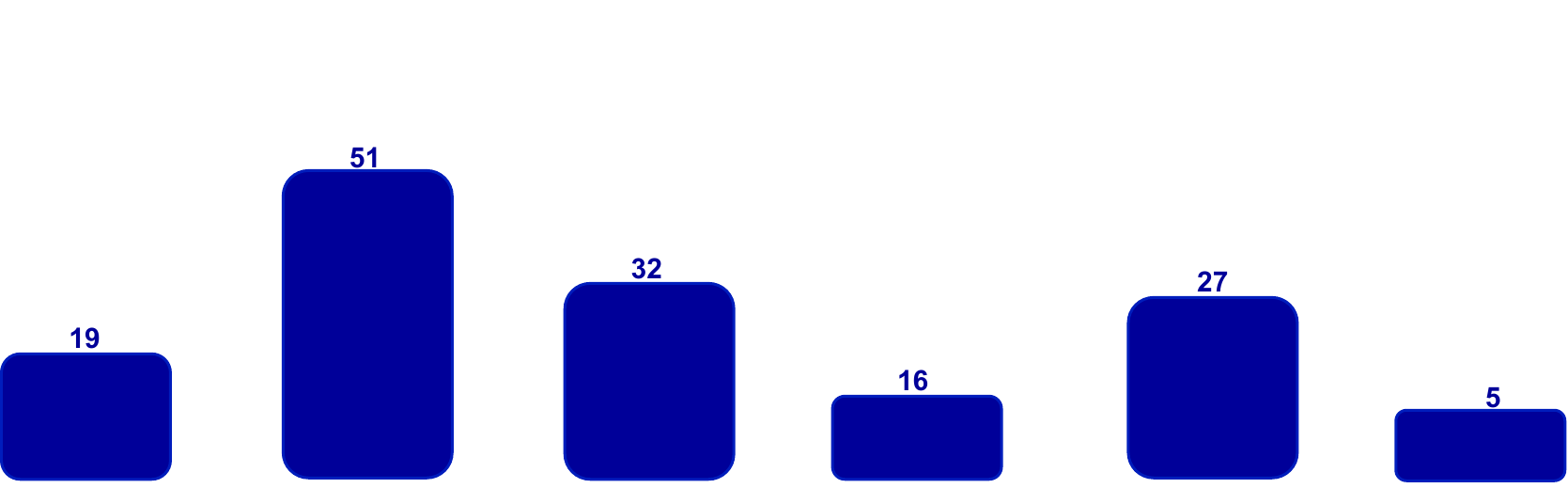}} &
 \multicolumn{1}{c|}{3.03} &
 \multicolumn{1}{c|}{0.12} &
 \multicolumn{1}{c|}{0.38}\\\hline
% \multicolumn{1}{c|}{0.53} &
 %0.18 \\ \hline

\multicolumn{1}{|r|}{Organizational Issue} &
 \multicolumn{1}{c|}{IC19} &
 \multicolumn{1}{c|}{\includegraphics[width = 0.6cm, height = 0.19cm]{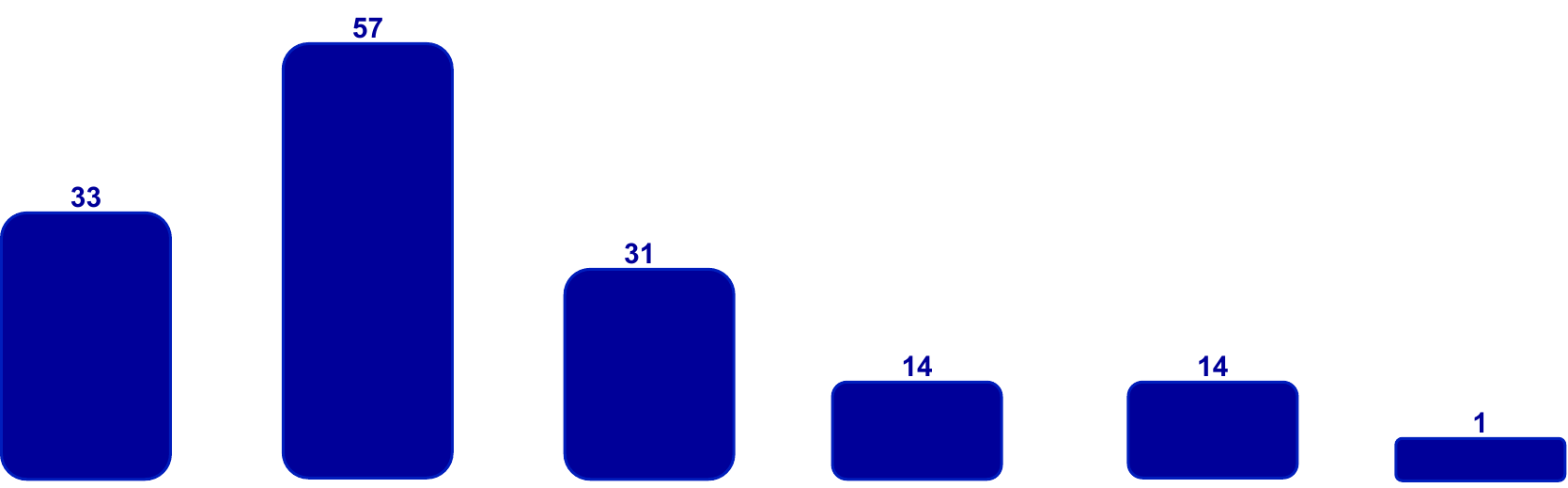}} &
 \multicolumn{1}{c|}{3.52} &
 \multicolumn{1}{c|}{0.14} &
 \multicolumn{1}{c|}{0.11} \\\hline
 %\multicolumn{1}{c|}{0.60} &
 %0.29 \\ \hline

\multicolumn{6}{|l|}{\textbf{Causes of Issues}} \\ \hline
\multicolumn{1}{|r|}{General Programming Error} &
 \multicolumn{1}{c|}{CC1} &
 \multicolumn{1}{c|}{\includegraphics[width = 0.6cm, height = 0.19cm]{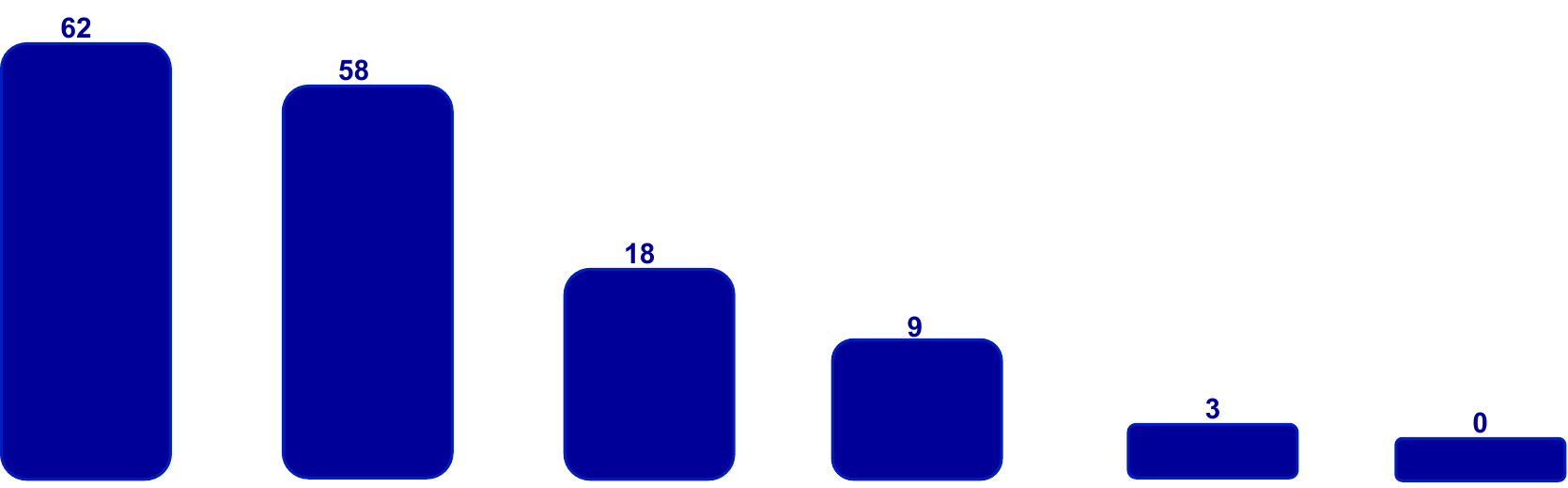}} &
 \multicolumn{1}{c|}{4.11} &
 \multicolumn{1}{c|}{0.53} &
 \multicolumn{1}{c|}{0.17} \\\hline
% \multicolumn{1}{c|}{0.92} &
% -0.10 \\ \hline

\multicolumn{1}{|r|}{Missing Features and Artifacts} &
 \multicolumn{1}{c|}{CC2} &
 \multicolumn{1}{c|}{\includegraphics[width = 0.6cm, height = 0.19cm]{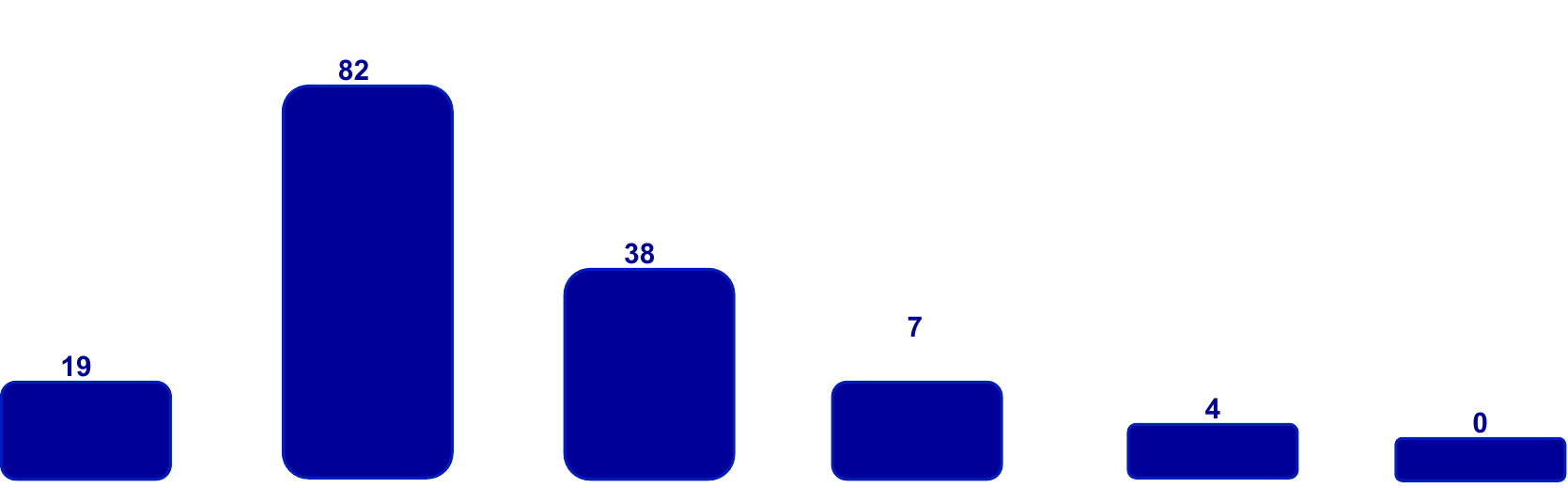}} &
 \multicolumn{1}{c|}{3.70} &
 \multicolumn{1}{c|}{0.47} &
 \multicolumn{1}{c|}{-0.04}\\\hline
% \multicolumn{1}{c|}{0.92} &
 %-0.20 \\ \hline

\multicolumn{1}{|r|}{Invalid Configuration and Communication Problem} &
 \multicolumn{1}{c|}{CC3} &
 \multicolumn{1}{c|}{\includegraphics[width = 0.6cm, height = 0.19cm]{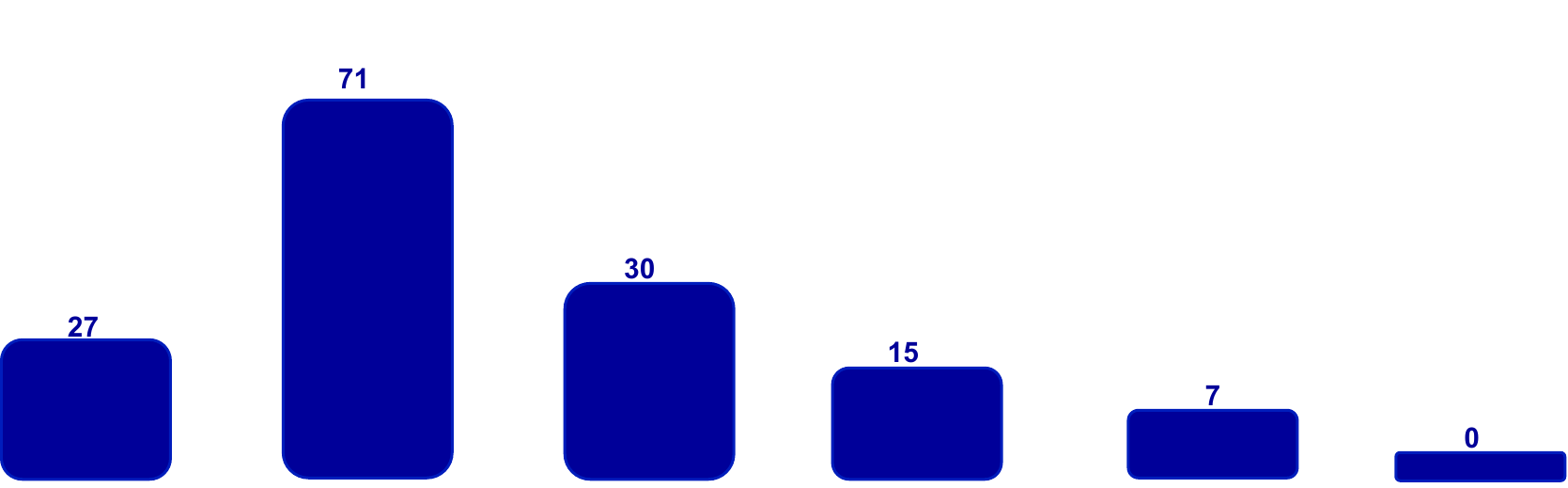}} &
 \multicolumn{1}{c|}{3.64} &
 \multicolumn{1}{c|}{0.25} &
 \multicolumn{1}{c|}{-0.31}\\\hline
% \multicolumn{1}{c|}{0.67} &
% 0.07 \\ \hline

\multicolumn{1}{|r|}{Legacy Versions, Compatibility, and Dependency Problem} &
 \multicolumn{1}{c|}{CC4} &
 \multicolumn{1}{c|}{\includegraphics[width = 0.6cm, height = 0.19cm]{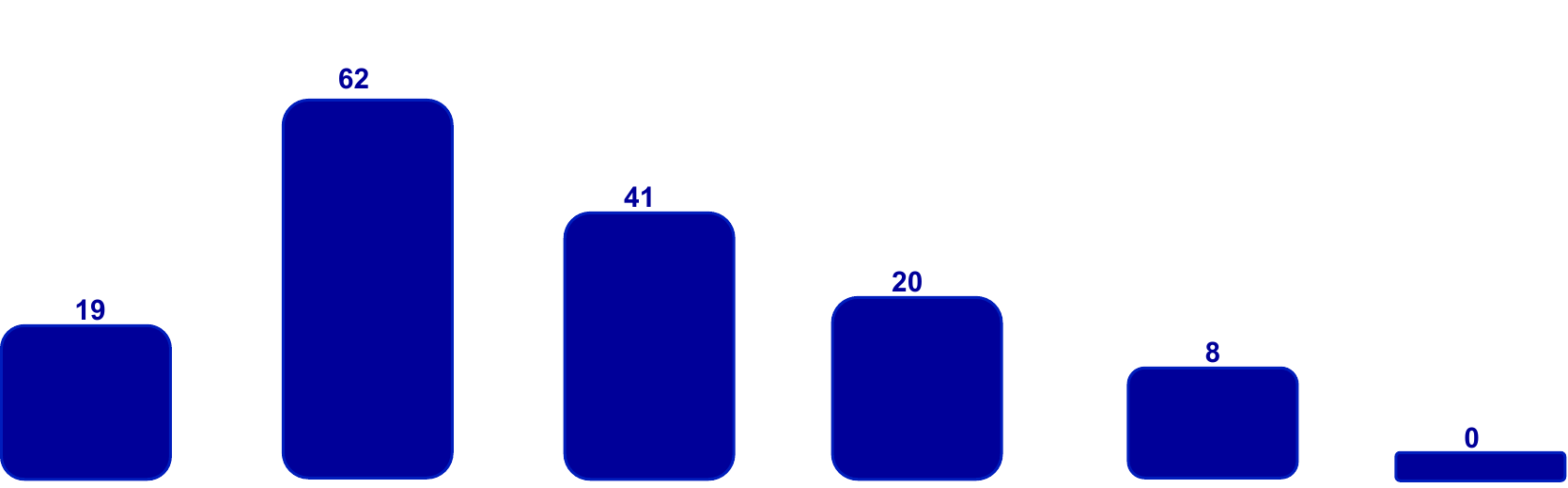}} &
 \multicolumn{1}{c|}{3.43} &
 \multicolumn{1}{c|}{0.25} &
 \multicolumn{1}{c|}{0.20}\\\hline
% \multicolumn{1}{c|}{1.00} &
% -0.22 \\ \hline

\multicolumn{1}{|r|}{Service Design and Implementation Anomaly} &
 \multicolumn{1}{c|}{CC5} &
 \multicolumn{1}{c|}{\includegraphics[width = 0.6cm, height = 0.19cm]{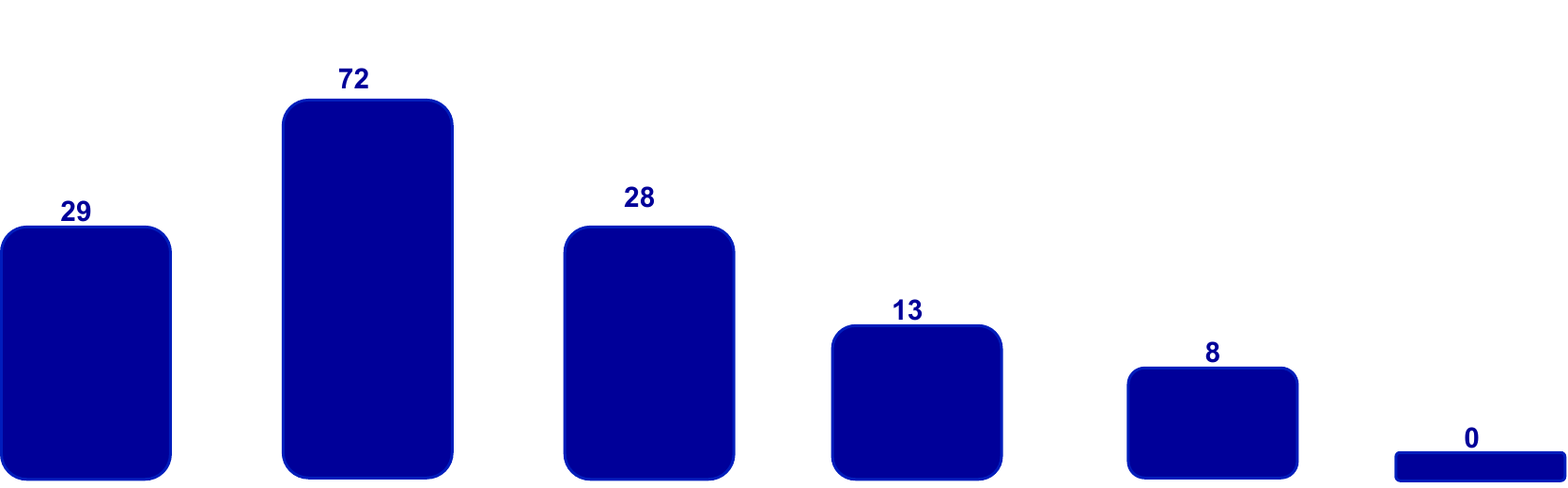}} &
 \multicolumn{1}{c|}{3.67} &
 \multicolumn{1}{c|}{0.35} &
 \multicolumn{1}{c|}{0.16}\\\hline
% \multicolumn{1}{c|}{0.67} &
% -0.05 \\ \hline

\multicolumn{1}{|r|}{Poor Security Management} &
 \multicolumn{1}{c|}{CC6} &
 \multicolumn{1}{c|}{\includegraphics[width = 0.6cm, height = 0.19cm]{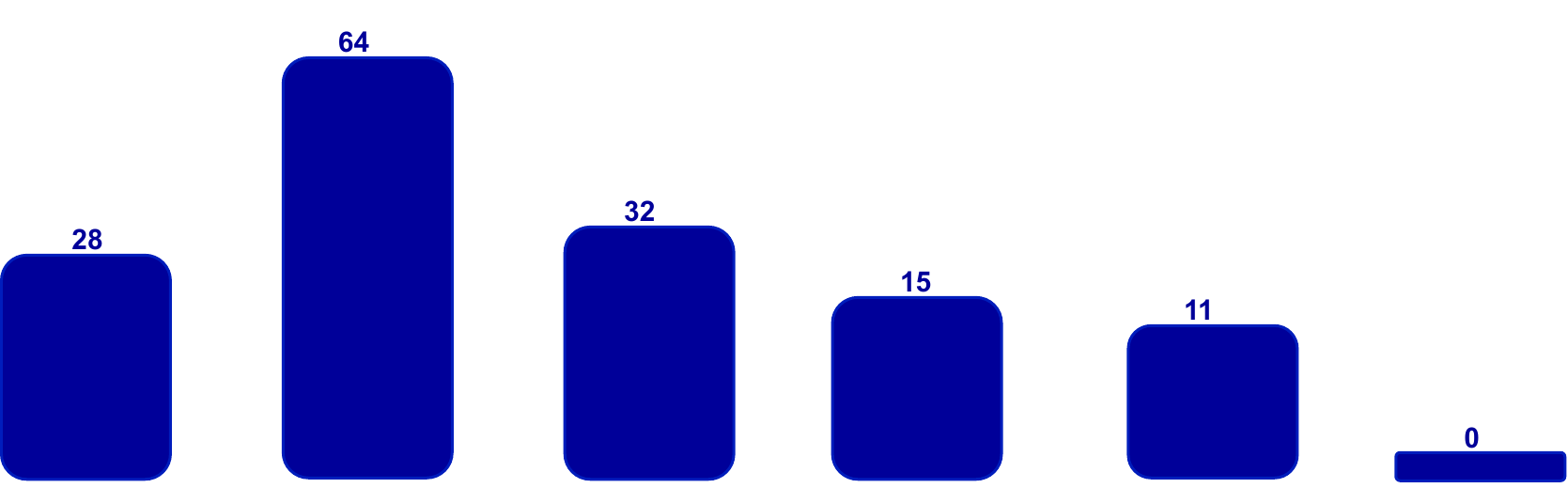}} &
 \multicolumn{1}{c|}{3.55} &
 \multicolumn{1}{c|}{0.21} &
 \multicolumn{1}{c|}{-0.20}\\\hline
% \multicolumn{1}{c|}{0.92} &
% 0.04 \\ \hline

\multicolumn{1}{|r|}{Insufficient Resources} &
 \multicolumn{1}{c|}{CC7} &
 \multicolumn{1}{c|}{\includegraphics[width = 0.6cm, height = 0.19cm]{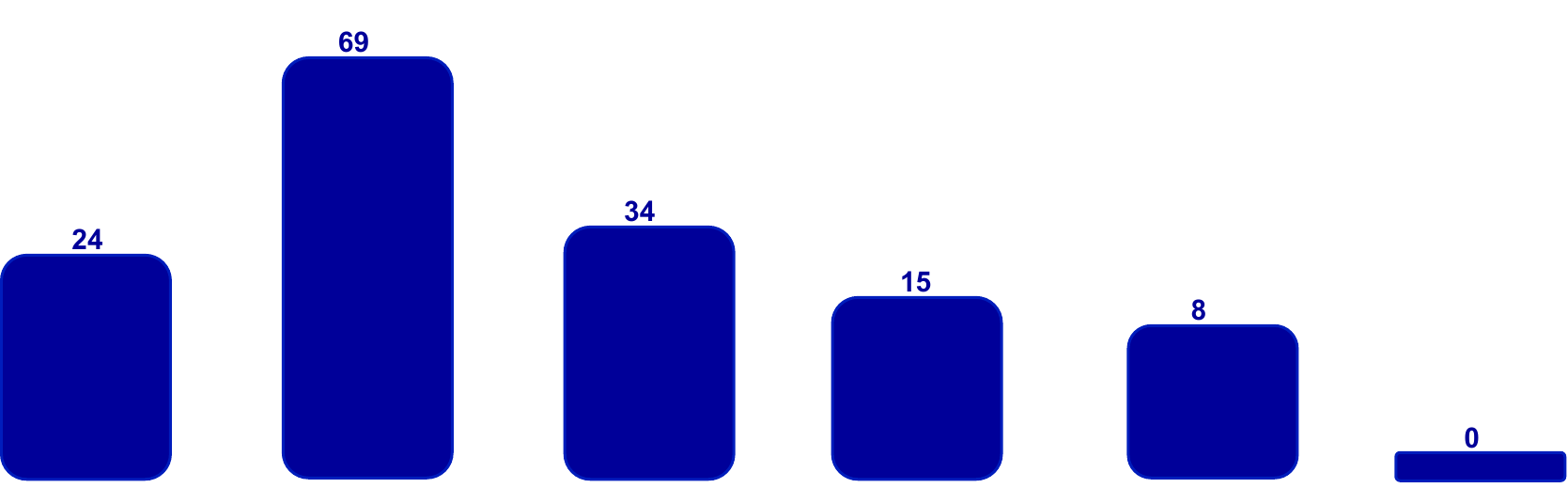}} &
 \multicolumn{1}{c|}{3.57} &
 \multicolumn{1}{c|}{0.40} &
 \multicolumn{1}{c|}{0.02} \\\hline
% \multicolumn{1}{c|}{0.76} &
% 0.03 \\ \hline

\multicolumn{1}{|r|}{Fragile Code} &
 \multicolumn{1}{c|}{CC8} &
 \multicolumn{1}{c|}{\includegraphics[width = 0.6cm, height = 0.19cm]{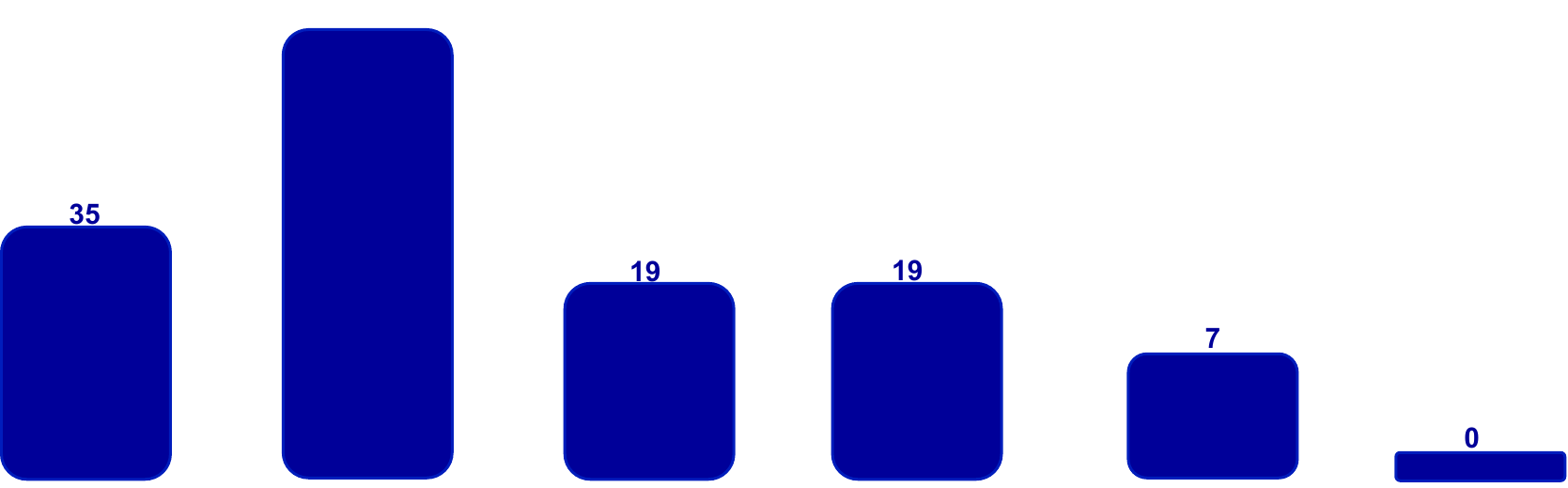}} &
 \multicolumn{1}{c|}{3.71} &
 \multicolumn{1}{c|}{0.25} &
 \multicolumn{1}{c|}{-0.12}\\\hline
% \multicolumn{1}{c|}{0.67} &
% 0.28 \\ \hline

\multicolumn{6}{|l|}{\textbf{Solutions for Issues}} \\ \hline
\multicolumn{1}{|r|}{Add Artifacts} &
 \multicolumn{1}{c|}{SC1} &
 \multicolumn{1}{c|}{\includegraphics[width = 0.6cm, height = 0.19cm]{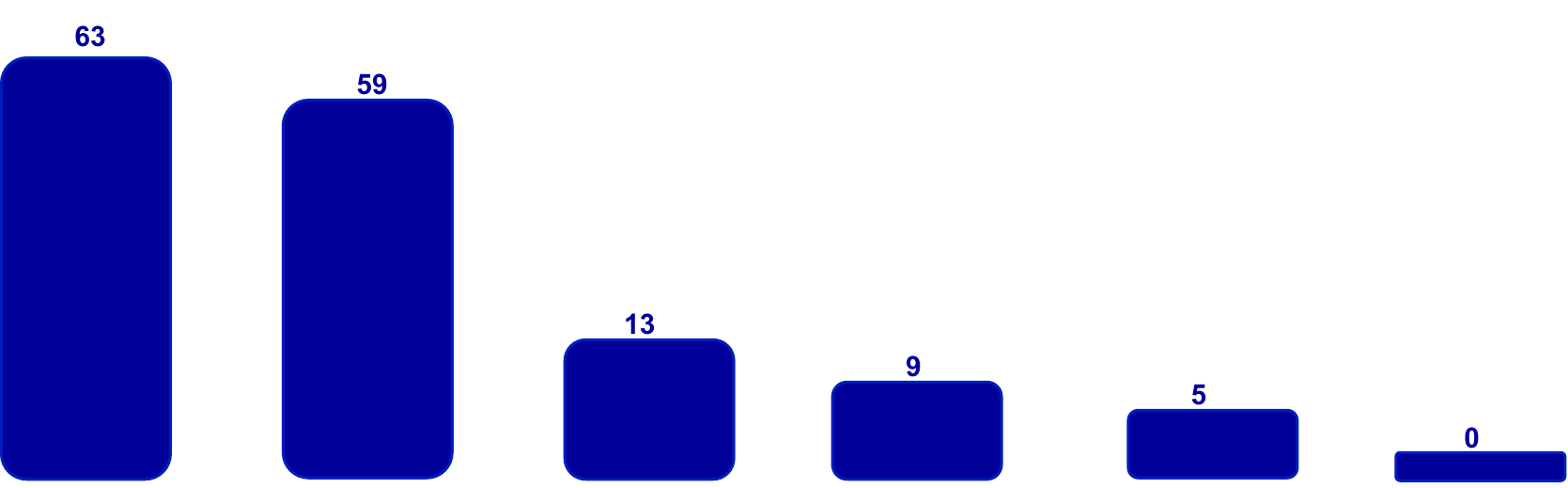}} &
 \multicolumn{1}{c|}{4.11} &
 \multicolumn{1}{c|}{0.40} &
 \multicolumn{1}{c|}{0.20}\\\hline
% \multicolumn{1}{c|}{0.53} &
% 0.25 \\ \hline

\multicolumn{1}{|r|}{Remove Artifacts} &
 \multicolumn{1}{c|}{SC2} &
 \multicolumn{1}{c|}{\includegraphics[width = 0.6cm, height = 0.19cm]{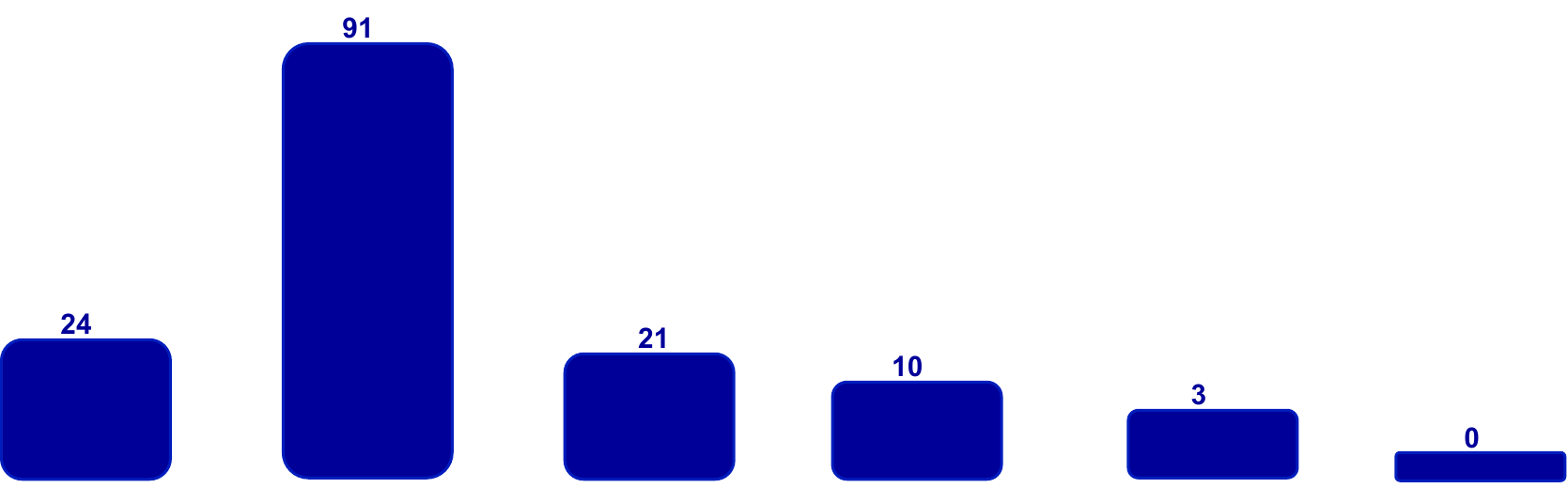}} &
 \multicolumn{1}{c|}{3.83} &
 \multicolumn{1}{c|}{0.60} &
 \multicolumn{1}{c|}{-0.01}\\\hline
% \multicolumn{1}{c|}{0.67} &
% -0.09 \\ \hline

\multicolumn{1}{|r|}{Modify Artifact} &
 \multicolumn{1}{c|}{SC3} &
 \multicolumn{1}{c|}{\includegraphics[width = 0.6cm, height = 0.19cm]{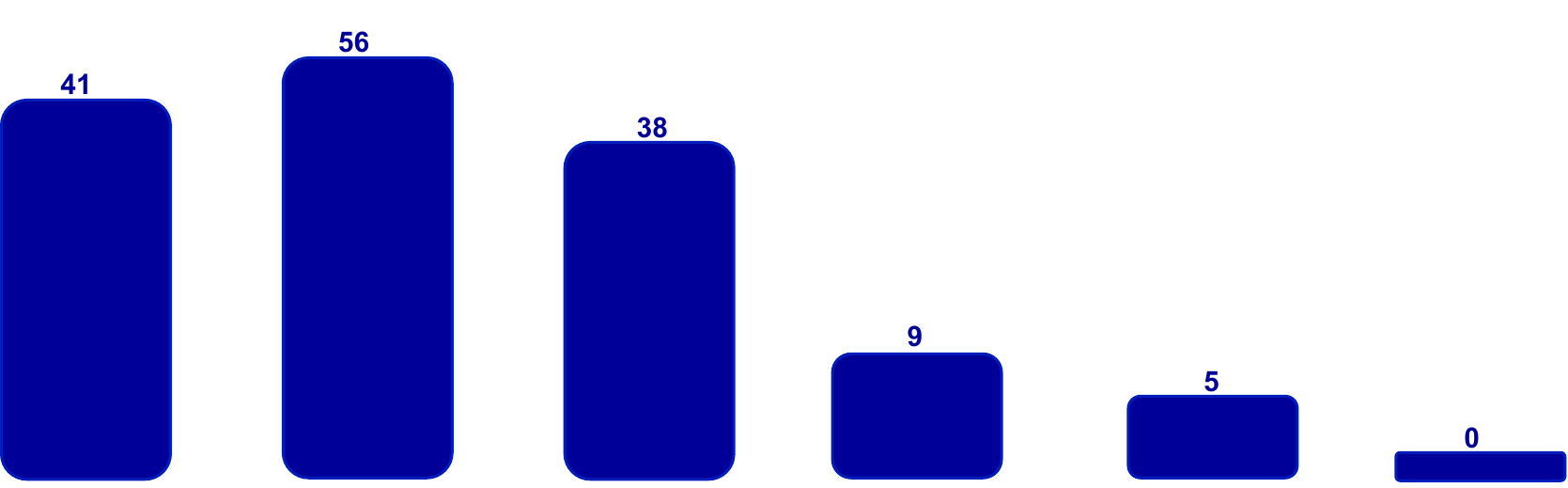}} &
 \multicolumn{1}{c|}{3.80} &
 \multicolumn{1}{c|}{0.21} &
 \multicolumn{1}{c|}{0.10} \\\hline
 %\multicolumn{1}{c|}{0.83} &
 %0.18 \\ \hline

\multicolumn{1}{|r|}{Manage Infrastructure} &
 \multicolumn{1}{c|}{SC4} &
 \multicolumn{1}{c|}{\includegraphics[width = 0.6cm, height = 0.19cm]{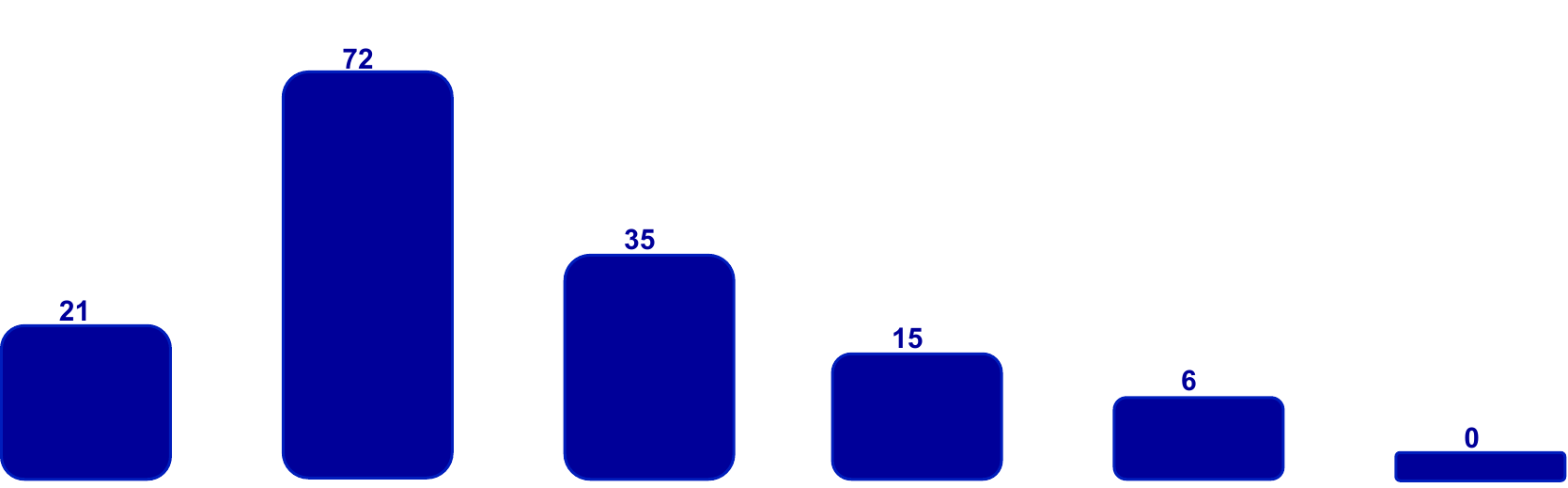}} &
 \multicolumn{1}{c|}{3.58} &
 \multicolumn{1}{c|}{0.30} &
 \multicolumn{1}{c|}{-0.12}\\\hline
% \multicolumn{1}{c|}{1.00} &
% 0.13 \\ \hline

\multicolumn{1}{|r|}{Fix Artifacts} &
 \multicolumn{1}{c|}{SC5} &
 \multicolumn{1}{c|}{\includegraphics[width = 0.6cm, height = 0.19cm]{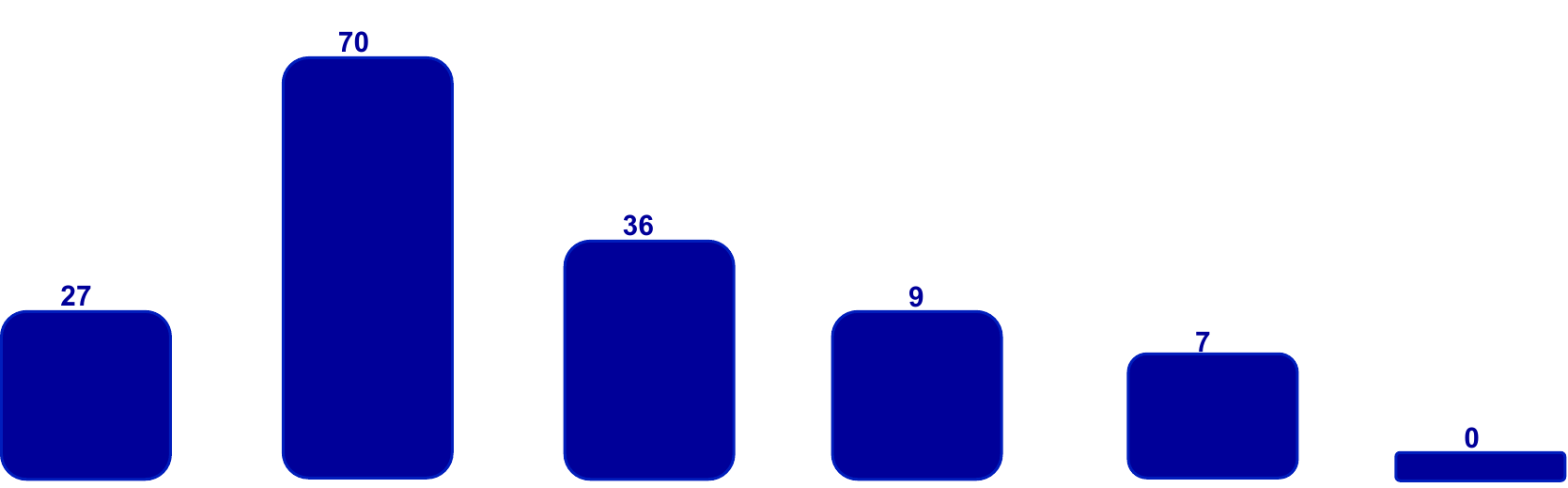}} &
 \multicolumn{1}{c|}{3.68} &
 \multicolumn{1}{c|}{0.30} &
 \multicolumn{1}{c|}{0.20}\\\hline
 %\multicolumn{1}{c|}{0.92} &
 %-0.17 \\ \hline

\multicolumn{1}{|r|}{Manage Configuration and Execution} &
 \multicolumn{1}{c|}{SC6} &
 \multicolumn{1}{c|}{\includegraphics[width = 0.6cm, height = 0.19cm]{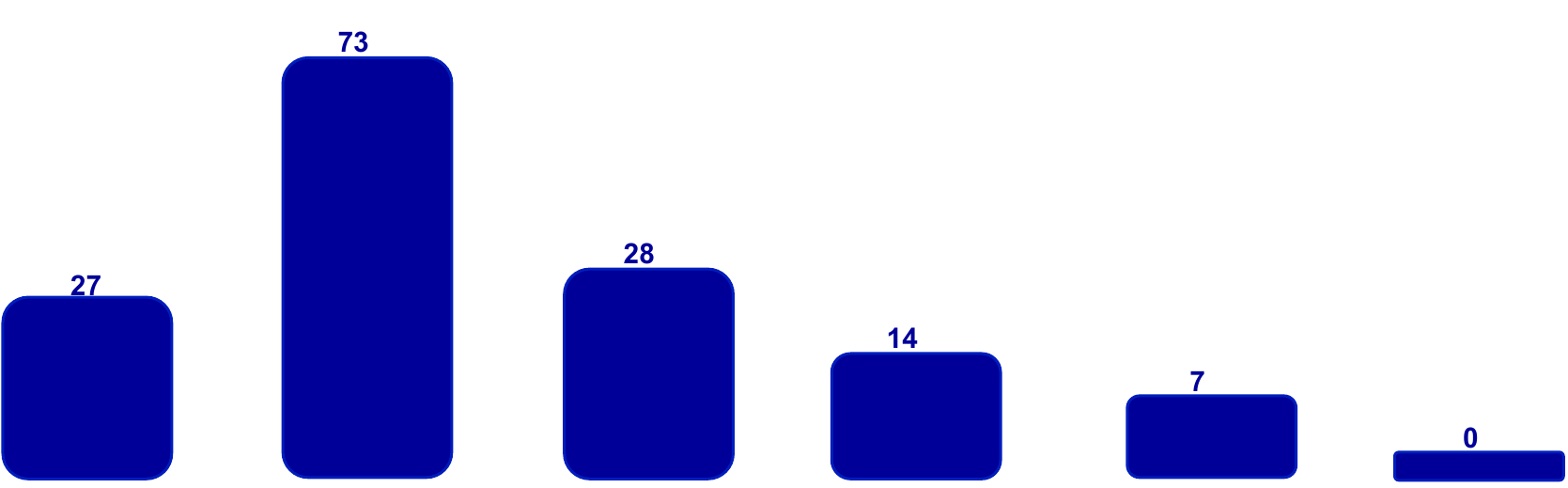}} &
 \multicolumn{1}{c|}{3.66} &
 \multicolumn{1}{c|}{0.35} &
 \multicolumn{1}{c|}{0.12}\\\hline
% \multicolumn{1}{c|}{0.76} &
 %-0.10 \\ \hline

\multicolumn{1}{|r|}{Upgrade Tools and Platforms} &
 \multicolumn{1}{c|}{SC7} &
 \multicolumn{1}{c|}{\includegraphics[width = 0.6cm, height = 0.19cm]{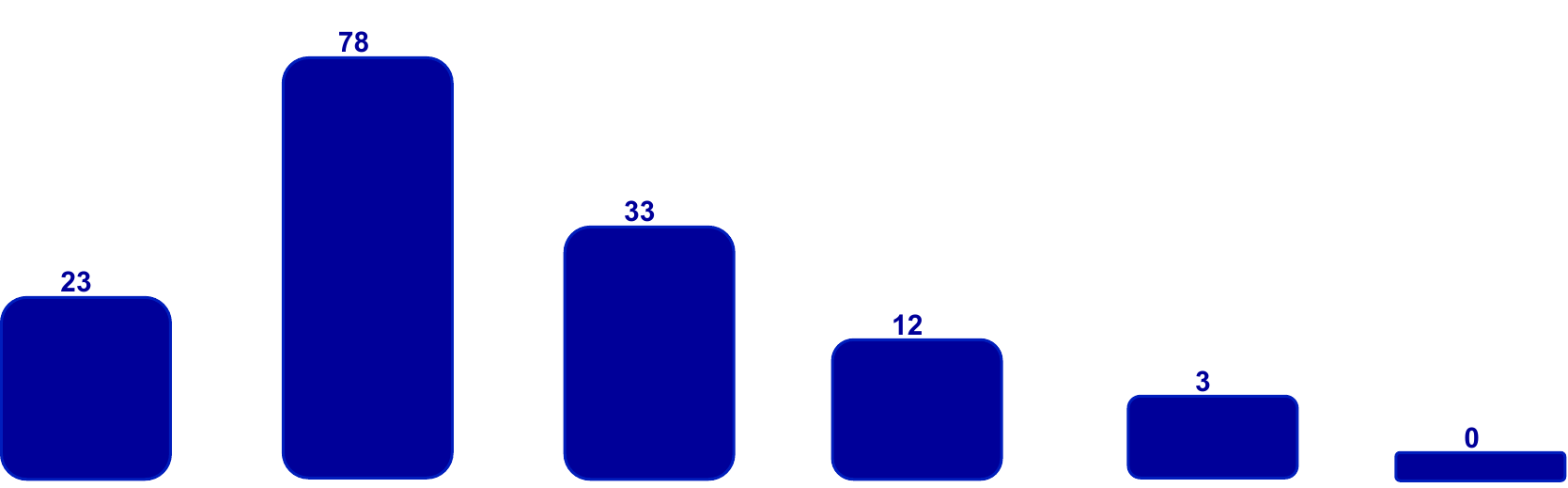}} &
 \multicolumn{1}{c|}{3.71} &
 \multicolumn{1}{c|}{0.35} &
 \multicolumn{1}{c|}{-0.27}\\\hline
% \multicolumn{1}{c|}{0.92} &
% -0.02 \\ \hline

\multicolumn{1}{|r|}{Import/Export Artifacts} &
 \multicolumn{1}{c|}{SC8} &
 \multicolumn{1}{c|}{\includegraphics[width = 0.6cm, height = 0.19cm]{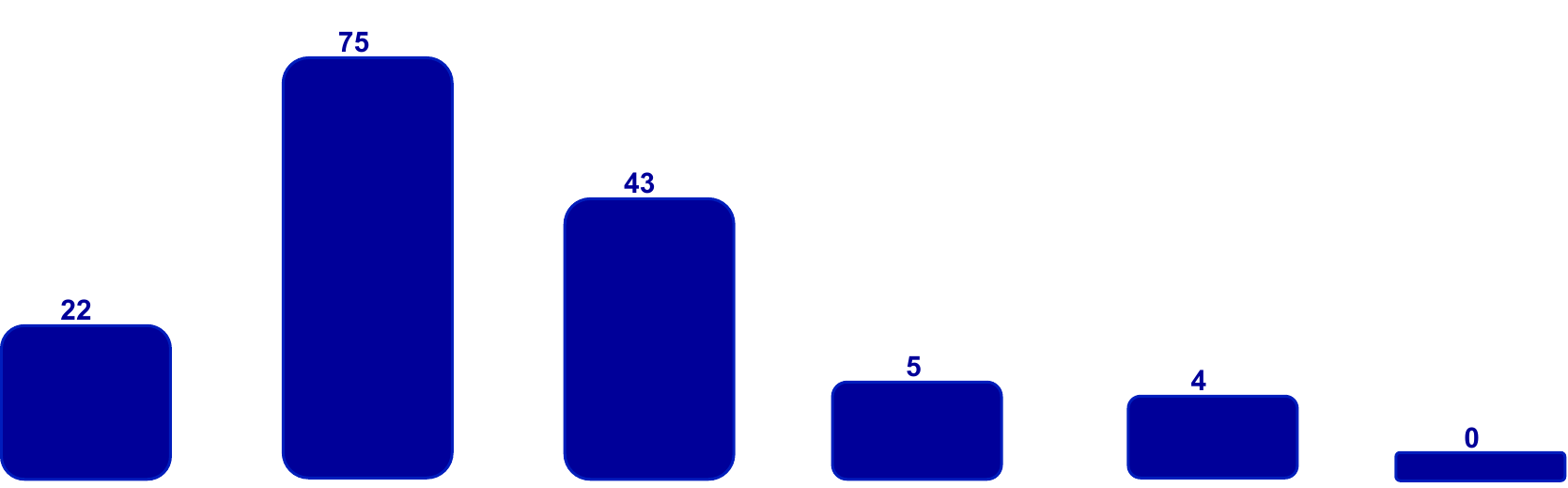}} &
 \multicolumn{1}{c|}{3.71} &
 \multicolumn{1}{c|}{0.47} &
 \multicolumn{1}{c|}{0.22}\\\hline
% \multicolumn{1}{c|}{0.92} &
 %-0.18 \\ \hline

\end{tabular}
\end{table*}}

\section{\textcolor{black}{Discussion and Needs for Future Research}}
\label{Discussion}

\textcolor{black}{
In this section, we analyze the relationships between the identified issues, causes, and solutions (Section~\ref{Sec:MappingIssueCausesSolutions}), and interpret these issue--cause--solution links derived from OSS issue discussions to explain dominance patterns and guide practitioners in prioritizing mitigation strategies. We then present the research implications (Section~\ref{Sec:Implications}), which synthesize the study results, discuss implications for research and practice, and identify directions for future research (Section~\ref{researchagenda}) on MSAs.}

%\textcolor{black}{In this section, we analyze the relationships between the identified issues, causes, and solutions (Section \ref{Sec:MappingIssueCausesSolutions}), followed by the research implications (Section \ref{Sec:Implications}), which synthesize the study results, discuss implications for research and practice, and identify directions for future research (\ref{Sec:Dimensionsoffuture}) on MSAs.}

%We analyze in this section the relationship between the issues, causes, and solutions (Section \ref{Sec:MappingIssueCausesSolutions}) followed by presenting the implications of the research (Section \ref{Sec:Implications})provides a synthesised discussion of study results, presents implications of the results for research and practice, and pinpoints potential dimensions of future research on MSAs.

\subsection{Relationship between issues, causes, solutions}
\label{Sec:MappingIssueCausesSolutions}
\begin{figure*}[!htbp]
\centering
\includegraphics[width=0.8\textwidth]{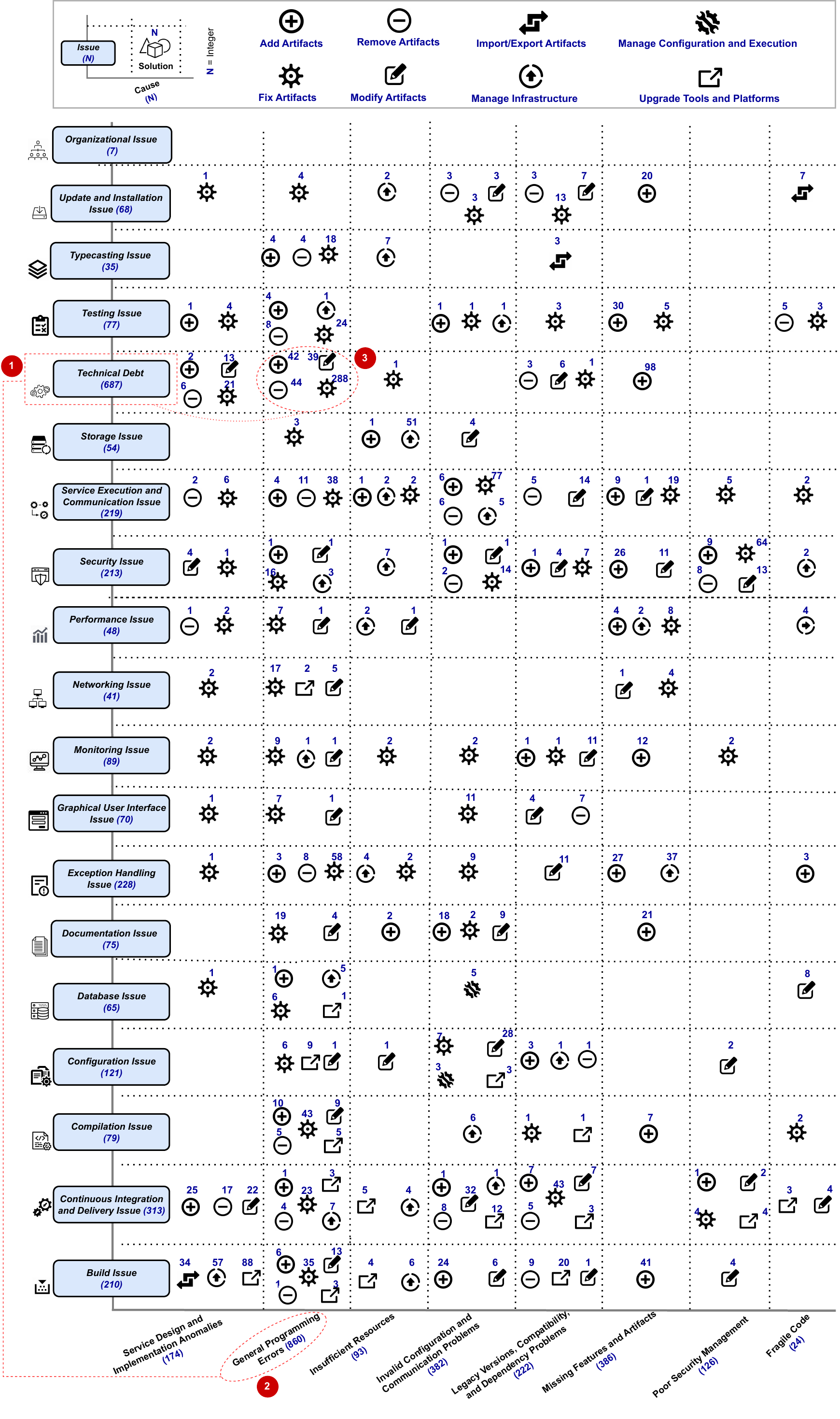}
 \caption{Mapping between issues, causes, and solutions in microservices systems}
 \label{fig:mapping}
\end{figure*}

While the taxonomy in Figure \ref{fig:Taxonomy} provides a categorization of issues in microservices systems, a mapping between the issues, their causes, and solutions is presented in Figure \ref{fig:mapping}. Mapping diagrams are frequently used in systematic mapping studies---relying on bubble plots---to correlate data or concepts along different dimensions \cite{petersen2008systematic}. We have chosen the mapping diagram to depict the issues on the Y-axis and their causes on the X-axis, and to present the solutions that can address these issues at the intersections of the X and Y axes. Eight different types of symbols, labeled at the top of the figure, are used to represent various types of solutions. The interpretation of the mapping in Figure \ref{fig:mapping} involves locating a given issue on the Y-axis, associating this issue with its cause(s) on the X-axis, and identifying the possible solutions to address the issue, as elaborated and exemplified below. Figure \ref{fig:mapping} indicate that it is possible that one issue could cause other issues. Similarly, one cause may be the reason for several issues, or several causes may contribute to one issue.

\begin{itemize}
\item \textit{Issues in microservices systems (Y-axis)}: A total 19 categories of issues, adopted from Figure \ref{fig:Taxonomy}, are presented on the Y-axis. For example, one of the issue categories Technical Debt has a total of 687 instances of issues as presented in Figure \ref{fig:Taxonomy}.
 
\item \textit{Causes of the issues (X-axis)}: A total of 8 categories of causes are presented on the X-axis, mapped with the corresponding issues. For example, General Programming Errors, such as \textit{incorrect naming and data type} (157, 6.92\%), \textit{testing error} (25, 1.10\%), and \textit{poor documentation} (22, 0.97\%), are the predominant causes of Technical Debt issues. In comparison, the causes like Poor Security Management have no impact on Technical Debt issues.
 
\item \textit{Solutions to resolve the issues}: A total of 8 categories of solutions are presented at the intersection of the issues and their causes. For example, solutions, such as fixing an artifact (code, GUI, errors in build files) or removing an artifact (code dependency, empty tag, unnecessary documentation) can help with fixing a majority of the Technical Debt issues. 
\end{itemize}

The mapping in Figure \ref{fig:mapping} can have a diverse interpretation based on the intent of the analysis that may include but is not limited to frequency analysis and data correlations. It is virtually impossible to elaborate on all possible interpretations, however; to exemplify some of the possible interpretations can be as follows:

\begin{itemize}
\item \textit{What are the most and least frequently occurring issues in microservices systems}? As per the mapping, the most frequent (top 3) issues are related to Technical Debt, CI/CD, and Exception Handling, representing a total of 764 identified issues. On the other hand, the least frequent issues relate to Organizational, Update and Installation, and Typecasting issues. \textcolor{black}{The dominance of Technical Debt and CI/CD in Figure~\ref{fig:mapping} should be interpreted as a recurring pattern in OSS issue discussions, rather than as isolated implementation mistakes. In our issue--cause--solution links, Technical Debt frequently appears when changes propagate across multiple services and accumulate into Code Debt and Service Design Debt, while CI/CD reappears when pipelines, dependencies, and execution environments must continuously co-evolve across many services. This is consistent with the mapping evidence, where the dominant cause categories (e.g., General Programming Error, Invalid Configuration and Communication Problem, and Legacy Versions, Compatibility and Dependency Problem) repeatedly link to these issue categories, and where the associated fixes are mainly artifact-oriented (Fix Artifacts, Add Artifacts, and Modify Artifacts). In contrast, low-frequency issue categories (e.g., Organizational, Update and Installation, and Typecasting) should not be interpreted as negligible. Rather, they may be reported less consistently, absorbed into broader issue categories (e.g., upgrade problems reported under CI/CD/build or dependency issues), or discussed indirectly through symptoms that are recorded under higher-frequency issue categories. Therefore, frequency in our dataset captures what is most visible and repeatedly discussed in OSS issue trackers, while some low-frequency issue categories may still have high impact when they occur.}%\textcolor{black}{The dominance of Technical Debt and CI/CD in Figure~\ref{fig:mapping} should be interpreted as a systemic phenomenon rather than isolated implementation mistakes. Technical Debt accumulates because microservices amplify distributed change and service-design drift, while CI/CD instability recurs because pipelines, dependencies, and environments co-evolve across many services. This interpretation is consistent with our mapping evidence, where cross-cutting cause families (e.g., General Programming Errors, Invalid Configuration and Communication, and Legacy/Compatibility/Dependency problems) repeatedly link to these dominant categories, and where solutions are mainly artifact-centric (Fix/Add/Modify Artifacts), reflecting reactive stabilization rather than preventive governance.} 
%\textcolor{black}{
%In contrast, low-frequency categories (e.g., Organizational, Update and Installation, and Typecasting) should not be interpreted as negligible. Rather, they are likely to be (i) reported less consistently in issue trackers, (ii) absorbed into broader categories (e.g., upgrade problems reported as CI/CD/build or dependency issues), or (iii) manifested as latent risks that surface indirectly through other high-frequency categories. Therefore, frequency in our dataset captures what is most visible and repeatedly encountered in OSS issue discussions, while some low-frequency categories may still have high impact when they occur.}

\item \textit{What are the most and least common causes of a specific category of issues}? The mapping of the causes suggests that General Programming Errors, Invalid Configuration and Communication Problems along with Legacy Versions, Compatibility and Dependency Problem are the most common causes of the issues. Similarly, Fragile Code represents the least common cause for issues in microservices systems. \textcolor{black}{
A key interpretive insight from the cause distribution is that the most common causes are not narrowly tied to one development phase; instead, they represent recurring breakdowns in integration and evolution across services. General Programming Errors frequently occur because microservices multiply implementation points and boundary conditions, so local coding problems more often surface as system-level symptoms. Invalid Configuration and Communication Problems recur because microservices externalize behavior into configuration and networked interactions, where inconsistencies across services and environments can trigger runtime and CI/CD failures. Legacy Versions, Compatibility and Dependency Problem is common because microservices systems rely heavily on libraries, container images, and platform/tool versions that evolve asynchronously across services. Taken together, these dominant cause categories recur across multiple issue categories in Figure~\ref{fig:mapping}, which helps explain why Technical Debt, CI/CD, build, and service execution and communicating issues remain prominent in OSS issue discussions.}

\item \textit{What are the most and least recurring solutions to fix a specific category of issues}? Fixing Artifacts, Add Artifacts, and Modify Artifacts represent the most recurring solutions to address microservices issues, whereas importing/Export Artifacts, Upgrade Tools and Platforms, and Manage Configuration and Execution categories represent the least recurring solutions to address microservices issues.  \textcolor{black}{
The distribution of solutions in Figure~\ref{fig:mapping} suggests that OSS microservices maintenance is predominantly reactive and artifact-centric. Fixing Artifacts, Add Artifacts, and Modify Artifacts are the most recurring solutions because they are typically fast to apply, localized to specific code/configuration/build artifacts, and immediately verifiable through builds, tests, or pipeline reruns (e.g., patching source code, updating build scripts, adjusting configuration files, fixing deployment descriptors). In contrast, the least recurring solution categories (Import/Export Artifacts, Upgrade Tools and Platforms, and Manage Configuration and Execution) often require broader coordination across services and toolchains, can introduce new compatibility and dependency risks, and may affect multiple services simultaneously, which makes them harder to implement and validate in OSS settings. Overall, this gap indicates that while recurring issues are frequently resolved via local artifact updates, there is a need for research and tooling that better supports broader, system-level solutions (e.g., safer upgrade strategies and systematic configuration/execution management) to reduce repeated issue recurrence across microservices systems.}

\item \textcolor{black}{
\noindent\textit{How practitioners can prioritize mitigation?} Figure~\ref{fig:mapping}, together with the solution taxonomy (Table~4), supports an evidence-based prioritization that practitioners can apply as follows. \emph{First}, prioritize the highest-frequency Issue Categories (RQ1) to focus effort where issues occur most often. \emph{Second}, within those Issue Categories, target the most common Cause Categories (RQ2) because they recur across multiple Issue Categories and therefore represent recurring drivers of problems (e.g., General Programming Errors, Invalid Configuration and Communication Problems, and Legacy Versions, Compatibility and Dependency Problem). \emph{Third}, choose an initial mitigation strategy using the most recurring Solution Categories (RQ3), such as Fixing Artifacts, Add Artifacts, and Modify Artifacts, and then complement these with less recurring but broader Solution Categories (e.g., Upgrade Tools and Platforms, Manage Configuration and Execution) when the mapping shows repeated cross-service occurrence. Overall, the issue--cause--solution links enable practitioners to decide \emph{what} to address first (high-frequency Issue Categories), \emph{where} to intervene (dominant Cause Categories), and \emph{how} to proceed (recurring Solution Categories).}

\end{itemize}

\subsection{Implications}
\label{Sec:Implications}
%\subsection{Issues in microservices systems} 

This section discusses the key findings of our study, using a \faLeanpub \hspace{0.5mm} icon to indicate implications for researchers and a \faMale \hspace{0.5mm} icon to denote the implications for practitioners. 

\textbf{Technical Debt}: The results of this study indicate that more than one-fourth (25.59\%) of the issues are related to TD, spreading across a plethora of microservices systems development activities, such as design, coding, refactoring, and configuration \cite{freire2022software}. The detailed analysis of the causes reveals that most TD issues occur due to GPE, including \textit{compile time errors}, \textit{erroneous method definition and execution}, and \textit{incorrect naming and data type}. We observed that TD issues are mainly addressed by fixing, adding, and removing the artifacts. We also observed that TD in microservices systems is growing at a higher rate than other types of issues identified in this study. Recently published studies that investigated TD in microservices systems (e.g. \cite{de2021identifying, de2021reducing, 2020Does, bogner2018limiting, A2-michael2023empirical, A69-bacchiega2022microservices}) have discussed several aspects of TD, such as architectural TD \cite{de2021identifying}, repaying architectural TD \cite{de2021reducing}, TD before and after the migration to microservices \cite{2020Does}, and limiting TD with maintainability assurance \cite{bogner2018limiting}. However, majority of TD in our study is related to \textit{code} and \textit{service design} debt, i.e., the architecture and implementation level of TD in microservices systems. Our study provides in-depth details about the types of TD, their causes, and solutions that can raise awareness of microservices practitioners to manage TD issues before they become too costly. Based on the study findings, \faLeanpub \hspace{0.5mm} we assert that future studies can investigate several other aspects of TD in microservices systems, such as (i) controlling TD through the design of microservices systems, (ii) investigating TD of microservices systems (e.g., \textsc{service dependencies}) at the code, design, and communication level, and (iii) proposing dedicated techniques and tools to identify, measure, prioritize, monitor, and prevent TD (e.g., \textsc{deprecated flags}, \textsc{data race}) in microservices systems. 

% \FrameSep=3pt\FrameRule=0.5pt\begin{framed} 
% \noindent \textit{Practitioners should prioritize managing technical debt, including code debt and service design debt, in microservices systems to mitigate potential risks and ensure the long-term health of the systems. This involves addressing source code issues like refactoring and duplication, and adhering to successful practices in microservices system design.}
% \end{framed}

% \begin{tcolorbox} [sharp corners, boxrule=0.1mm,]
% \small
% \textcolor{black}{\textit{Practitioners should prioritize managing technical debt, including code debt and service design debt, in microservices systems to mitigate potential risks and ensure the long-term health of the systems. This involves addressing source code issues like refactoring and duplication, and adhering to successful practices in microservices system design.}}
% \end{tcolorbox}

\textbf{CI/CD Issues}: CI and CD rely on a number of software development practices, such as rapid prototyping, and sprinting to enable practitioners with frequent integration and delivery of software systems and applications \cite{shahin2017continuous}. The combination of CI/CD and microservices systems enables practitioners to gain several benefits, including maintainability, deployability, and cohesiveness \cite{o2017continuous}. This study reports various CI/CD issues (55 types), their causes, and solutions mainly related to delivery pipelines and establishing cloud infrastructure management platforms (e.g., Google Cloud, AWS) for microservices systems. The primary causes behind the CI/CD issues are related to SD\&IA (e.g., \textsc{wrong dependencies chain}), GPE (e.g., \textsc{long message chain}), and ICC (e.g., \textsc{incorrect configuration setting }) categories. Most of the CI/CD issues are addressed by fixing artifacts and upgrading tools and platforms. Various issues related to continuous deployment, delivery, and integration of microservices in continuous software engineering (e.g., CI/CD, DevOps) have been discussed in the literature (e.g., \cite{jamshidi2018microservices, chen2018microservices, waseem2020systematic}). However, none of the above mentioned studies provide fine-grain details about these issues. Based on the findings of this study, future work can investigate several aspects of combining microservices systems with CI/CD, \faLeanpub \hspace{0.5mm} such as (i) proposing general guidelines and strategies for preventing and addressing \textsc{cd pipeline errors} in microservices systems, and (ii) enriching the issue, cause, and solution knowledge for microservices systems in the multi-cloud (e.g., AWS, Google Cloud) containerized environment.

% \FrameSep=3pt\FrameRule=0.5pt\begin{framed}
% \noindent \textit{Practitioners should pay attention to correct configuration settings, updating and managing dependencies, and keeping abreast of best practices in tools and platforms being used (like Docker, Kubernetes, AWS, Google Cloud). They should also consider investing in upgrading tools and platforms, and fixing artifacts as a common resolution to many CI/CD issues. This will ensure a smoother and more reliable CI/CD process.}
% \end{framed}

% \begin{tcolorbox} [sharp corners, boxrule=0.1mm,]
% \small
% \textcolor{black}{\textit{Practitioners should pay attention to correct configuration settings, updating and managing dependencies, and keeping abreast of best practices in tools and platforms being used (like Docker, Kubernetes, AWS, Google Cloud). They should also consider investing in upgrading tools and platforms, and fixing artifacts as a common resolution to many CI/CD issues. This will ensure a smoother and more reliable CI/CD process.}}
% \end{tcolorbox}

%\begin{noteBox}
% \textcolor{black}{\textit{Practitioners should pay attention to correct configuration settings, updating and managing dependencies, and keeping abreast of best practices in tools and platforms being used (like Docker, Kubernetes, AWS, Google Cloud). They should also consider investing in upgrading tools and platforms, and fixing artifacts as a common resolution to many CI/CD issues. This will ensure a smoother and more reliable CI/CD process.}}
%\end{noteBox}

\textbf{Security Issues}: Microservices systems are vulnerable to a multitude of security threats due to their distributed nature and availability over the public clouds, making them a potential target for cyber-attacks. Our study results indicate that 7.99\% of the issues are related to the security of microservices systems, mainly due to PSM (e.g., \textsc{security dependencies}), GPE (e.g., \textsc{long message chain}), and ICC (e.g., \textsc{wrong connection closure}), and most of the security issues can be addressed by fixing, adding, and modifying artifacts. The issues related to \textsc{handling authorization header}, \textsc{shared authentication}, and \textsc{OAuth token error} indicate that most security issues occur during the authorization and authentication process of microservices. It also indicates that microservices have poor security at the application level.  Some of our Other issues related to \textit{access control} and \textit{secure certificate and connection} also confirm that microservices systems have a much larger attack surface area than traditional systems (e.g., monolithic systems). Current literature (e.g., \cite{A71-beahan2025characterizing}) also lists several security-related vulnerabilities in microservices systems. The security issues, causes, and solutions identified in this study can help practitioners better understand why and where specific security issues may occur in microservices systems. For instance, practitioners might want to avoid writing unsafe code to prevent access control issues. Our findings suggest that security issues are multi-faceted, and security problems can be raised at different levels of microservices systems. Therefore, it is valuable to \faMale \hspace{0.5mm} (i) develop dedicated strategies and guidelines to address security vulnerability and related risks at various levels, such as data centers, cloud providers, virtualization, communication, orchestration, and (ii) propose multi-layered security solutions for fine-grained security management in microservices systems.

% \FrameSep=3pt\FrameRule=0.5pt\begin{framed}
% \noindent \textit{To ensure robust security in microservices systems, practitioners should implement multi-layered security solutions. This involves strategies for data centers, virtualization, communication, and orchestration. Measures such as secure certificates, encryption, and access control policies are essential for achieving strong security management in microservices.}
% \end{framed}

% \begin{tcolorbox} [sharp corners, boxrule=0.1mm]
% \small
%      \textcolor{black}{\textit{To ensure robust security in microservices systems, practitioners should implement multi-layered security solutions. This involves strategies for data centers, virtualization, communication, and orchestration. Measures such as secure certificates, encryption, and access control policies are essential for achieving strong security management in microservices.}}
% \end{tcolorbox}

\textbf{Build Issues}: Build is a process that compiles code, runs unit tests, and produces artifacts that are ready to deploy as a working program for the software release. The build process may consist of several activities, such as parsing, dependency resolution, resource processing, and assembly \cite{lou2020}. Our study results indicate that 7.87\% of the issues are related to the build process of microservices systems, mainly due to GPE and SD\&IA, and most of the build issues are addressed by upgrading tools and platforms, managing infrastructure, and fixing artifacts. The types of build issues, their causes, and solutions indicate that most build problems of microservices systems occur during the parsing, resource processing, and assembly activities. These results can help practitioners to avoid various types of build issues. For example, practitioners should not add unnecessary dependencies in Docker build files and introduce outdated Kubernetes versions while establishing build and deployment pipelines for microservices systems. The most frequently reported build issues are mainly related to compilation and linking phases of build process. We identified one prior study that proposes a taxonomy of build-specific issues, and its results are partially aligned with our findings \cite{A45-lou2020understanding}. However, It would be interesting to further explore the build process of microservices systems in the perspective of \faLeanpub \hspace{0.5mm} (i) code analysis and artifact generation for the build issues, and (ii) the effort required to fix the build issues in microservices systems.

% \FrameSep=3pt\FrameRule=0.5pt\begin{framed}
% \noindent \textit{The majority of build errors are related to build scripts, plugin compatibility, and docker build failures. It is crucial for practitioners to focus on these aspects and ensure that build scripts are error-free, plugins are compatible, and Docker build files do not contain unnecessary dependencies.}
% \end{framed}

% \begin{tcolorbox} [sharp corners, boxrule=0.1mm]
% \small
% \textcolor{black}{\textit{The majority of build errors are related to build scripts, and plugin compatibility, and docker build failures. It is crucial for practitioners to focus on these aspects and ensure that build scripts are error-free, plugins are compatible, and Docker build files do not contain unnecessary dependencies.}}
% \end{tcolorbox}

\textbf{Service Execution and Communicating Issues}: Generally, microservices systems communicate through synchronous (e.g., HTTP/HTTPS) and asynchronous (e.g., AMQP) protocols to complete the business process. The taxonomy of our study shows that 8.03\% of the issues are related to the execution and communication of microservices mainly because of ICC (e.g., \textsc{incorrect configuration setting}) and GPE (e.g., \textsc{syntax error in code}). Most of the service execution and communication issues are addressed by fixing, adding, and modifying artifacts. Microservices systems may have hundreds of services and their instances that frequently communicate with each other. Service execution and communication in microservices systems can also exacerbate the issues of resiliency, load balancing, distributed tracing, high coupling, and complexity \cite{newman2020building}. Several studies (e.g., \cite{yu2019survey}) also confirm that poor communication between microservices and their instances poses significant challenges for deployment, security, performance, fault tolerance, and monitoring of microservices systems. The identified issues, causes, and solutions can help practitioners to (i) to identify the problem areas of service execution and communication and (ii) adopt the strategies to prevent microservices execution and communication issues. Moreover, we argue that future studies can \faLeanpub \hspace{0.5mm} (i) propose architecture techniques for microservices systems with a particular focus on highly resilient and low coupled microservices systems in order to address service discovery issues, and \faMale \hspace{0.5mm} (ii) propose solutions to trace and isolate service communication issues to increase fault tolerance.

% \FrameSep=3pt\FrameRule=0.5pt\begin{framed}
% \noindent \textit{Practitioners need to be cautious of how microservices communicate with each other. Poor implementation of communication between microservices can lead to insecure communication, latency, lack of scalability, and difficulties in fault identification at runtime. By focusing on effective communication protocols and configurations, practitioners can ensure smoother service interactions and mitigate the challenges related to security, performance, and fault tolerance.}
% \end{framed}

% \begin{tcolorbox} [sharp corners, boxrule=0.1mm,]
% \small
%     \textcolor{black}{\textit{Practitioners need to be cautious of how microservices communicate with each other. Poor implementation of communication between microservices can lead to insecure communication, latency, lack of scalability, and difficulties in error and fault identification at runtime. By focusing on effective communication protocols and configurations, practitioners can ensure smoother service interactions and mitigate the challenges related to security, performance, and fault tolerance.}}
% \end{tcolorbox}

\textbf{Configuration Issues}: Microservices systems can have a large number of services and their instances to configure and manage with third-party systems, deployment platforms, and log templates \cite{avritzer2020scalability}. It is essential that microservices systems should have the ability to track and manage the code and configuration changes. Our study results indicate that 4.65\% of the issues are related to the configuration of microservices systems, mainly due to ICC (e.g., \textsc{incorrect configuration setting}) and GPE (e.g., \textsc{syntax error in code}), and most of the configuration issues are addressed by modifying and fixing artifacts. There has been considerable research conducted on configuring traditional software systems. However, we found only a few studies (e.g., \cite{avritzer2020scalability, schaffer2018configuration, kehrer2018autogenic}) that investigated configuration for microservices systems. By considering the configuration issues identified from our study and existing literature, it would be interesting to further investigate and propose techniques and algorithms for \faLeanpub \hspace{0.5mm} (i) dynamically optimizing configuration settings, \faMale \hspace{0.5mm} (ii) identifying critical paths for improving performance and removing performance bottlenecks due to configuration issues, and \faLeanpub \hspace{0.5mm} (iii) detecting security breach points during the configuration of microservices systems.

% \FrameSep=3pt\FrameRule=0.5pt\begin{framed}
% \noindent \textit{In microservices systems, proper configuration is vital due to the extensive service instances and third-party integrations. Practitioners must carefully manage configurations to avoid increased latency and slow microservice calls. This involves diligently handling settings and files, and avoiding errors such as configuration mismatches and incorrect file paths, especially during development, implementation, and deployment phases.}
% \end{framed}

% \begin{tcolorbox} [sharp corners, boxrule=0.1mm,]
% \small
%     \textcolor{black}{\textit{In microservices systems, proper configuration is vital due to the extensive service instances and third-party integrations. Practitioners must carefully manage configurations to avoid increased latency and slow microservice calls. This involves diligently handling settings and files, and avoiding errors such as configuration mismatches and incorrect file paths, especially during development, implementation, and deployment phases.}}
% \end{tcolorbox}

\textbf{Monitoring Issues}: The dynamic nature of microservices systems needs monitoring infrastructure to diagnose and report errors, faults, failures, and performance issues \cite{waseem2021design}. Our study result shows that 3.18\% of the issues are related to the monitoring of microservices systems, mainly due to LC\&D (e.g., \textsc{compatibility error}) and GPE (e.g., \textsc{inconsistent package used}), and most of the monitoring issues are addressed by fixing artifacts and upgrading tools and platforms. The monitoring of microservices systems is fascinating to researchers from several perspectives, including tracing, real-time monitoring, and monitoring tools \cite{cinque2019microservices, phipathananunth2018synthetic, shiraishi2020real}. Future research can design and develop intelligent systems for \faMale \hspace{0.5mm} (i) monitoring hosts, processes, network, and real-time performance of microservices systems, and (ii) identifying the root causes of container issues.

% \FrameSep=3pt\FrameRule=0.5pt\begin{framed}
% \noindent \textit{The dynamic and distributed nature of microservices systems necessitates robust monitoring infrastructures. Since traditional tools may not suffice due to their focus on specific components or overall health, it is crucial for practitioners to employ a combination of tools and strategies that effectively monitor various aspects such as errors, faults, performance issues, and the health of the system across distributed environments.}
% \end{framed}

% \begin{tcolorbox} [sharp corners, boxrule=0.1mm,]
% \small
%     \textcolor{black}{\textit{The dynamic and distributed nature of microservices systems necessitates robust monitoring infrastructures. Since traditional tools may not suffice due to their focus on specific components or overall health, it is crucial for practitioners to employ a combination of tools and strategies that effectively monitor various aspects such as errors, faults, performance issues, and the health of the system across distributed environments.}}
% \end{tcolorbox}

\textbf{Testing Issues}: Testing poses additional challenges in microservices systems development, such as the polyglot code base in multiple repositories, feature branches, and databases per service \cite{waseem2021design, waseem2020testing}. Our study results indicate that 2.86\% of the issues are related to testing of microservices systems, mainly due to GPE (e.g., \textsc{incorrect test case}) and MFA (e.g., \textsc{missing essential system feature}), and most of the testing issues are addressed by adding missing features and fixing syntax and semantic errors in test cases (see Figure \ref{fig:mapping}). We identified multifaceted issues, causes, and solutions regarding microservices testing in this study that highlight several problematic areas for microservices systems, such as \textsc{faulty test cases}, \textsc{debugging}, and \textsc{load test cases} (see Figure \ref{fig:Taxonomy}). We also found several primary (e.g., \cite{camilli2022automated, heorhiadi2016gremlin}) and secondary studies (e.g., \cite{waseem2020testing}) that explore testing of microservices systems. By considering the testing issues identified from our study and existing literature, \faLeanpub \hspace{0.5mm}  it is worthwhile to propose and develop the strategies to automatically test APIs, load, and application security for microservices systems.

% \FrameSep=3pt\FrameRule=0.5pt\begin{framed}
% \noindent \textit{Testing in microservices systems is complex and multifaceted, and practitioners must be cognizant of specific challenges, such as coordinating tests among multiple services, ensuring adequate test coverage, and dealing with polyglot code bases and network communications. Special attention should be paid to issues like faulty test cases, debugging, and load test cases.}
% \end{framed}

% \begin{tcolorbox} [sharp corners, boxrule=0.1mm,]
% \small
% \textcolor{black}{\textit{Testing in microservices systems is complex and multifaceted, and practitioners must be cognizant of specific challenges, such as coordinating tests among multiple services, ensuring adequate test coverage, and dealing with polyglot code bases and network communications. Special attention should be paid to issues like faulty test cases, debugging, and load test cases.}}
% \end{tcolorbox}

\textbf{Storage Issues}: One major problem that practitioners encounter is related to memory management, as per the build, execution, and deployment requirements of microservices \cite{fazio2016open, soldani2018pains}. It is argued in \cite{soldani2018pains} that ``\textit{storage-related pains started decreasing in 2017}'', and our study results indicate that storage issues still exist. The results show that 2.02\% of the issues are related to storage, mainly due to limited memory for process execution, and most of the storage issues are addressed by upgrading the memory size for process execution (see Figure \ref{fig:mapping}). However, scale-up memory could pose an additional burden on managing energy, cost, performance, and required algorithms. Storage issues could also stimulate other issues like performance, reliability, compliance, backup, data recovery, and archiving. \faLeanpub \hspace{0.5mm} Future studies can propose techniques for dynamically assigning storage platforms (e.g., containers, virtual machines) according to the requirements of microservices, which can bring efficiency to the utilization of storage platforms.

% \FrameSep=3pt\FrameRule=0.5pt\begin{framed}
% \noindent \textit{Practitioners should explore dynamic storage platform assignment techniques to optimize storage utilization. By assigning storage platforms (e.g., containers or virtual machines) based on microservice requirements, practitioners can improve resource allocation, enhance system performance, and efficiently manage energy, cost, and compliance considerations.}
% \end{framed}

% \begin{tcolorbox} [sharp corners, boxrule=0.1mm,]
% \small
%     \textcolor{black}{\textit{Practitioners should explore dynamic storage platform assignment techniques to optimize storage utilization. By assigning storage platforms (e.g., containers or virtual machines) based on microservice requirements, practitioners can improve resource allocation, enhance system performance, and efficiently manage energy, cost, and compliance considerations.}}
% \end{tcolorbox}

\textbf{Database Issues}: Another key challenge for microservices systems is managing their databases. Our study identified 2.21\% database issues (e.g., \textit{Database Query}, \textit{Database Connectivity}), mostly due to \textit{GPE} (e.g., \textsc{wrong query parameters}), and most of the database issues are addressed by fixing artifacts and managing infrastructure. Several studies (e.g., \cite{viennot2015synapse, laigner2021data, laigner2021distributed}) explored the use of databases from the perspectives of heterogeneous and distributed database management for event-based microservices systems. However, we did not find any study that explored the issues related to databases in microservices systems. Based on the study results, future studies can propose database patterns and strategies for \faLeanpub \hspace{0.5mm} (i) organizing polyglot databases for efficient read-and-write operations by considering performance, and (ii) storing and accessing decentralized and shared data without losing the independence of individual microservices. 

% \FrameSep=3pt\FrameRule=0.5pt\begin{framed}
% \noindent \textit{The distributed nature of microservices systems leads to distributed ownership of databases, with each microservice having its private data store. This distribution poses challenges related to database implementation, data accessibility, and database connectivity. Practitioners need to be aware of these challenges and consider strategies to effectively manage databases in microservices architectures.}
% \end{framed}

% \begin{tcolorbox} [sharp corners, boxrule=0.1mm,]
% \small
%     \textcolor{black}{\textit{The distributed nature of microservices systems leads to distributed ownership of databases, with each microservice having its private data store. This distribution poses challenges related to database implementation, data accessibility, and database connectivity. Practitioners need to be aware of these challenges and consider strategies to effectively manage databases in microservices architectures.}}
% \end{tcolorbox}

\textbf{Networking Issues}: The network infrastructure for microservices systems consists of many components (both hardware and software), including but not limited to hosting servers, network protocols, load balancer, firewall, hardware devices, series of containers, public and private clouds, and a set of common APIs for accessing different components. Our study identified 1.65\% issues related to networking (e.g., \textsc{hosting and protocols}, \textsc{service accessibility}) during the development of microservices systems, mostly due to GPE (e.g., \textsc{content delivery networks (CDN) deployment error}), and most of the network issues are addressed by managing infrastructure and modifying artifacts. We found a few studies (e.g., \cite{luo2018high, kratzke2017microservices, bhattacharya2019smart}) that mainly focus on networking for containerized microservices and smart proxying for microservices. However, these studies do not report networking issues, causes, and solutions for microservices systems. Future research can \hspace{0.5mm}(i) provide deeper insights into networking issues in the context of microservices systems and \faMale \hspace{0.5mm} (ii) propose automatic correction methods to fix networking issues, like \textsc{Localhost}, \textsc{IP address}, and \textsc{Webhook} errors.

% \FrameSep=3pt\FrameRule=0.5pt\begin{framed}
% \noindent \textit{Addressing networking issues in microservices systems involves managing network infrastructure and modifying artifacts. Prioritizing infrastructure management, including protocols, ports, topologies, and service accessibility, enables practitioners to mitigate networking issues and ensure smooth operations.}
% \end{framed}

% \begin{tcolorbox} [sharp corners, boxrule=0.1mm,]
% \small
%     \textcolor{black}{\textit{Addressing networking issues in microservices systems involves managing network infrastructure and modifying artifacts. Prioritizing infrastructure management, including protocols, ports, topologies, and service accessibility, enables practitioners to mitigate networking issues and ensure smooth operations.}}
% \end{tcolorbox}

\textbf{Performance Issues}: The performance of microservices systems is one of the highly discussed topics, and the existing studies (e.g., \cite{amaral2015performance, heinrich2017performance}) mainly focus on performance evaluation, monitoring, and workload characterization of microservices systems. We also found one study \cite{WuNoms} that presents a ``\textit{system to locate root causes of performance issues in microservices}''. We identified 1.65\% performance issues (e.g., \textit{Service Response Delay}, \textit{Resource Utilization}) discussed by practitioners in our study. These issues mainly occur due to MFA (e.g., \textsc{missing resource}) and SD\&IA (e.g., \textsc{wrong dependencies chain}), and are addressed by adding new features and fixing design anomalies. The identified performance issues, causes, and solutions can help \faMale \hspace{0.5mm} (i) further explore performance differences for containerization platforms and combinations of configurations, (ii) propose strategies and techniques for reducing the service response delay when multiple microservices are accessing shared resources, (iii) propose techniques for optimizing resource utilization (i.e., CPU and GPU usage), and (iv) develop a framework for improving scalability to increase microservices performance.

% \FrameSep=3pt\FrameRule=0.5pt\begin{framed}
% \noindent \textit{The decentralized nature of microservices, with numerous independent services deployed on different infrastructures and platforms, can lead to performance overhead and increased resource consumption. Practitioners need to be aware of these challenges and take them into account when designing, deploying, and managing microservices systems. They should consider strategies and techniques to optimize resource utilization and minimize performance overhead to ensure efficient and scalable microservices performance.}
% \end{framed}

% \begin{tcolorbox} [sharp corners, boxrule=0.1mm,]
% \small
%     \textcolor{black}{\textit{The decentralized nature of microservices, with numerous independent services deployed on different infrastructures and platforms, can lead to performance overhead and increased resource consumption. Practitioners need to be aware of these challenges and take them into account when designing, deploying, and managing microservices systems. They should consider strategies and techniques to optimize resource utilization and minimize performance overhead to ensure efficient and scalable microservices performance.}}
% \end{tcolorbox}

\textbf{Typecasting Issues}: Practitioners frequently use multiple programming languages and technologies simultaneously to develop microservices systems. Each of the used programming languages has its own syntax, structure, and semantics. It is most often the case that developers need to convert variables from one datatype to another datatype (e.g., double to string) for one or multiple microservices, a process referred to as typecasting. Our study identified 1.31\% issues related to networking during the development of microservices systems, mostly due to GPE (e.g., \textsc{wrong data conversion}), and most of the typecasting issues are addressed by managing infrastructure and modifying artifacts. To the best of our knowledge, no study has been conducted that investigates typecasting in the context of microservices systems. Based on the study results, \faLeanpub \hspace{0.5mm} future studies can propose and implement the techniques for converting code of one language to another used in microservices systems.

% \FrameSep=3pt\FrameRule=0.5pt\begin{framed}
% \noindent \textit{Practitioners in microservices systems should be aware of typecasting issues when using multiple programming languages and technologies. Understanding language differences enables effective anticipation and mitigation of typecasting problems.}
% \end{framed}

% Preamble (add once if needed):
% \usepackage{booktabs}
% \usepackage{tabularx}
% \usepackage{ragged2e}
% \newcolumntype{Y}{>{\RaggedRight\arraybackslash}X}

% Preamble (add once if needed):
% \usepackage{booktabs}
% \usepackage{tabularx}
% \usepackage{ragged2e}
% \usepackage[table]{xcolor}
% \newcolumntype{Y}{>{\RaggedRight\arraybackslash}X}

\begin{table*}[H]
\centering
{\color{black}
\scriptsize
\renewcommand{\arraystretch}{1.25}
\setlength{\tabcolsep}{6pt}
\caption{Practitioner implications derived from the issue--cause--solution mapping.}

\label{tab:implications_messages}
\begin{tabularx}{\textwidth}{p{3.2cm} Y}
\toprule
\textbf{Category} & \textbf{Practitioner Takeaway} \\
\midrule

\textbf{Technical Debt} &
Practitioners should prioritize managing technical debt, including code debt and service design debt, in microservices systems to mitigate potential risks and ensure the long-term health of the systems. This involves addressing source code issues like refactoring and duplication, and adhering to successful practices in microservices system design. \\

\textbf{CI/CD Issues} &
Practitioners should pay attention to correct configuration settings, updating and managing dependencies, and keeping abreast of best practices in tools and platforms being used (like Docker, Kubernetes, AWS, Google Cloud). They should also consider investing in upgrading tools and platforms, and fixing artifacts as a common resolution to many CI/CD issues. This will ensure a smoother and more reliable CI/CD process. \\

\textbf{Security Issues} &
To ensure robust security in microservices systems, practitioners should implement multi-layered security solutions. This involves strategies for data centers, virtualization, communication, and orchestration. Measures such as secure certificates, encryption, and access control policies are essential for achieving strong security management in microservices. \\

\textbf{Build Issues} &
The majority of build errors are related to build scripts, plugin compatibility, and docker build failures. It is crucial for practitioners to focus on these aspects and ensure that build scripts are error-free, plugins are compatible, and Docker build files do not contain unnecessary dependencies. \\

\textbf{Service Execution and Communication Issues} &
Practitioners need to be cautious of how microservices communicate with each other. Poor implementation of communication between microservices can lead to insecure communication, latency, lack of scalability, and difficulties in fault identification at runtime. By focusing on effective communication protocols and configurations, practitioners can ensure smoother service interactions and mitigate the challenges related to security, performance, and fault tolerance. \\

\textbf{Configuration Issues} &
In microservices systems, proper configuration is vital due to the extensive service instances and third-party integrations. Practitioners must carefully manage configurations to avoid increased latency and slow microservice calls. This involves diligently handling settings and files, and avoiding errors such as configuration mismatches and incorrect file paths, especially during development, implementation, and deployment phases. \\

\textbf{Monitoring Issues} &
The dynamic and distributed nature of microservices systems necessitates robust monitoring infrastructures. Since traditional tools may not suffice due to their focus on specific components or overall health, it is crucial for practitioners to employ a combination of tools and strategies that effectively monitor various aspects such as errors, faults, performance issues, and the health of the system across distributed environments. \\

\textbf{Testing Issues} &
Testing in microservices systems is complex and multifaceted, and practitioners must be cognizant of specific challenges, such as coordinating tests among multiple services, ensuring adequate test coverage, and dealing with polyglot code bases and network communications. Special attention should be paid to issues like faulty test cases, debugging, and load test cases. \\

\textbf{Storage Issues} &
Practitioners should explore dynamic storage platform assignment techniques to optimize storage utilization. By assigning storage platforms (e.g., containers or virtual machines) based on microservice requirements, practitioners can improve resource allocation, enhance system performance, and efficiently manage energy, cost, and compliance considerations. \\

\textbf{Database Issues} &
The distributed nature of microservices systems leads to distributed ownership of databases, with each microservice having its private data store. This distribution poses challenges related to database implementation, data accessibility, and database connectivity. Practitioners need to be aware of these challenges and consider strategies to effectively manage databases in microservices architectures. \\

\textbf{Networking Issues} &
Addressing networking issues in microservices systems involves managing network infrastructure and modifying artifacts. Prioritizing infrastructure management, including protocols, ports, topologies, and service accessibility, enables practitioners to mitigate networking issues and ensure smooth operations. \\

\textbf{Performance Issues} &
The decentralized nature of microservices, with numerous independent services deployed on different infrastructures and platforms, can lead to performance overhead and increased resource consumption. Practitioners need to be aware of these challenges and take them into account when designing, deploying, and managing microservices systems. They should consider strategies and techniques to optimize resource utilization and minimize performance overhead to ensure efficient and scalable microservices performance. \\

\textbf{Typecasting Issues} &
Practitioners in microservices systems should be aware of typecasting issues when using multiple programming languages and technologies. Understanding language differences enables effective anticipation and mitigation of typecasting problems. \\

\bottomrule
\end{tabularx}
} % end blue color
\end{table*}

\subsection{\textcolor{black}{Evidence-driven future research agenda}}
\label{researchagenda}
\textcolor{black}{To address the empirically observed challenges in microservices systems, we synthesize our findings into a structured future research agenda. Table~\ref{tab:future_framework} consolidates this agenda into thematically grouped research problems, study strategies, and expected outcomes. The discussion below follows the same thematic grouping to help researchers quickly locate where each challenge belongs, how it should be studied, and what types of research artifacts would provide practical value, while staying consistent with our Issue Categories, Cause Categories, and Solution Categories.}

\subsubsection{\textcolor{black}{Research problems}}
\textcolor{black}{\textbf{Predictive governance and system evolution.} The first theme highlights that many recurring challenges arise because microservices systems evolution is frequently addressed through repeated Fixing/Adding/Modifying Artifacts rather than through service-aware governance that prevents recurrence. This theme is grounded in dominant Issue Categories such as \emph{Technical Debt} and \emph{Continuous Integration and Delivery}, and in recurring Cause Categories such as \emph{General Programming Errors}, \emph{Missing Features and Artifacts}, and \emph{Invalid Configuration and Communication}. A common gap is the lack of mechanisms to quantify risk, forecast degradation, and prioritize interventions as microservices, dependencies, and environments change over time. Consequently, corrective actions are often applied locally, while the payoff of interventions (e.g., reduced recurrence or improved stability) remains difficult to measure and compare across projects.}

\textcolor{black}{\textbf{Runtime behavior.} The second theme concerns issues whose causes and consequences span multiple microservices rather than remaining confined within a single microservice. The recurring challenges in this theme are reflected in Issue Categories such as \emph{Exception Handling} and \emph{Service Execution and Communication}, where failures can propagate across service boundaries and complicate root-cause identification. Local debugging is therefore insufficient because it rarely captures end-to-end execution context across microservices and heterogeneous runtime environments. Progress in this area requires approaches that explicitly model and monitor cross-service execution and communication, detect and contain cascading failures, and support diagnosis and prevention at the level of microservices systems rather than individual services.}

\textcolor{black}{\textbf{Co-evolution and non-functional correctness.} The third theme reflects that microservices systems correctness is repeatedly affected by misalignment across co-evolving artifacts and concerns. Configuration drift, dependency drift, and deployment and runtime changes can surface as recurring \emph{Invalid Configuration and Communication} problems and can also trigger CI/CD and build failures. The research problem is not only detecting such drift, but also attributing failures and regressions to the combination of code, configuration, dependencies, and runtime behavior so that interventions can be targeted and evaluated.}

\clearpage
\onecolumn
{\color{black}
\renewcommand{\arraystretch}{1.22}
\setlength{\tabcolsep}{6pt}
\scriptsize
\begin{longtable}{L{2.5cm} L{5.0cm} L{5.0cm} L{4.5cm}}
\caption{Future research framework for OSS microservices systems: problems, study strategies, and expected outcomes}
\label{tab:future_framework}\\
\toprule
\textbf{Dimension} &
\textbf{Problem} &
\textbf{How to study it} &
\textbf{What the research should produce}\\
\midrule
\endfirsthead

\toprule
\textbf{Dimension} &
\textbf{Problem} &
\textbf{How to study it} &
\textbf{What the research should produce}\\
\midrule
\endhead

\midrule
\multicolumn{4}{r}{\emph{Continued on next page}}\\
\endfoot

\bottomrule
\endlastfoot

% =========================
\rowcolor{lightgrayrow}
\multicolumn{4}{c}{\textbf{Predictive governance and system evolution}}\\
\midrule

\textbf{Technical debt governance} &
There is a lack of service-aware mechanisms to quantify, forecast, and prioritize Technical Debt as systems evolve; practitioners therefore rely on repeated Fixing/Adding/Modifying Artifacts whose long-term payoff is unclear. &
Longitudinal mining of issue--commit--PR traces; service dependency and ownership graph reconstruction; quasi-experiments around refactorings and architectural changes; learning-to-rank models for debt repayment choices. &
A validated Technical Debt indicator suite for microservices; forecasting models for Technical Debt growth; decision support that ranks repayment actions with measurable payoff (lower recurrence, lower maintenance effort). \\

\textbf{CI/CD and build reliability} &
Continuous Integration and Delivery failures recur because pipelines, dependencies, and environments co-evolve across services, but models that explain failure mechanisms and provide transferable repair guidance across projects are missing. &
Joint analysis of pipeline logs, build manifests, container specifications, and dependency graphs; controlled pipeline mutation and replay; counterfactual evaluation of pre-flight checks; explainable failure classification tied to fixes. &
Predictors for CI/CD failure risk; explainable triage recommending evidence-backed repairs; benchmark datasets linking failure$\rightarrow$cause$\rightarrow$fix; reduced rollback rate and downtime. \\

\textbf{Dependency and upgrade risk management} &
Upgrades fail because dependency ecosystems and polyglot toolchains evolve unevenly across services, while upgrade risk is rarely predicted at the service-graph level. &
Dependency evolution mining; compatibility break prediction; service-graph impact analysis; automated upgrade planning validated via staged rollout and rollback traces. &
Upgrade-risk predictors; planning tools for safer upgrades; fewer upgrade failures and shorter MTTR during upgrade incidents. \\

\textbf{Missing features and artifacts management} &
Issue resolution is repeatedly affected by incomplete or missing features and supporting artifacts (e.g., missing configuration, build, documentation, or test assets), yet systematic detection and prevention of these gaps across services is limited. &
Mining traces of feature requests and missing-asset reports; linking issues to commits, build/test failures, and documentation changes; controlled comparisons of projects that adopt templates/guides versus ad-hoc practices; evaluation using recurrence and lead time. &
Detectors for missing/insufficient artifacts; templates and checklists for feature completeness across services; recommendation support for what artifacts to add and where. \\

% =========================
\rowcolor{lightgrayrow}
\multicolumn{4}{c}{\textbf{Runtime behavior}}\\
\midrule

\textbf{Exception propagation and fault containment} &
Exception Handling problems propagate across service boundaries and obscure root causes, yet empirically validated propagation models and containment tactics for real-world conditions are lacking. &
Exception-flow mining from code and telemetry; call-graph and trace-based propagation modeling; chaos and fault-injection experiments; evaluation of containment tactics under controlled scenarios. &
Propagation models identifying cascade triggers; validated fault-containment playbooks; tooling that reduces blast radius and improves MTTR. \\

\textbf{Service interaction and contract stability} &
Service contracts drift over time (schemas, APIs, versions), creating backward-compatibility failures that are difficult to predict and detect before release. &
Mining and semantic differencing of OpenAPI/gRPC schemas; consumer-driven contract testing at scale; change-impact analysis over service interaction graphs; validation using historical breakages. &
Automated contract drift detection; compatibility checks integrated into release pipelines; empirically validated versioning strategies. \\

\textbf{Networking robustness and incident forensics} &
Network-induced faults manifest as application errors, while diagnostic practices lack repeatable models to distinguish network causes from service-level causes. &
Network fault injection; verification of service-mesh policies against invariants; correlation of infrastructure logs with service telemetry; topology-aware diagnosis models. &
Validated network resilience patterns; automated incident forensics pipelines; faster root-cause identification and fewer network-induced outages. \\

\textbf{Polyglot interoperability and typecasting errors} &
Cross-language boundary failures persist because interfaces are weakly specified and tests fail to capture semantic mismatches across services. &
Mining mismatch bugs and fixes; schema-first and contract-verified interfaces; boundary test generation; static and dynamic serialization guards. &
Interoperability patterns and tools; benchmark datasets of boundary bug classes; reduced runtime type and serialization failures. \\

% =========================
\rowcolor{lightgrayrow}
\multicolumn{4}{c}{\textbf{Co-evolution and non-functional correctness}}\\
\midrule

\textbf{Configuration co-evolution} &
Configuration evolves alongside code across services and environments, while risky configuration changes are rarely detected before deployment and misconfiguration patterns recur (often surfacing as Invalid Configuration and Communication). &
Config-as-code static analysis with cross-service invariants; drift detection from historical diffs; staged validation (canary and differential testing); incident-based evaluation. &
Preventive configuration verification tools; configuration quality metrics; safer rollout procedures with fewer config-induced incidents. \\

\textbf{Data consistency and migrations} &
Schema evolution and data migrations remain a recurring source of incidents, with limited empirical guidance on safe strategies and consistency trade-offs at scale. &
Mining schema and migration histories; change-impact analysis over data dependencies; simulation-based evaluation; derivation and validation of migration playbooks. &
Migration safety checks; validated consistency and migration patterns; reduced data-related incidents and faster recovery. \\

\textbf{Performance regression localization} &
Performance regressions arise from cross-service interactions and runtime effects, yet attribution remains local and incomplete, delaying remediation. &
Workload-controlled benchmarking; trace-based critical-path analysis; change-coupling and performance attribution across services; longitudinal evaluation. &
Regression risk predictors; attribution reports localizing true bottlenecks; benchmark suites improving latency and resource efficiency. \\

% =========================
\rowcolor{lightgrayrow}
\multicolumn{4}{c}{\textbf{Observability, validation, and socio-technical alignment}}\\
\midrule

\textbf{Observability quality and actionable diagnosis} &
Observability is often reduced to data collection, while objective measures of observability quality and reliable diagnosis pipelines are missing. &
Definition of observability quality metrics; topology-aware anomaly detection; explainable root-cause localization; controlled fault scenarios. &
Observability scorecards; diagnosis pipelines with explainability; standardized evaluation protocols and improved MTTR. \\

\textbf{Testing of distributed microservices} &
Existing testing practices under-approximate distributed behavior, leading to flakiness, integration gaps, and limited behavioral coverage. &
Trace-guided test generation; contract and property-based testing; mutation testing at API boundaries; systematic flakiness diagnosis. &
Behavioral coverage metrics; reduced flakiness and regression leakage; reusable testing strategies for multi-service releases. \\

\textbf{Documentation traceability and knowledge drift} &
Documentation drifts from APIs, configurations, and deployments, creating hidden maintenance costs and onboarding friction without systematic detection. &
Traceability graphs linking documentation, APIs, configurations, and tests; drift detection from change histories; quality metrics tied to outcomes. &
Documentation drift detectors; traceability tooling; improved onboarding and reduced incidents caused by outdated documentation. \\

\textbf{Organizational and socio-technical coordination} &
Recurring issues reflect misalignment between system structure and organizational practices, yet validated indicators and interventions remain scarce. &
Socio-technical network mining; congruence analysis; evaluation of governance interventions using longitudinal outcomes. &
Socio-technical risk indicators; evidence-based governance practices; reduced defect recurrence with quantified coordination trade-offs. \\

\end{longtable}
}

\twocolumn
\textcolor{black}{\textbf{Observability, validation, and socio-technical alignment.} The fourth theme highlights that diagnosis, assurance, and long-term maintainability depend on both technical and organizational conditions. Issue Categories such as \emph{Monitoring} and \emph{Testing}, along with documentation and coordination challenges, suggest that practitioners may collect operational data without ensuring diagnostic usefulness, test distributed behavior without robust coverage notions, and maintain documentation and practices that diverge from implementation reality. These gaps increase onboarding friction, reduce release confidence, and amplify recurrence by weakening the ability to detect, explain, and prevent issues over time.}

\subsubsection{\textcolor{black}{Methodological directions}}
\textcolor{black}{The thematic structure in Table~\ref{tab:future_framework} implies methodological directions that strengthen empirical rigor and practical relevance. First, longitudinal and evolutionary analysis should be treated as a baseline, particularly for challenges where recurrence and drift are central. Mining OSS issue discussions and linking them to commits, pull requests, configuration changes, dependency updates, and release histories over time is necessary to characterize recurring patterns, identify leading indicators, and evaluate whether interventions (e.g., Technical Debt repayment, CI/CD hardening, configuration validation) yield measurable payoff in stability and maintainability.}

\textcolor{black}{Second, future work should adopt cross-artifact and cross-layer study designs, because many recurring problems span services and connect software artifacts with infrastructure artifacts. Effective designs combine multiple evidence sources including logs and traces, CI/CD pipeline records, build manifests, dependency graphs, configuration histories, and service communication traces, to capture relationships that single-artifact studies miss. This direction is essential for explaining issue recurrence across CI/CD, build, configuration, and Service Execution and Communication, where the causes are distributed rather than local.}

\textcolor{black}{Third, controlled experimentation in realistic settings is needed to move from correlation toward explanation and prevention. Techniques such as fault injection experiments, pipeline mutation and replay, and workload-controlled performance analysis allow researchers to test hypotheses about failure propagation, attribution, and resilience under known conditions. When integrated with longitudinal mining and modeling, these experiments enable reproducible evaluation of mitigation strategies and their impact on recurrence. Overall, the agenda motivates hybrid empirical methods that combine longitudinal mining, cross-artifact analysis, controlled experimentation, and validation to deliver results that are both scientifically defensible and practically actionable.}

\subsubsection{\textcolor{black}{Research outcomes and artifacts}}
\textcolor{black}{The proposed agenda emphasizes research outcomes that move beyond descriptive analysis and directly support engineering practice. A first outcome class consists of \emph{measurable indicators and metrics} that make system-level risks explicit and comparable. Examples include indicators for Technical Debt in microservices systems, CI/CD and build reliability indicators, configuration-risk measures, and monitoring/observability quality measures. Such indicators enable principled comparison of alternatives, support risk-aware planning, and reduce reliance on intuition-driven decisions. These metrics should be designed for longitudinal use so that researchers and practitioners can quantify recurrence reduction, maintenance effort changes, and the payoff of interventions over time.}

\textcolor{black}{A second outcome class comprises \emph{predictive and decision-support tools} that operationalize empirical insights from Issue Categories, Cause Categories, and Solution Categories. Representative examples include predicting CI/CD failures and recommending Fixing/Adding/Modifying Artifacts actions, estimating upgrade risk across dependencies, supporting diagnosis for Service Execution and Communication issues, and assisting practitioners in prioritizing Technical Debt repayment. To ensure practical relevance, such tools should be evaluated against realistic baselines using outcome-oriented measures such as recurrence, rollback rate, and mean time to recovery.}

\textcolor{black}{A third outcome class focuses on \emph{reusable empirical resources and validated guidance}. Datasets that link Issue Categories to Cause Categories and Solution Categories, controlled scenarios for evaluating failures and fixes, and empirically validated strategies for preventing recurrence enable cumulative research and fair comparison across studies. By producing artifacts that are reusable, evaluable, and grounded in real microservices systems, future work can accelerate progress while strengthening empirical foundations and external validity.}

\subsubsection{\textcolor{black}{Agenda for future research}}
\textcolor{black}{Taken together, Table~\ref{tab:future_framework} encourages a shift from isolated, problem-specific studies toward integrated research programs that address prediction, explanation, and prevention of failures in microservices systems. By explicitly aligning research problems with suitable empirical methods and outcome-oriented artifacts, the agenda supports cumulative knowledge building and increases the likelihood that research results translate into sustained improvements in reliability, evolvability, and trustworthiness.}

\textcolor{black}{For researchers, the agenda provides a structured lens for positioning contributions by clarifying which problem is addressed, why it matters in microservices settings, how it should be studied, and what concrete artifacts should result. For the broader community, it offers a roadmap toward research that not only explains observed phenomena but also delivers actionable metrics, tools, datasets, and reference practices that can be adopted, evaluated, and evolved across diverse microservices ecosystems.}

\section{Threats to Validity}
\label{sec:threats}
This section reports the potential threats to the validity of this research and its results, along with mitigation strategies that could help minimize the impacts of the outlined threats based on \cite{easterbrook2008selecting}. The threats are broadly classified across internal, construct, external, and conclusion validity. 

\subsection{Internal Validity}
Internal validity examines the extent to which the study design, conduct, and analysis answer to the research questions without bias \cite{easterbrook2008selecting}. We discuss the following threats to internal validity.

\textit{Improper project selection}. The first internal validity threat to our study is an improper selection of OSS projects for executing our research plan. \textit{Mitigation}. We used a multi-step project selection approach to control the possible threat associated with subject system selection (see Phase 1 in Section \ref{sec:RMPhase1} and Figure \ref{fig:researchmethod}). A step-wise and criteria-driven approach for the project selection has been used that helped us to include the relevant (see Table \ref{tab:selectedProjects}) and eliminate irrelevant open-source microservices projects.

\textit{Instrument understandability}. The participants of interviews and surveys may have a different understanding of the interview and survey instruments. \textit{Mitigation}. We adopted Kitchenham and Pfleeger’s guidelines for conducting surveys \cite{kitchenham2008personal}. We also piloted both interview and survey instruments to ensure understandability (see Section \ref{InterviewsProtocol} and Section \ref{PilotSurvey}). Our questionnaire for the survey were in English. However, during the interviews, we found that a few participants could not conveniently convey their answers in English. Therefore, we requested them to answer in their native languages, and for the latter we translated the answers into English.
 
\textit{Participants selection and background}: The survey participants may not have sufficient expertise to answer questions. \textit{Mitigation}. We searched for microservices practitioners through personal contacts and relevant platforms (e.g., LinkedIn and GitHub). We explicitly made participants’ characteristics (e.g., roles and responsibilities) in the interview and survey preamble, and we selected only those practitioners with sufficient experience designing, developing, and operationalizing microservices systems. For example, the average experience of the interview participants is 5.33 years, and they mainly have responsibilities for designing and developing microservices systems (see Table \ref{tab:IntervieweesDemographics} and Figure \ref{fig:demography}).

\textit{Interpersonal bias in extracting and synthesizing data}. (i) The interpersonal bias in the mining developer discussions and analysis process may threaten the internal validity of the study findings. \textit{Mitigation}. To address this threat, we defined explicit data collection criteria (e.g., exclude the issues of general questions). We had regular meetings with all the authors in data labelling, coding, and mapping. The conclusions were made based on the final consensus made by all the authors. (ii) The survey and interview participants may be slightly biased in providing actual answers due to company policies, work anxiety, or other reasons. \textit{Mitigation}. To mitigate this threat, we highlighted the anonymity of the participants and their companies in the interview and survey instruments preamble. (iii) The interviewer of the study may be biased toward getting the favourite answers. \textit{Mitigation}. We sent our interview questions 3 to 4 days before the interview to the interviewees. Hence, they had sufficient time to understand the context of the study and could provide the required feedback on the completeness and correctness of the developed taxonomies. 

\subsection{Construct Validity} 
Construct validity focuses on whether the study constructs (e.g., interview protocol, survey questionnaire) are correctly defended~\cite{easterbrook2008selecting}. Microservices issues, causes, and solutions are the core constructs of this study. Having said this, we identified the following threats.

\textit{Inadequate explanation of the constructs}: This threat refers to the fact that the study constructs are not sufficiently described. \textit{Mitigation}. To deal with this threat, we prepared the protocols for mining developer discussions and conducting interviews and surveys, and these protocols were continuously improved during the internal meetings, feedback, and taxonomy refinements. Mainly, the authors had meetings (i) to establish a common understanding of issues, causes, and solutions (see Figure \ref{fig:IssuesBackground}), (ii) for extracting the required data to answer the RQs, and (iii) for evaluating interview and survey question format, understandability, and consistency. We also invited two survey-based research experts to check the validity and integrity of the survey questions. Based on their feedback, we included Figure \ref{fig:Taxonomy}, Table \ref{tab:CausesTaxnomey}, and Table \ref{tab:SolutionsTaxnomey} as a part of the interview and survey questions for improving the participants' understandability of the issues, their causes and solutions in microservices systems.

\textit{Data extraction and survey dissemination platforms}: This threat refers to the authenticity and reliability of the platforms we used for data collection and survey dissemination. \textit{Mitigation}. (i) To mitigate this threat for data collection, we identified 2,641 issues from the issue tracking systems of 15 open-source microservices systems on GitHub which were confirmed by the developers. (ii) We disseminated the survey at social media and professional networking groups. The main threat to survey dissemination platforms is identifying the relevant groups. We addressed this threat by reading the group discussion about microservices system development and operations. After ensuring that the group members frequently discussed various microservices aspects, we posted our survey invitation.
 
\textit{Inclusion of valid issues and responses}: The threat is mainly related to the inclusion and exclusion of issues from the issue tracking systems and responses from survey participants. \textit{Mitigation}. We defined explicit criteria for including and excluding issues from the issue tracking systems (see Section \ref{sec:Ext&Syn}) and responses from survey participants (see Section \ref{WebSurvey}). For example,  when screening issues from the issue tracking systems, we excluded those issue discussions consisting of general questions, ideas, and proposals. Similarly, we excluded those responses that were either randomly filled or filled by research students and professors who were not practitioners.

\subsection{External Validity} 
External validity refers to the extent to which the study findings could be broadly generalized in other contexts \cite{easterbrook2008selecting}. The sample size and sampling techniques might not provide a strong foundation to generalize the study results. It is the case for all three data collection methods (mining developer discussions from open-source microservices projects, interviews, and surveys) used in this study. \textit{Mitigation}. (i) To minimize this threat, we derived the taxonomies of issues, causes, and solutions from a relatively large number of issues from 15 sampled open-source microservices projects belonging to different domains by involving multiple researchers. (ii) The taxonomies have been evaluated and improved by taking the feedback of experienced microservices practitioners through the interviews. (iii) A cross-sectional survey was conducted based on the derived and evaluated taxonomies. Overall, we received 150 valid responses from 42 countries of 6 continents (see Figure~\ref{fig:demography}(a)) having varying experience (from less than one year to more than ten years, see Figure~\ref{fig:demography}(b)) with different roles (see Figure~\ref{fig:demography}(c)), and working with diverse domains (see Figure~\ref{fig:demography}(d)) and programming languages and technologies (see Figure~\ref{fig:demography}(e)) to develop microservices systems. Our study is primarily based on analyzing qualitative data, which presents potential threats, such as subjectivity bias, low reproducibility, data overload, and a lack of transparency that researchers may encounter. \textit{Mitigation}. To mitigate these threats during data analysis, we employed thematic analysis by following the guidelines (see Section \ref{sec:Ext&Syn}) and involved all the researchers to reduce subjectivity bias, enhance replicability, manage data overload, and ensure transparency.

textit{Interview data saturation}: This threat arises from the possibility that the 15 interviews we conducted to validate and expand our proposed taxonomies may not be sufficient to achieve data saturation in our study. \textit{Mitigation}.  To address this threat, we conducted interviews with practitioners who had varied backgrounds (i.e., responsibilities, skills in programming languages, work domains, experience, countries) in microservices systems (see Table \ref{tab:IntervieweesDemographics}). Second, we asked open-ended questions, which allowed the interviewees to spontaneously express their views on the issues, causes, and solutions (see the Interview Questionnaire sheet in \cite{replpack}). Third, we presented the interviewees with pre-existing taxonomies that were derived from analyzing developer discussions in 15 open-source microservices systems. We asked the interviewees to pinpoint any issue, cause, and solution they felt absent. Fourth, while we were analyzing interview data, we started observing recurring themes and patterns in the responses, which was an indication of data saturation. As we did not encounter any new significant issues, causes, or solutions after the 15th interview, we decided to stop conducting further interviews. However, we encourage further research involving more interviews or additional data sources to provide more insights and refinement to our findings.

We acknowledge that the findings of this study may not be generalizable, fully replicable, or representative of the issues, causes, and solutions for all types of microservices systems. However, the size of the investigated issues, the number of microservices systems, interviews with microservices practitioners, the survey population, and systematic data analysis methods (e.g., thematic analysis) can partially strengthen the overall generalizability of the study results.

\subsection{Conclusion Validity}
Conclusion validity is related to dealing with threats that affect the correct conclusions in empirical studies \cite{easterbrook2008selecting}. Concluded findings of the results may be based on a single author's understanding and experience, and conflicts on the conclusions between authors may not be sufficiently discussed or resolved. \textit{Mitigation}. To address this threat, the first author extracted and analyzed study data (i.e., from the developer discussions, interviews, and survey). All other authors comprehensively reviewed the data through multiple meetings. Conflicts on data analysis results were resolved through mutual discussions and brainstorming among all the authors. Different researchers can interpret the inclusion and exclusion criteria differently which influences the conclusions of the study. \textit{Mitigation}. To minimize the effect of this issue, we applied the explicitly defined inclusion and exclusion criteria during the study data screening (see Section \ref{sec:Ext&Syn} - Step B). Finally, the interpreted results and conclusions have been confirmed by arranging several brainstorming sessions among all the authors.
\section{Related Work}
\label{RelatedWork}
We performed a thorough literature search via eight major databases, including Google Scholar, IEEE Xplore, ACM Digital Library, ScienceDirect, Scopus, Web of Science, SpringerLink, and Wiley Online Library, to ensure that no relevant research study has been overlooked during the literature review. The literature search revealed that no study (except our previous work \cite{waseem2021nature}) has empirically investigated the issues faced, causes reported, and solutions adopted by practitioners to develop microservices systems. We discuss the existing research that can be broadly classified into two categories including (i) mining OSS repositories to extract reusable knowledge for MSA (Section \ref{miningoss}) and (ii) empirical studies on issues in microservices systems and software systems (Section \ref{empiricalstudiesonissues}). A conclusive summary and comparative analysis (Section \ref{conclusivesummary}) position the proposed research in the context of mining issues from OSS repositories and justify the scope and contributions of this research.

\subsection{Mining OSS Repositories to Extract Reusable Knowledge for MSA}
\label{miningoss}

\subsubsection{Mining (Anti-)Patterns and Tactics to Architect MSA}
\textcolor{black}{From the MSA perspective, patterns \cite{R1} and tactics \cite{A66-genfer2025understanding} represent empirically grounded knowledge, that promote reuse of best practices, as recurring solutions to address frequently occurring issues during architecture-centric development of service-driven software systems \cite{A10-daniel2023towards}. To derive empirically-grounded reuse knowledge, i.e., discovering patterns and tactics for architecting MSA, a number of studies such as \cite{A10-daniel2023towards, A67-soldani2021mutosca, A68-fritzsch2022towards, A69-bacchiega2022microservices} have focused on analyzing open-source code and project repositories such as mining version controls, searching change logs, and exploring design documents etc. to investigate historical data that can reveal recurring solutions as patterns. The investigation of historical data involves postmortem analysis to detect anti-patterns and bad smells \cite{A70-cerny2023catalog} and mining repositories such as GitHub \cite{A71-beahan2025characterizing} to identify vulnerabilities in MSA systems. Daniel et al. \cite{A10-daniel2023towards} employed a metric-based approach and used two case studies to discover microservice patterns, while also evaluating the accuracy and applicability of the approach based on developers’ feedback.  In a similar work, Soldani et al. \cite{A67-soldani2021mutosca} have developed a TOSCA toolchain to detect and resolve architectural smells and to ensure that complex microservice ecosystems remain aligned with core design principles. Compared to the metric and tool-based approaches for mining MSA specific knowledge, Fritzsch et al. \cite{A68-fritzsch2022towards} employed a mixed-method approach, synthesising the findings of literature review and industry expert interviews to derive a framework for architectural refactoring and migration in the context of microservice systems. }

\textcolor{black}{Compared to the pattern-based solutions discussed above, architectural tactics represent design decisions that focus on improving one specific quality aspect of MSA, such as service availability and fault avoidance for security critical systems \cite{A66-genfer2025understanding} . Amoroso et al.\cite{A72-amoroso2024dataset} curated and manually labeled a dataset of 378 Dockerized open-source microservice projects from an initial pool of 389,559 candidates. The dataset includes project metadata such as contributors, dependencies, and application purposes, supporting future empirical research on microservice practices and evolution. Lercher et al. \cite{lercher2024managing} investigated API evolution in microservice systems through interviews with practitioners, identifying six strategies and six challenges in handling REST API changes. They proposed automated mechanisms for extracting structural and behavioral API changes and notifying consumer teams. Their findings highlight the sociotechnical complexities of API evolution in microservices, complementing prior work on migration and evolvability. Lubas et al. \cite{lubas2025microservice} mined GitHub to curate two datasets: one of 553 microservice applications and another of 8 representative workload repositories. Their analysis provides insights into programming languages, frameworks, and architectural components used in practice, supporting benchmarking and empirical studies on workload characterization in microservice systems. }

\textcolor{black}{While patterns promote best practices to develop MSA, anti-patterns represent a class of patterns that have been perceived as best practices and commonly used but are proven to be ineffective and/or counterproductive, such as bad smells in service design \cite{A14-avritzer2025architecture}. Cerny et al. \cite{A70-cerny2023catalog} conducted a tertiary study synthesising secondary studies on poor design practices in microservice architectures – referred to as anti-patterns. The study systematically identifies recurring anti-patterns and bad smells, classifies them into coherent categories, and summarizes existing approaches for their detection. The results offer a structured reference for developers to design higher-quality microservice systems and for researchers seeking to assess system quality through anti-patterns.  Tighilt et al. \cite{tighilt2023maintenance} proposed a specification and detection approach for microservice anti-patterns. Their study defined formal detection rules, validated them on open-source systems, and demonstrated their utility for supporting maintenance and quality improvement. This work goes beyond descriptive taxonomies by providing automated mechanisms for identifying recurring bad practices. In the context of repository-based knowledge mining approaches that investigate code and architectural-centric artifacts available on GitHub such as \cite{A2-michael2023empirical, A10-daniel2023towards, A66-genfer2025understanding, A69-bacchiega2022microservices}, Beahan et al. \cite{A71-beahan2025characterizing} empirically analyzed security vulnerabilities in microservice-based systems by examining 30 open-source projects using multiple detection tools. The results indicate that vulnerabilities primarily arise from application code, dependencies, and container configurations. While microservices do not introduce fundamentally new vulnerability types, the findings of the study highlight the need for stronger practices in dependency management, container configuration, and secure data handling.}

\subsubsection{Analyzing Issues for Evolving Systems to MSA}
\textcolor{black}{In recent years, research and practices on the evolution of microservices systems have gained significant attention based on analysis of methods and techniques that enable a systematic evolution of MSAs \cite{A40-cerny2025analyzing, A76-nogueira2024insights}, service extraction from legacies \cite{A77-mohottige2025reengineering, A78-abdellatif2025identifying}, and identifying the motivations and challenges for legacy migration to MSA \cite{A2-michael2023empirical}. For example, Nogueira et al \cite{A76-nogueira2024insights} investigated a practitioner-led migration from monolithic legacy systems to microservice architectures, focusing on motivations, activities, and data consistency strategies. The study conducted a questionnaire-based survey with 53 participants and highlighted that while technical benefits like scalability and maintenance drive adoption, significant challenges remain in testing, monitoring, and managing decentralized data consistency. Beyond the practitioners’ survey-based studies  to enable and enhance legacy migration to service-driven systems, Abdellatif et al \cite{A78-abdellatif2025identifying} evaluated the transition of legacy systems to service-oriented architectures by comparing academic and industrial service identification approaches. The comparison identified gaps between the two. To bridge this gap, the authors introduce ServiceMiner, a bottom-up identification approach that utilizes source-code analysis and structural patterns to detect reusable functionalities when documentation is unavailable. Several recently published empirical studies, such as \cite{A2-michael2023empirical, A66-genfer2025understanding, A76-nogueira2024insights} have focused on experimental analysis, case studies, and practitioners’ feedback to understand the processes, motivations, and challenges related to legacy system migration in general and service extraction in particular. The results of these studies provide empirically-grounded recommendations and guidelines related to service design, maintainability and scalability as the prime motivations for the enterprise-scale adoption of MSAs.}

\subsection{Empirical Studies on Issues in Microservices Systems and Software Systems}
\label{empiricalstudiesonissues}
\subsubsection{\textcolor{black}{Detecting Bad Smells and Performance Issues}}
\textcolor{black}{Code smells and architectural smells, often collectively referred to as bad smells, represent an anti-pattern that reflect symptoms of poorly designed microservice system that decrease code understandability and increase effort for maintainability [70, 14]. Bad smells represent low-level design problems and poor coding decisions that often indicate the presence of anti-patterns with a negative impact on software quality \cite{A14-avritzer2025architecture}. In absence of a systematic process and tool support for automation, smell detection can be a laborious task that is time-consuming and error-prone. In the context of automating the detection of bad smells, Monsef et al. \cite{A79-monsef2025detecting} have proposed an automated detection approach using graph neural networks, modeling microservices as graphs to identify four architectural anti-patterns. To mitigate data scarcity, large language models are employed to generate and augment architectural graph datasets. In addition, a number of other prevalent issues in microservices systems, such as faults, bugs, performance issues, and service decomposition problems, are detailed in \cite{A15-zhang2024trace, A80-wizenty2023model}. More specifically, Zhang et al.\cite{A15-zhang2024trace} have proposed an approach namely TraceContrast that applies a trace-based root cause localization to address the complexity of microservice failures by exploiting fine-grained distributed traces. Empirical evaluation on a standard microservice benchmark demonstrates that TraceContrast achieves significantly higher accuracy than existing methods at both multidimensional and instance-level localization of microservice faults. The work by Wizenty et al. \cite{A80-wizenty2023model} tackles the challenge of automated detection of security bad smells and apply refactoring of microservice systems using a model-driven detection approach. }

% \subsubsection{Smells \& Performance Issues in MSA Systems}
% Code smells and architectural smells (a.k.a bad smells) are often synonyms as a microservices anti-pattern reflecting symptoms of poorly designed microservices that decrease code understandability and increase effort for maintainability \cite{2019Microservices, taibi2018definition}. According to practitioners' suggestions and industrial case studies, detecting bad smells in microservices systems is critical for large-scale microservice systems \cite{taibi2018definition}. Walker \textit{et al}. \cite{C-2020Automated} provided tool support for the automatic detection of bad smells in microservices systems, and the tool MSANose can detect up to eleven microservices specific bad smells within microservices applications using bytecode and/or source code analysis throughout the development process or even before its deployment to production. In addition, a number of other prevalent issues in microservices systems, such as faults, bugs, performance issues, and service decomposition problems, are detailed in \cite{ZhouIEEE, WuNoms, matias2020determining}. More specifically, Wu \textit{et al}. \cite{WuNoms} proposed a solution named Microservice Root Cause Analysis (MicroRCA), which works by inferring the root causes of performance issues by correlating application performance symptoms with corresponding system resource utilization. Their proposed solution MicroRCA can addresses performance related issues in microservices systems by analyzing resource utilization and throughput of the services.

\subsubsection{Taxonomies of Issues \& Faults in Software Systems}

\textcolor{black}{While exploring issues from the microservices system point of view, it is important not to overlook the most recent taxonomies, empirical studies, and proposed solutions that address a multitude of issues, errors, and faults in non-MSA systems such as, deep learning-based bug classification \cite{Humbatova20} and application build systems \cite{A45-lou2020understanding}. In particular, a deep learning approach to classify bugs for code inspection \cite{Humbatova20} and the types of build issues, their symptoms, and fix patterns \cite{A45-lou2020understanding} also inspired our work on microservices issues, causes, and solutions. From the MSA perspective, a recently conducted study \cite{A15-zhang2024trace} focuses on localization of root cause analysis to resolve the issues related to microservice bad smells. Our proposed research draws inspiration from empirically derived taxonomies \cite{Humbatova20, A45-lou2020understanding} and goes beyond issue categorization to investigate their causes and proposed solutions as resolution strategies to fix multi-faceted issues related to Security, Testing, and Configuration that impact architecting and implementing microservices systems.}
%While exploring issues from the microservices system point of view, it is important not to overlook the most recent taxonomies, empirical studies, and proposed solutions that address a multitude of issues, errors, and faults in non-MSA systems such as, deep learning \cite{Humbatova20} and application build systems \cite{lou2020}. In particular, a taxonomy of the types of faults in deep learning systems \cite{Humbatova20} and the types of build issues, their symptoms, and fix patterns \cite{lou2020} inspired our work on microservices issues, causes, and solutions. From the MSA perspective, a recently conducted study \cite{ZhouIEEE} identified typical faults occurring in microservices systems, practices of service debugging, and challenges faced by developers while addressing these faults. For example, a fault such as ``transactional service failure'' is due to overloaded requests to a third-party (payment gateway) service, ultimately leading to denial of service issues. Our proposed research draws inspiration from empirically derived taxonomies \cite{Humbatova20, lou2020} and goes beyond issue categorization to investigate their causes and proposed solutions as resolution strategies to fix multi-faceted issues related to Security, Testing, and Configuration that impact architecting and implementing microservices systems. 

\subsection{Conclusive Summary}
\label{conclusivesummary}

\textcolor{black}{The studies reported in \cite{A2-michael2023empirical, A66-genfer2025understanding, A70-cerny2023catalog, A76-nogueira2024insights} are grounded in the empirical analysis of microservices systems to identify a multitude of issues, such as faults, bad smells, and performance issues faced by practitioners while designing, developing, and deploying microservices systems. To complement empiricism in microservices research and development, our proposed study mined the social coding platform (i.e., 15 open-source microservices systems on GitHub) and identified issues faced by developers with improvement and validation by microservices practitioners. While there is work on mining reusable knowledge \cite{A10-daniel2023towards, A67-soldani2021mutosca} - patterns and best practices – from software repositories to develop microservices systems, there is no research on mining knowledge, overseeing the broader microservices system development life cycle to streamline the plethora of issues, their causes and fixing strategies. Our study primarily focuses on survey-driven validation of the issues, causes, and solutions of microservices systems by practitioners, and complements the body of research comprising some of the recent empirical studies on evolvability \cite{A77-mohottige2025reengineering, A78-abdellatif2025identifying}, migration \cite{A40-cerny2025analyzing}, and bad smells \cite{A15-zhang2024trace} of microservices systems.}

%The studies reported in \cite{C-2020Automated, ZhouIEEE, WuNoms} are grounded in the empirical analysis of microservices systems to identify a multitude of issues, such as faults, bad smells, and performance issues faced by practitioners while designing, developing, and deploying microservices systems. To complement empiricism in microservices research and development, our proposed study mined the social coding platform (i.e., 15 open-source microservices systems on GitHub) and identified issues faced by developers with improvement and validation by microservices practitioners. While there is work on mining reusable knowledge \cite{1-marquez2018actual, R3} - patterns and best practices – from software repositories to develop microservices systems, there is no research on mining knowledge, overseeing the broader microservices system development life cycle to streamline the plethora of issues, their causes and fixing strategies. Our study primarily focuses on survey-driven validation of the issues, causes, and solutions of microservices systems by practitioners, and complements the body of research comprising some of the recent industrial studies on evolvability \cite{assuring2019}, migration \cite{2020Does}, and debugging \cite{ZhouIEEE} of microservices systems.

We provide a comparative analysis of the most relevant existing research and our proposed study in Table \ref{tab:Comparativeanalysis} based on (i) focus of investigation: specific aspects to be empirically investigated, (ii) research method: specific approaches employed for investigation, (iii) number of issues: specific number of the identified issues, and (iv) method for evaluation: specific method for evaluation. Additionally, the study reference, year of publication, and contribution of the research study are provided as complementary information for a fine-grained presentation of the results. \textcolor{black}{For example, an empirical study reported in \cite{A14-avritzer2025architecture} published in 2025 focuses on the detection of issues and anti-patterns of microservice performance by adopting metric-based analysis that investigates the call graphs between microservices. The main contribution of this study is to a tool-supported analysis of performance anti-patterns and issues in microservice systems.}

{\renewcommand{\arraystretch}{1}
\begin{table*}[t]
\centering
\scriptsize
\caption{Comparative analysis and summary of existing vs. proposed research}
\label{tab:Comparativeanalysis}
\begin{tabular}{|p{1.5cm}|p{1.5cm}|p{4.5cm}|p{4.5cm}|p{4cm}|}
\hline
\rowcolor[HTML]{D9D9D9} 
\textbf{Reference} & \textbf{Year} & \textbf{Focus of Investigation} & \textbf{Research Method} & \textbf{\# of Issues}\\ \hline

\textcolor{black}{\cite{A14-avritzer2025architecture}}&	\textcolor{black}{2025}&	\textcolor{black}{Detection of issues and anti-patterns of microservice performance}&	\textcolor{black}{Empirical study (metric analysis)}& 	\textcolor{black}{Service call graph}\\ \hline
\rowcolor[HTML]{E2EFD9} 
%\cite{ZhouIEEE} & 2021 & Fault analysis, Debugging & Industry survey & Not explicit mentioned\\ \hline

\multicolumn{5}{|p{\dimexpr\linewidth-2\tabcolsep-2\arrayrulewidth}|}{\cellcolor[HTML]{E2EFD9}\parbox[t]{\linewidth}{\textcolor{black}{\textbf{Contribution}: Tool-based assessment of Train Ticket benchmark, showing correlations between performance and architecture issues via service call graph analysis.}} } \\ \hline

%\multicolumn{5}{|p{\dimexpr\linewidth-2\tabcolsep-2\arrayrulewidth}|}{\cellcolor[HTML]{E2EFD9}\parbox[t]{\linewidth}{\textbf{Contribution}: An industrial survey to learn typical faults, current practice of debugging, and the challenges faced by developers for microservices systems.}} \\ \hline

\textcolor{black}{\cite{A66-genfer2025understanding}} &
\textcolor{black}{2025} &
\textcolor{black}{Identification and evaluation of security tactics for microservice systems} &
\textcolor{black}{Controlled experiments with 70 participants to identify security features} &
\textcolor{black}{Security tactics applied in an MSA system} \\ \hline

\rowcolor[HTML]{E2EFD9} 
%\cite{WuNoms} & 2020 & Performance issues & MSA application monitoring & Not explicit mentioned\\ \hline
\multicolumn{5}{|p{\dimexpr\linewidth-2\tabcolsep-2\arrayrulewidth}|}{%
\cellcolor[HTML]{E2EFD9}\parbox[t]{\linewidth}{%
\textcolor{black}{\textbf{Contribution}: Automatic generation of security models and the utilization of security-based metrics to guide software architectures through the assessment of security tactics employed within microservice systems.}%
}} \\ \hline

%\multicolumn{5}{|p{\dimexpr\linewidth-2\tabcolsep-2\arrayrulewidth}|}{\cellcolor[HTML]{E2EFD9}\parbox[t]{\linewidth}{\textbf{Contribution}: MicroRCA, as a solution to identify root causes of performance issues in MSAs by monitoring application performance and resource utilization.}} \\ \hline

%\cite{assuring2019} & 2019 & Evolvability analysis & Structured interviews & 16 types of challenges\\ \hline

\textcolor{black}{\cite{A10-daniel2023towards}} &
\textcolor{black}{2023} &
\textcolor{black}{Detection of patterns as best practices to architect microservice systems} &
\textcolor{black}{Empirical study (data mining)} &
\textcolor{black}{Architecture of open source system, i.e., MSA-based platform} \\ \hline

\rowcolor[HTML]{E2EFD9} 
\multicolumn{5}{|p{\dimexpr\linewidth-2\tabcolsep-2\arrayrulewidth}|}{%
\cellcolor[HTML]{E2EFD9}\parbox[t]{\linewidth}{%
\textcolor{black}{\textbf{Contribution}: Detection of microservice patterns and two case studies with real-world applications to evaluate the accuracy and applicability of the proposed solution.}%
}} \\ \hline

%\multicolumn{5}{|p{\dimexpr\linewidth-2\tabcolsep-2\arrayrulewidth}|}{\cellcolor[HTML]{E2EFD9}\parbox[t]{\linewidth}{\textbf{Contribution}: A qualitative study to explore the tools, metrics, patterns, and practitioners’ view on practices and challenges of microservices system evolvability.}} \\ \hline

\cite{C-2020Automated} & 2020 & Code smells & Static code analysis & 11 types code smells\\ \hline
\rowcolor[HTML]{E2EFD9} 

\multicolumn{5}{|p{\dimexpr\linewidth-2\tabcolsep-2\arrayrulewidth}|}{\cellcolor[HTML]{E2EFD9}\parbox[t]{\linewidth}{\textbf{Contribution}: MSANose tool to detect microservice-specific code smells using bytecode and/or source code analysis.}} \\ \hline

\cite{nasab2022ess} & 2023 & Security points & Repository mining & 
\\ \hline
\rowcolor[HTML]{E2EFD9} 

\multicolumn{5}{|p{\dimexpr\linewidth-2\tabcolsep-2\arrayrulewidth}|}{\cellcolor[HTML]{E2EFD9}\parbox[t]{\linewidth}{\textbf{Contribution}: An empirical study that investigates 10 GitHub projects and 306 Stack Overflow posts to identify security challenges and practices for MSA systems.}} \\ \hline

\cite{A15-zhang2024trace}& 	2024& 	Root-cause analysis and localisation of performance issues in microservice systems& 	Controlled experiments based on sequential pattern mining and spectrum analysis of microservices& 	Service interaction and coordination calls\\ \hline

\rowcolor[HTML]{E2EFD9} 
\multicolumn{5}{|p{\dimexpr\linewidth-2\tabcolsep-2\arrayrulewidth}|}{\cellcolor[HTML]{E2EFD9}\parbox[t]{\linewidth}{\textbf{Contribution}: TraceContrast is a root cause localization tool that identifies bugs by comparing failed sequences against successful ones to pinpoint the specific combination of attributes causing the error.}} \\ \hline

\cite{wu2022way} & 2022 & Microservice problems and solutions & Repository mining & 14 problems,
47 solutions\\ \hline

\rowcolor[HTML]{E2EFD9} 
\multicolumn{5}{|p{\dimexpr\linewidth-2\tabcolsep-2\arrayrulewidth}|}{\cellcolor[HTML]{E2EFD9}\parbox[t]{\linewidth}{\textbf{Contribution}: An empirical study that analyzes 17522 Stack Overflow posts to explore problems and solutions corresponding to microservice development process.}} \\ \hline

\cite{zhong2022impacts} & 2022 & Architectural smells & Interviews, Repository mining & 6 types of architectural smells\\ \hline

\rowcolor[HTML]{E2EFD9} 
\multicolumn{5}{|p{\dimexpr\linewidth-2\tabcolsep-2\arrayrulewidth}|}{\cellcolor[HTML]{E2EFD9}\parbox[t]{\linewidth}{\textbf{Contribution}: An industrial case study that collects repository data and practitioners' views on the impacts, causes, and solutions of architectural smells in MSA.}} \\ \hline

\cite{tighilt2023maintenance} & 2023 & Antipattern specification and detection & Repository mining, Static analysis & Catalog of antipatterns \\ \hline
\rowcolor[HTML]{E2EFD9} 
\multicolumn{5}{|p{\dimexpr\linewidth-2\tabcolsep-2\arrayrulewidth}|}{\cellcolor[HTML]{E2EFD9}\parbox[t]{\linewidth}{\textbf{Contribution}: This study formalizes the specification of microservice antipatterns and provides automated detection rules validated on open-source systems to support maintenance and quality assurance.}} \\ \hline

\cite{amoroso2024dataset} & 2024 & Dataset of OSS microservice projects & Repository mining, Manual curation & 378 projects \\ \hline
\rowcolor[HTML]{E2EFD9} 
\multicolumn{5}{|p{\dimexpr\linewidth-2\tabcolsep-2\arrayrulewidth}|}{\cellcolor[HTML]{E2EFD9}\parbox[t]{\linewidth}{\textbf{Contribution}: A curated dataset of 378 Dockerized open-source microservice projects.}} \\ \hline

\cite{lercher2024managing} & 2024 & API evolution strategies, challenges, and automation & Practitioner interviews (17), Proposed automation & 6 strategies, 6 challenges \\ \hline
\rowcolor[HTML]{E2EFD9} 
\multicolumn{5}{|p{\dimexpr\linewidth-2\tabcolsep-2\arrayrulewidth}|}{\cellcolor[HTML]{E2EFD9}\parbox[t]{\linewidth}{\textbf{Contribution}: An empirical study on strategies and pitfalls in microservice API evolution, with proposals for automated change extraction and notification.}} \\ \hline

\cite{lubas2025microservice} & 2025 & Dataset of microservice applications and workloads & Repository mining, Empirical analysis & 553 applications, 8 workloads \\ \hline
\rowcolor[HTML]{E2EFD9} 
\multicolumn{5}{|p{\dimexpr\linewidth-2\tabcolsep-2\arrayrulewidth}|}{\cellcolor[HTML]{E2EFD9}\parbox[t]{\linewidth}{\textbf{Contribution}: Curated two GitHub-based datasets of 553 microservice applications and 8 workload repositories, offering insights into languages, frameworks, and components for benchmarking and workload studies.}} \\ \hline

Our previous study \cite{waseem2021nature} & 2022 & Issues and Causes in microservices systems & Repository mining & 138 types of issues, 109 types of causes\\ \hline

\rowcolor[HTML]{E2EFD9} 
\multicolumn{5}{|p{\dimexpr\linewidth-2\tabcolsep-2\arrayrulewidth}|}{\cellcolor[HTML]{E2EFD9}\parbox[t]{\linewidth}{\textbf{Contribution}: 
This study presents a taxonomy of issues, and identifies and maps to their causes in microservices systems.}} \\ \hline

\rowcolor[HTML]{DEEAF6} 
\textcolor{black}{This study} &
\textcolor{black}{2025} &
\textcolor{black}{Issues, Causes, and Solutions in microservices systems} &
\textcolor{black}{Repository mining (2,641 issues), Structured interviews (15 interviews), Practitioners’ survey (150 responses)} &
\textcolor{black}{402 types of issues, 228 types of causes, 177 types of solutions} \\ \hline

\multicolumn{5}{|p{\dimexpr\linewidth-2\tabcolsep-2\arrayrulewidth}|}{%
\cellcolor[HTML]{E2EFD9}\parbox[t]{\linewidth}{%
\textcolor{black}{\textbf{Contribution}: This study presents empirically validated taxonomies for issues, causes, and solutions in microservices systems, maps the relationships between them, and provides the dataset for further research.}%
}} \\ \hline

\end{tabular}
\end{table*}}

\section{Conclusions}
\label{sec:conclusion}
This paper empirically investigates the issues, causes, and solutions of microservices systems by employing a mixed-methods approach, combining 2,641 GitHub issues from 15 open-source microservices systems, 15 interviews, and an online survey completed by 150 practitioners. The primary contribution of this work is the empirically grounded taxonomies of \emph{Issues}, \emph{Causes}, and \emph{Solutions} in microservices systems, together with the mapping between them. Overviewing our results, the major Issue Categories are Technical Debt, Continuous Integration and Delivery, and Exception Handling, while the most common Cause Categories include General Programming Errors, Missing Features and Artifacts, and Invalid Configuration and Communication. These issues are mainly addressed through Fixing, Adding, and Modifying artifacts.

Finally, by linking Issue Categories to recurring Cause Categories and Solution Categories, our results provide an evidence base for both researchers and practitioners to understand where microservices problems concentrate and how they are typically addressed in OSS practice. Directions for future research are summarised in Section~\ref{researchagenda}.

\section*{Data availability and ethics statement}
\textcolor{black}{The replication package (Ref.~\citep{replpack}) contains the study material and aggregated data used in this paper (e.g., survey questionnaire, interview and survey supporting artifacts). Participation in the interviews and survey was voluntary and consent-based. We applied data minimization: the instruments do not collect direct identifiers (e.g., names, email addresses, IP addresses, or organization names), and results are reported only in aggregated or anonymized form.}

\appendix
% \appendices
% See Table \ref{tab:Abbreviations}
\label{sec:Appendix}

\section*{Acknowledgments}
This work is partially sponsored by the National Natural Science Foundation of China (NSFC) with Grant Nos. 92582203 and 62172311, and the Major Science and Technology Project of Hubei Province under Grant No. 2024BAA008. The numerical calculations in this paper have been done on the supercomputing system in the Supercomputing Center of Wuhan University. The authors also thank the participants of the interviews and online survey.

\section*{Declaration of AI Assistance}
During the preparation of this work, the author(s) used ChatGPT to refine grammar, improve sentence structure, and resolve formatting issues. After utilizing this tool, the author(s) thoroughly reviewed and edited the content as needed, taking full responsibility for the final publication.

\printcredits

% ---- References ----
\bibliographystyle{elsarticle-num} % CAS numeric style
\bibliography{References}

\balance
\end{sloppypar}
\end{document}